\documentclass[12pt]{article}
\usepackage{amssymb, amsmath, amsthm, amsfonts, latexsym, mathabx}
\usepackage{multirow} \usepackage{epsfig} \usepackage{psfrag} 
\usepackage{color}    \usepackage{xcolor} \usepackage{array} 
\usepackage[pdftex,bookmarks, colorlinks, citecolor =blue, breaklinks]{hyperref}
\usepackage{enumitem}\usepackage{bm}

\usepackage[version=4]{mhchem}
\usepackage{siunitx}
\usepackage{longtable,tabularx}

\newtheorem{teor}{Theorem}     
\newtheorem{prop}{Proposition} 
\newtheorem{lema}{Lemma}       
\newtheorem{corol}{Corollary} 
 
\newtheorem{rema}{Remark}

\usepackage{graphicx}
\def\be{\begin{equation}}  \def\ee{\end{equation}}
\def\bea{\begin{eqnarray}} \def\eea{\end{eqnarray}}

\def\Sset{\mathbb{S}}

\def\Rset{\mathbb{R}}  
\def\Zset{\mathbb{Z}}

\def\Ccal{\mathcal{C}}

\def\Fcal{\mathcal{F}}
\def\Hcal{\mathcal{H}}
\def\Kcal{\mathcal{K}}

\def\Ocal{\mathcal{O}}

   \def\eps{\varepsilon}
      \def\vp{\varphi}
          \def\al{\alpha}

\def\bfi{{\bf i}}
\def\bfn{{\bf n}}

\def\bfu{{\bf u}}

\def\bfM{{\bf M}}
\def\bfF{{\bf F}}
\def\hPhi{\hat{\Phi}}
\def\hphi{\hat{\phi}}
\def\bPhi{\bar{\Phi}}
\def\bphi{\bar{\phi}}
\def\tPhi{\tilde{\Phi}}
\def\tphi{\tilde{\phi}}
\def\vth{\vartheta}

  \newcommand{\fin}{\hfill $\Box$}

\begin{document}
\begin{center}
\noindent {\bf\Large
Attitude and orbit coupling of planar helio-stable solar sails
}
\end{center}

\begin{center}
{\large Narc\'{\i}s Miguel and Camilla Colombo} 
\vspace*{1mm}

{\footnotesize
\noindent Dipartimento di Scienze e Tecnologie Aerospaziali\\
Politecnico di Milano\\
Via La Massa 34, 20156, Milano, Italia
}\\
 
\tt      {narcis.miguel@polimi.it}, 
\tt      {camilla.colombo@polimi.it} 
\end{center}

\begin{center}\today\end{center}

\begin{abstract}
The coupled attitude and orbit dynamics of solar sails is studied. The shape of
the sail is a simplified quasi-rhombic-pyramid that provides the structure
helio-stablility properties. After adimensionalisation, the system is put in the
form of a fast-slow dynamical system where the different time scales are
explicitely related to the physical parameters of the system. The orientation
of the body frame with respect to the inertial orbit frame is a fast phase that
can be averaged out. This gives rise to a simplified formulation that only
consists of the orbit dynamics perturbed by a flat sail with fixed attitude
perpendicular to the direction of the sunlight. The results are exemplified
using numerical simulations.
\end{abstract}

\tableofcontents

\section*{Nomenclature}

{\renewcommand\arraystretch{1.0} 
\noindent\begin{longtable}{@{}l @{\quad=\quad} l@{}}
$a$ & semi-major axis, km\\
$\alpha$ & aperture angle of the sail, deg or rad\\
$A_s$    & area of the panels, m$^2$\\
$\beta$  & frequency of the Sun-pointing direction, deg or rad\\
$C$ & inertia moment around the $\zeta$ axis, kg m$^2$\\
$d$      & center of mass-center of pressure offset, m\\
$e$ & eccentricity, [-]\\
$\eps$ & ratio between characteristic time scales, [-]\\
$\Fcal_b$ & body frame\\
$\Fcal_I$ & Earth centered inertial frame\\
$\bfF$ & force vector, N\\
$\gamma$ & difference $\omega-\Omega$, deg or rad\\
$h$      & height of the panels, m\\
$\bfi$ & unit vectors of a basis\\
$J_2$ & adimensional $J_2$ coefficient, [-]\\
$\lambda$ & angle between Sun position and $\bfi_x$, rad or deg\\
$m_s/2$  & mass of the panels, kg\\
$m_b$    & mass of the bus, kg\\
$\bfM, M$ & torque vector, component of torque vector, Nm\\
$\mu$ & earth's mass parameter, km$^3$/s$^2$\\
$\bfn$ & normal vectors to panels\\
$n$ & mean motion, deg or rad/s\\ 
$\Omega$ & right ascension of the ascending node, deg or rad\\
$\omega$ & argument of the perigee, deg or rad\\
$P$      & sail panel\\
$p_{\rm SR}$ & solar radiation pressure at 1 AU, N/m$^2$\\
$r$ & magnitude of the position vector of the spacecraft, km\\
$R$ & Earth's radius, km\\
$\Sigma$ & surface of section\\
$\theta$ & true anomaly, deg or rad\\
$\bfu$ & unit vector\\
$v$ & velocity components and its magnitude, km/s\\
$\vp$ & Euler angle, rad or deg\\
$\Phi$ & angular velocity of the attitude, rad or deg/s\\
$\phi$ & difference $\vp - \lambda$, rad or deg\\
$w$      & width of the panels, m\\
$\xi,\nu,\zeta$ & coordinates of $\Fcal_b$\\
$x, y, z$ & coordinates of $\Fcal_I$\\
$X$ & characteristic function of an interval\\
\multicolumn{2}{@{}l}{Subscripts}\\
$\pm$ & that refers to panels $+$ or $-$\\
$\odot$ & that refers to Sun\\
${\rm SRP}$  & that refers to solar radiation pressure\\
${\rm GG}$  & that refers to gravity gradient\\ 
$\xi,\nu,\zeta$ & in the direction of, in $\Fcal_b$\\
$x, y, z$ & in the direction of, in $\Fcal_I$
\end{longtable}}

\section{Introduction}\label{sect:intro}

Solar sails are a low-thrust propulsion system that takes the advantage of
Solar Radiation Pressure (SRP) to accelerate a probe using a highly reflective
surface. Even though the acceleration due to SRP is smaller than that achieved
by traditional thrusters, it is continuous and enables possibilities that
range from trajectory design to attitude control.

The effects of the SRP acceleration have been widely studied in the literature.
In works that deal with the effect of SRP acceleration on the dynamics of
spacecraft with a solar sail, its attitude is usually assumed to be fixed with
respect to the direction of sunlight. This assumption simplifies the equations
of motion and, in some cases, allows to deal with SRP acceleration as a
perturbative effect and even to write the system as if it had Hamiltonian
structure~\cite{JCFJ16}. But maintaining the attitude fixed with respect to any
direction, in particular that of sunlight, requires attitude control.

For the specific case of works that deal with flat sails whose surface is
(theoretically) assumed to be always perpendicular to the sunlight direction,
it is expected that in practice the attitude oscillates close to this direction
if an appropriate center of Mass - center of Pressure Offset (that in this
paper is referred to as MPO) is assumed.

In this work we investigate the possibility of considering a different
structure that consist of a number of flat panels oriented in a way that
cancels out some torque components and makes the Sun-pointing attitude stable.
The motivation is to foster the use of such sails as passive deorbiting devices,
that can be employed as end-of-life disposals that would reduce the attitude
control requirements. Here the term {\sl passive} is understood in the sense of
the so-called passive deorbiting strategy, as defined in~\cite{CBF16}, that
consists of using the idea of deorbiting ``outwards" on an elliptical orbit:
the increase of the eccentricity of the orbit causes the perigee radius to
progressively decrease. As justified in~\cite{LCI12, LCI13}, this can be
attained by orienting the sail panel always perpendicular to the sunlight
direction.  Other deorbiting strategies are the so-called ``active" approaches
(as opposed to ``passive") that consist of changing between maximal and minimal
SRP acceleration along the motion. This was first studied in~\cite{BT06}, where
the authors suggested maximizing the SRP acceleration when travelling towards
the Sun and minimizing it when travelling away from the Sun. A refined version
of this approach is to maximize (resp.  minimize) the SRP acceleration when the
first averaged variational equation of the eccentricity is positive (resp.
negative), see~\cite{CMG19}. This refinement reduces the number of attitude
change maneuvers from twice per turn around the Earth to twice per year,
see~\cite{CBF16}.

Apart from considering an adequate MPO, helio-stability can be enhanced by
means of a Quasi-Rhombic Pyramid (QRP) shape. This idea was first introduced
in~\cite{CHMR13}. The suggested structure consists of 4 reflective panels that
resemble the shape of a pyramid. In case the center line of the pyramid is
oriented close enough to the sunlight direction and has an adequate MPO, this
structure cancels out, on average along the motion, the components of the
acceleration in other directions.  For example, in~\cite{FCH16} the authors
study the linear stability of the Sun-pointing attitude; and this stability can
be further enhanced by assuming a moderate spin around the adequate axis of
inertia as proposed in~\cite{FHC17}.

A simplified version of the QRP that consists of a single triangular flat panel
and an appropriately positioned payload of the spacecraft was considered
in~\cite{CHM14}, and later exploited in~\cite{HC17} to design new periodic
orbits in the circular restricted three body problem. The suggested spacecraft
was shown to have undamped conservative oscillatory dynamics around the
Sun-pointing direction.

Despite the contributions~\cite{CHMR13, CHM14, FCH16, FHC17, HC17} provide
satisfactory results, there is, to the authors' knowledge, a lack of
understanding of 
\begin{enumerate}
	\item The attitude dynamics, especially close to the Sun-pointing
attitude, and 
	\item The attitude and orbit coupling: especially whether they can be
analytically separated taking into account the fact that these two components
have two characteristic time scales.
\end{enumerate}

These two questions are addressed in this paper by considering a sail structure
in between the single-panel considered in~\cite{CHM14, HC17} and the full
QRP~\cite{CHMR13,FCH16,FHC17} whose orbit dynamics (without any attitude
control) evolves strictly on the ecliptic plane if adequate initial attitude
conditions are chosen.  The structure consists of two panels with variable
aperture and variable position of the payload, as introduced in~\cite{MC18}.
The acceleration due to SRP assumes that photons are partially specularly
reflected and partially absorbed.  Building on previous contributions on the
usage of the SRP effect for the design of end-of-life disposals (see, e.g.,
\cite{LCI12,CLI12}), here the considered orbital dynamics are the
$J_2$-perturbed two-body problem\footnote{This is the motion of an artificial
satellite around an oblate planet keeping only the $J_2$ term of the expansion
of the perturbing potential in spherical harmonics.} always perturbed by the
SRP acceleration, that is, the spacecraft is considered to be always
illuminated so the effect of eclipses are not taken into account.  The SRP
acceleration depends on the attitude of the spacecraft. The attitude dynamics
are assumed to solely happen around an axis perpendicular to the ecliptic
plane, and to be  perturbed by the SRP and gravity-gradient torques.  The
effect of atmospheric drag is not taken into account.  Hence, this study is
relevant to higher Low Earth Orbits (LEO) (i.e. with altitude 700/800 km and
above). 

As numerically demonstrated in~\cite{MC18}, this structure has the advantage
that under some hypotheses related to the geometry of the sail -that are
discussed later in this contribution-, the dynamics close to the Sun-pointing
direction is close to a mathematical pendulum and hence the motion has a
quasi-integral (adiabatic invariant) of motion that is almost preserved over
some time interval. Also, the length of this time interval depends on the ratio
between time scales as usually described by theorems concerning the accuracy of
the averaging method.\\

The organization and presentation of the main results of this paper are as
follows. First of all, \S~\ref{sect:model} is devoted to the review of the
geometry of the spacecraft under consideration and to the derivation of the
equations of motion. Despite having two characteristic time scales, the
equations are not written in the form of fast-slow systems. The equations are
put as a fast-slow system of differential equations, where the variables that
evolve in different time scales are splitted, and related via a physical
parameter that represents the ratio between time scales that only depends on
the geometry of the spacecraft. In \S~\ref{sect:dynasp}, the dynamics of the
system are studied in the context of fast-slow systems, and this includes the
discussion of the possibilities of the separation of the motions. The system
obtained by direct averaging of the fast phase (after adequate changes of
variables) is related to the results of the averaging theorems. The section
finishes with an enumeration of the physical interpretations of the results.
The theoretical results and formulas of \S~\ref{sect:dynasp} are tested in
\S~\ref{sect:numerics} with special emphasis on the physical interpretations
just mentioned. The paper concludes in \S~\ref{sec:conclusions}, where the main
results of the contribution are summarized and different possible lines for
future research are suggested. 

\section{Model}\label{sect:model}

This section is devoted to providing the equations of motion of the planar
dynamics of a helio-stable solar sail. These are a set of differential
equations that govern the coupled attitude and orbit dynamics of the spacecraft
under consideration.  The content of \S~\ref{subsect:geom} and
\S~\ref{subsect:dyneq} is a summary of the derivation of the equations of
motion that is added for completeness.  For further details the reader is
referred to~\cite{MC18}.  The section ends by putting the equations of motion
in the context of dynamical systems with multiple time scales in
\S~\ref{subsect:prepeq}.

\subsection{Geometry of the sail structure}\label{subsect:geom}

\begin{figure}[h!]
\begin{center}
\includegraphics[width = 0.45\textwidth]{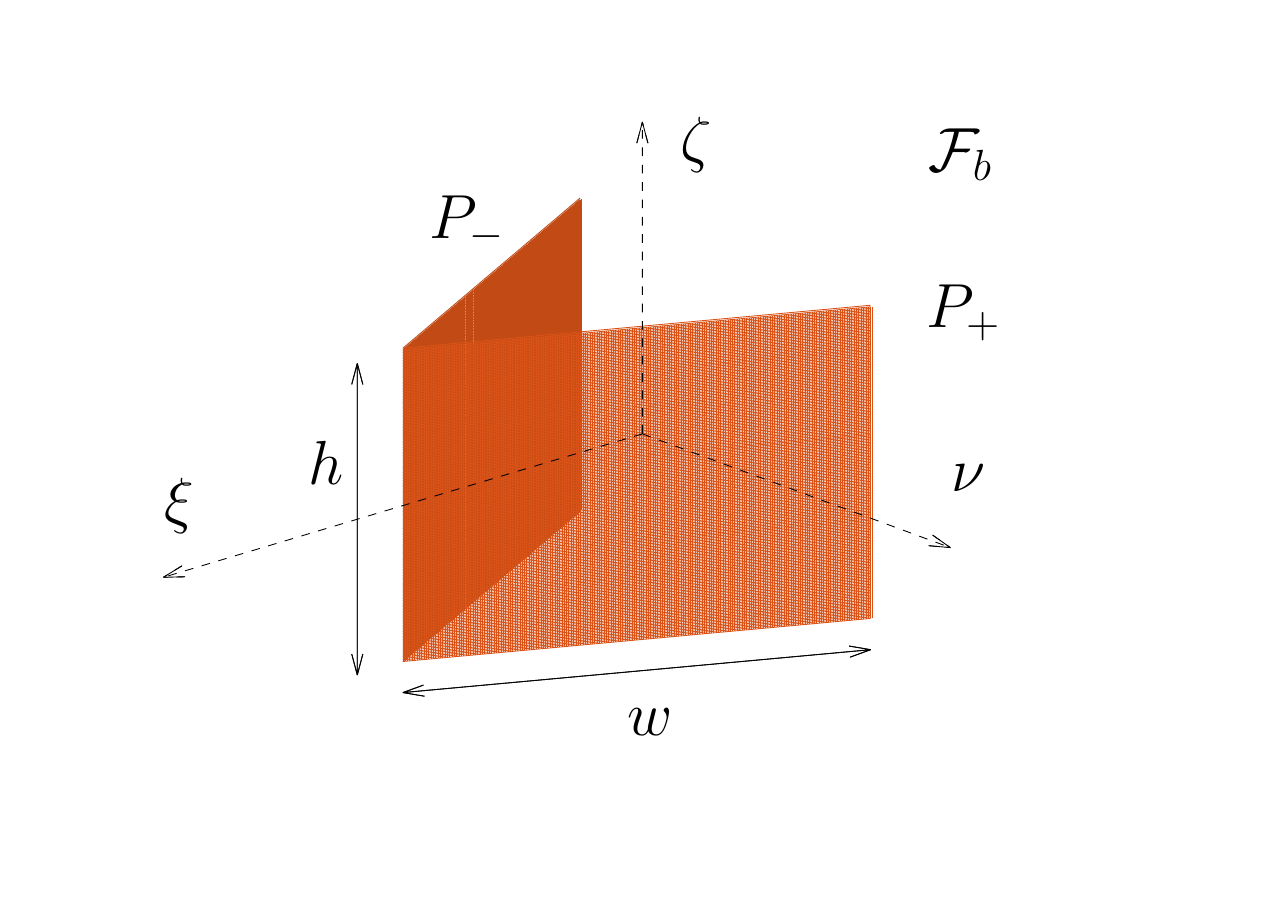}
\includegraphics[width = 0.45\textwidth]{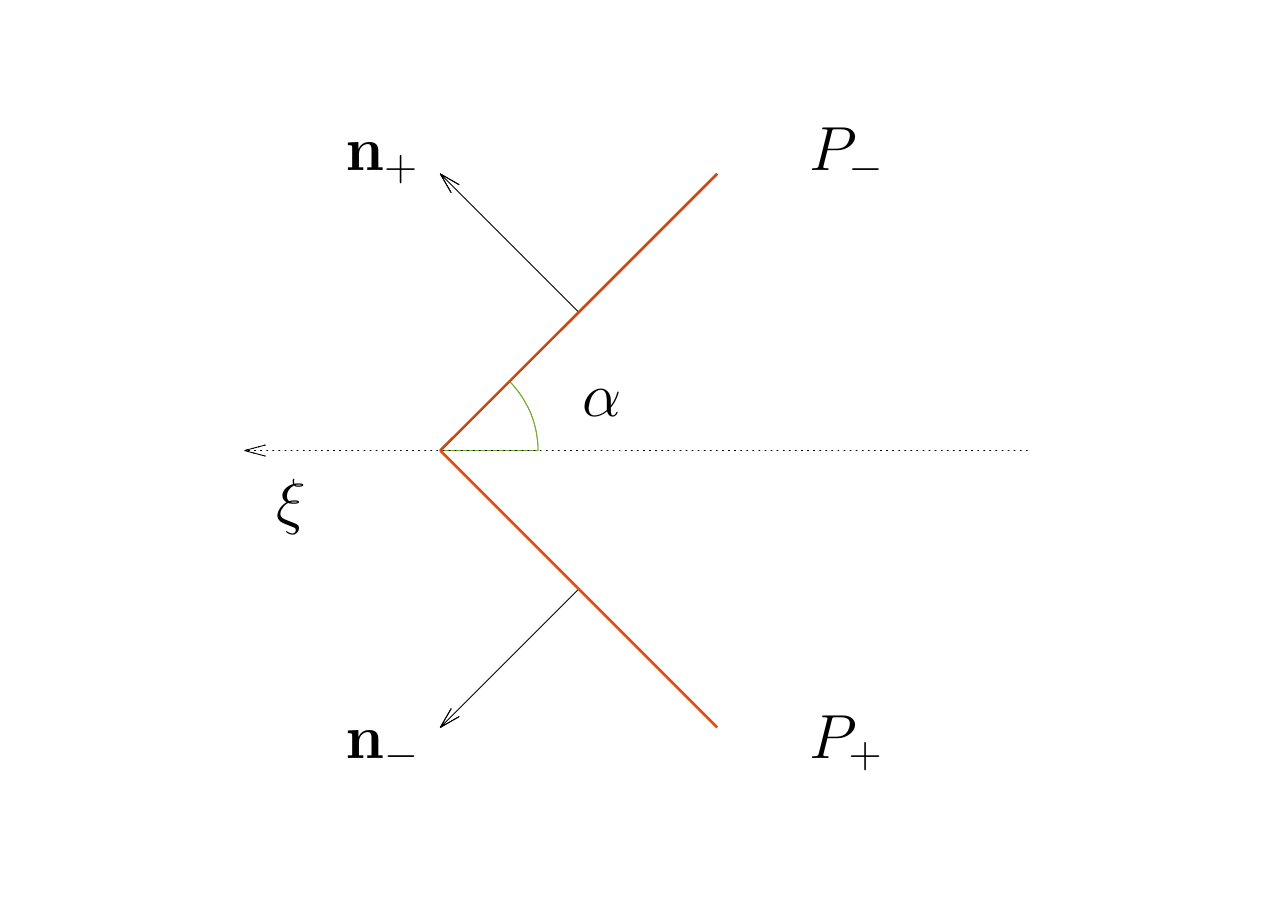}
\end{center}
\caption{Sketch of the sail structure. Left: 3D view. Right: top view.}
\label{fig:shapesail}
\end{figure}

The spacecraft under consideration consist of a payload or bus attached to two
panels forming an angle. To avoid out-of-plane motion, one is lead to consider
a simplification of a QRP~\cite{CHMR13} that consists of two panels of equal
size $P_\pm$; of height $h$, width $w$, and area $A_s=hw$. Assume that the mass
of each panel is $m_s/2$, so the mass of the whole sail structure is $m_s$. In
the left panel of Fig.~\ref{fig:shapesail} a sketch of the sail structure is
depicted. 

\begin{figure}[h!]
\begin{center}
\includegraphics[width = 0.32\textwidth]{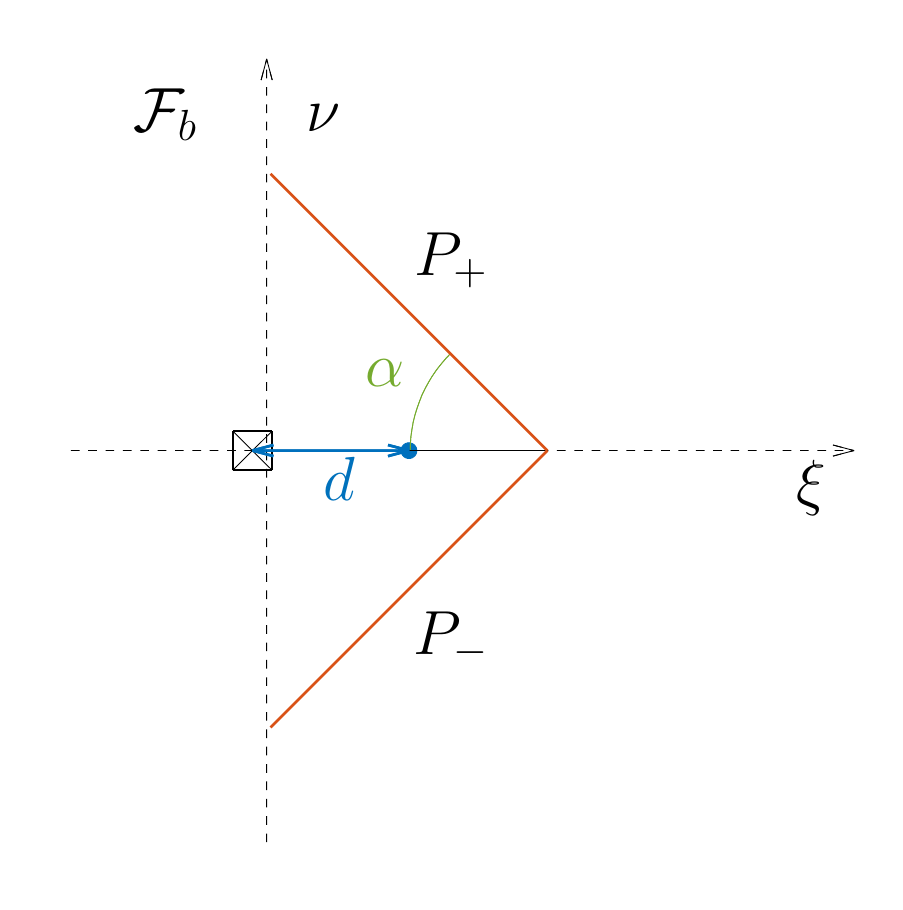}
\includegraphics[width = 0.32\textwidth]{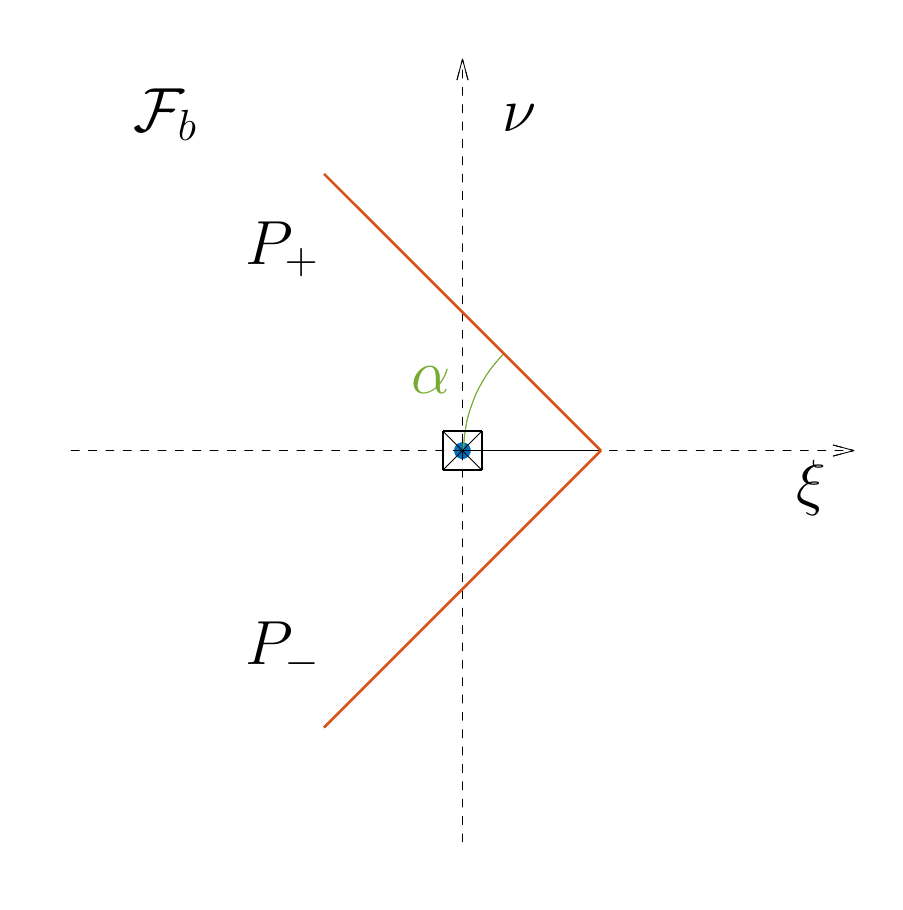}
\includegraphics[width = 0.32\textwidth]{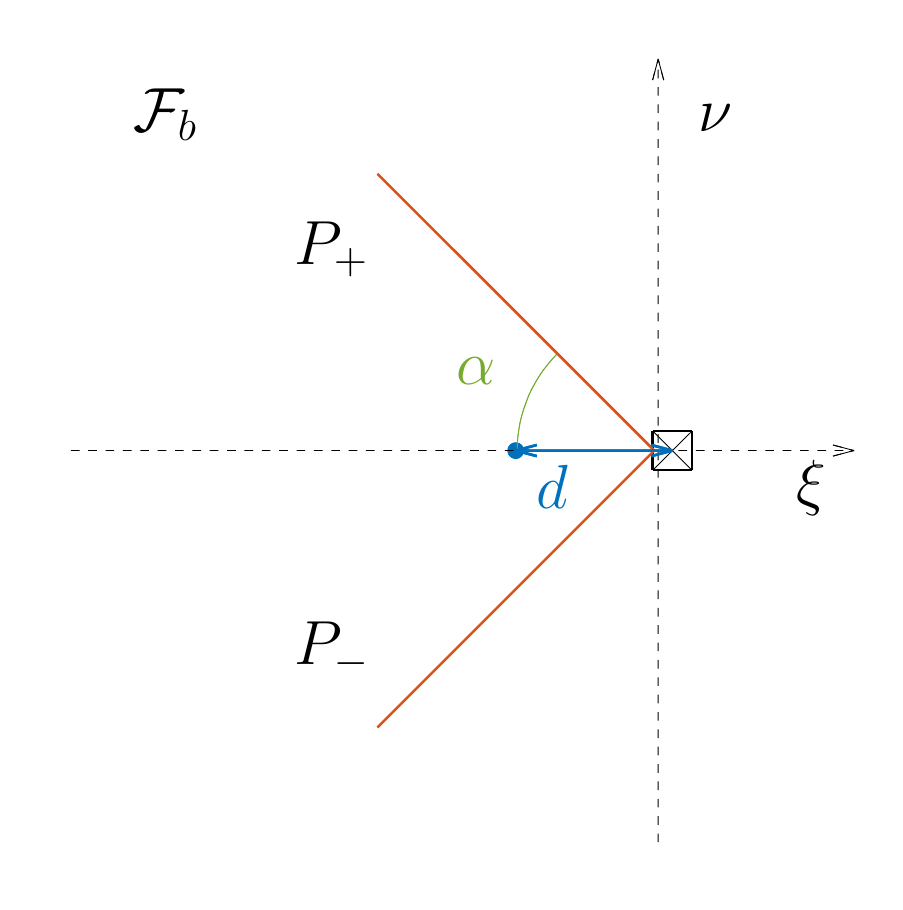}
\end{center}
\caption{Sketch of the top view of the spacecraft in $\Fcal_b$, where the bus
is depicted as a black square, and the center of mass of the sail
structure is depicted as a blue solid dot.  Left: $d < 0$. Center:
$d = 0$. Right: $d > 0$.}
\label{fig:shapesc}
\end{figure}

The attitude dynamics of the spacecraft occurs in a reference frame $\Fcal_b$
attached to it. The coordinates in this frame are referred to as $\xi,\nu$ and
$\zeta$ and the vectors of the basis $\bfi_\xi,\bfi_\nu,\bfi_\zeta$. The panels
are attached to each other along an $h$-long side, that lies on a line parallel
to the $\zeta$ axis, and they form an angle $\alpha$ with respect to the plane
$\nu = 0$. The payload, whose mass is denoted as $m_b$, is assumed to be on the
$\xi$ axis, at a distance $d$ from the center of mass of the two panels, see
Fig.~\ref{fig:shapesc}.  The parametrization of the panels is chosen so that
the center of mass of the spacecraft is at the origin of $\Fcal_b$.  The main
physical parameters of the system are: $\alpha$, the aperture angle; and $d$,
that accounts for the MPO. 

Sketches of top views of the spacecraft in $\Fcal_b$ can be seen in
Fig.~\ref{fig:shapesc}, where the bus is depicted as a black square, and the
center of mass of the sail structure is depicted as a blue solid dot,
added to visualize the parameter $d$.  The left, center and right panels are
sketches of spacecraft with $d<0$, $d=0$ and $d>0$, respectively. 

The aim of this contribution is to study the oscillatory attitude dynamics
close to the Sun-pointing direction of the spacecraft described in this
subsection. For the purposes of this article and to simplify the exposition we
considered that the back part of the panels (the side where the angle $\alpha$
is measured in Fig.~\ref{fig:shapesc}) did not produce any SRP acceleration.
In~\cite{MC19c} the attitude dynamics model is extended to take into account
the effect of the back side neglected here, and the dynamics close to the
Sun-pointing attitude is shown to be exactly the same as the one obtained in
this paper. Moreover, a numerical study of the consequences of considering
different reflectance properties is performed in~\cite{MC19c}.

\subsection{Equations of motion}\label{subsect:dyneq}

The considered planar orbit and attitude dynamics, are a coupled system of
differential equations in $(\Sset^1\times\Rset)\times\Rset^4$, where
$\Sset^1:=\Rset/(2\pi\Zset)$: orientation and angular velocity for the
attitude; and position and velocity of the spacecraft in an Earth centered
inertial reference frame $\Fcal_I$. 

Here SRP is considered to be the coupling effect between the attitude and orbit
dynamics. It is then necessary to study the attitude dynamics in relation to
the orbit dynamics that are considered to evolve in $\Fcal_I$.  The coordinates
of $\Fcal_I$ are denoted $x,y$ and $z$, and the vectors of the orthonormal
basis are denoted $\bfi_{x,y,z}$.  The vector $\bfi_x$ points towards an
arbitrarily chosen direction on the ecliptic (e.g.  J2000), and since we are
dealing with the planar problem, the vector $\bfi_z$ is parallel to
$\bfi_\zeta$, and they are also perpendicular to the ecliptic plane.  The triad
is completed by choosing $\bfi_y = \bfi_z\times \bfi_x$.

As the motion is planar the change of coordinates from $\Fcal_I$ to $\Fcal_b$
is done through $R_3(-\vp)$, where $\vp\in\Sset^1$ is an Euler angle and $R_3$
is the rotation matrix around the $z$ (and $\zeta$) axis.  The rotation matrix
$R_3$ reads, for any angle $\psi\in[0,2\pi)$,
$$
R_3(\psi) = 
\left(
\begin{array}{rr}
 \cos\psi&\sin\psi\\
-\sin\psi&\cos\psi
\end{array}
\right).
$$
The Euler attitude equations in the present situation reduce to
\begin{eqnarray}\label{eq:EulerSimp}
C\ddot{\vp} & = &M_\zeta
\end{eqnarray}
where $\bfM = (M_\xi, M_\nu, M_\zeta)$ is the
torque due to the external forces considered, and $C$ is the inertia moment
around the $\zeta$ axis in $\Fcal_b$. Denote $I_{\xi,b},I_{\nu,b},I_{\zeta,b}$
as the inertia moments of the bus. Then one can see that
\begin{eqnarray}\label{eq:inertiamom}
C & = & I_{\zeta,b} + D(\alpha,d),\qquad 
D(\alpha,d)  =  
\frac{1}{6}m_sw^2\cos^2\alpha +
\frac{d^2m_b^2(m_b+2m_s)}{(m_b+m_s)^2}.
\end{eqnarray}

\subsubsection{SRP model}\label{subsubsect:SRPmodel}

The force due to SRP exerted on each panel of the sail in $\Fcal_b$ is
considered to be modelled as~\cite{MC14}
\begin{eqnarray}\label{eq:forcesrp}
\bfF_{\rm SRP}^\pm & = & -p_{\rm SR}A_s (\bfn_\pm\cdot \bfu_\odot)
\left(
2 \eta (\bfn_\pm\cdot \bfu_\odot)\bfn_\pm + (1- \eta ) \bfu_\odot
\right),
\end{eqnarray}
where $\bfu_\odot$ is the unit vector in the Earth-Sun direction, and
$\bfn_\pm$ are the normal vectors to each panel, recall
Fig.~\ref{fig:shapesail}. Concerning the constants,  $\eta\in(0,1)$ is the
(dimensionless) reflectance of the sail and $p_{\rm SR} =
4.56\times10^{-6}\;{\rm N}/{\rm m}^2$ is the solar pressure at 1 AU which is
considered to be constant.

\subsubsection{Attitude dynamics}\label{subsubsect:attdyn}

\begin{figure}[h!]
\begin{center}
\includegraphics[width = 0.45\textwidth]{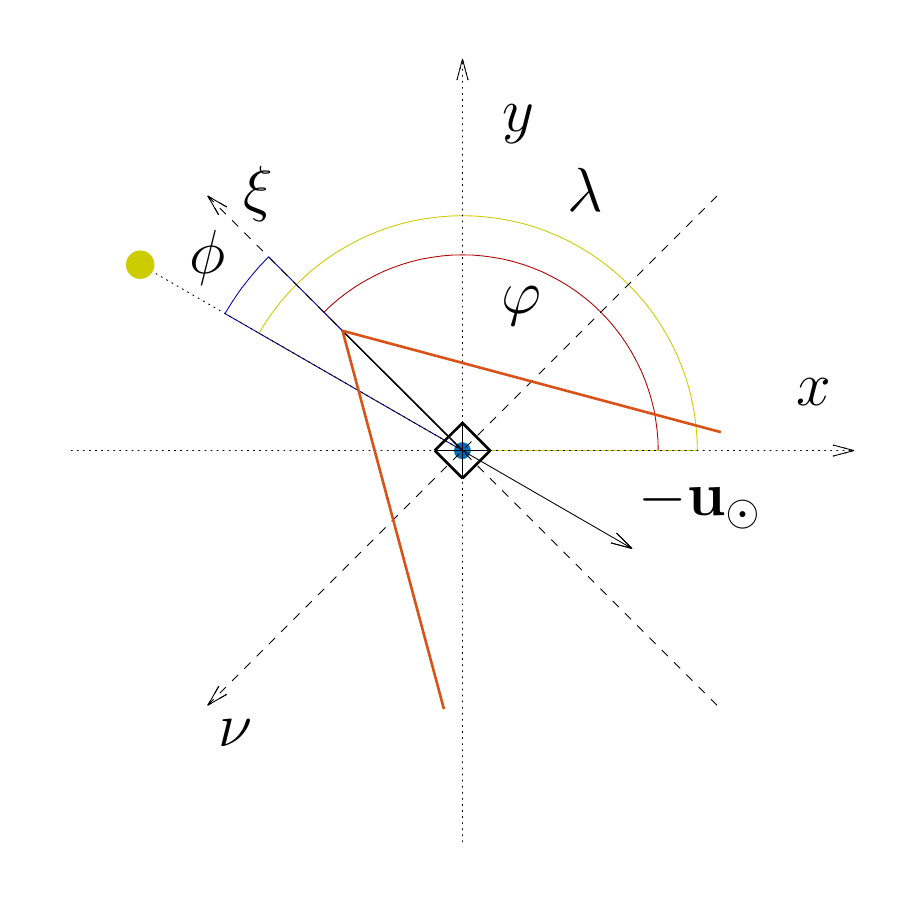}
\end{center}
\caption{
Sketch of the angles $\vp, \lambda$ and $\phi$. The angle $\vp$ informs about 
the orientation of $\Fcal_b$ relative to $\Fcal_I$, the angle $\lambda$ about
the position of the Sun in $\Fcal_I$, and $\phi=\vp-\lambda$ is the relative
orientation with respect to the position of the Sun in $\Fcal_I$.}
\label{fig:sketchangles}
\end{figure}

The effects taken into consideration are SRP and the non-symmetry of the
spacecraft, so the total torque is $\bfM = \bfM_{\rm SRP} + \bfM_{\rm GG}$, the
sum of the SRP and gravity gradient torques. Let $\lambda$ be the argument of
latitude of the apparent position of the Sun. The SRP torque has a different
representation depending on the orientation of the sail structure with respect
to the Sun, that is, it depends on the value of $\vp$ relative to $\lambda$, so
denote $\phi=\vp-\lambda$. These three angles are sketched in
Fig.~\ref{fig:sketchangles}.

\begin{figure}[h!]
\begin{center}
\begin{tabular}{ccc}
(a) & (b) & (c)\\
\includegraphics[width = 0.32\textwidth]{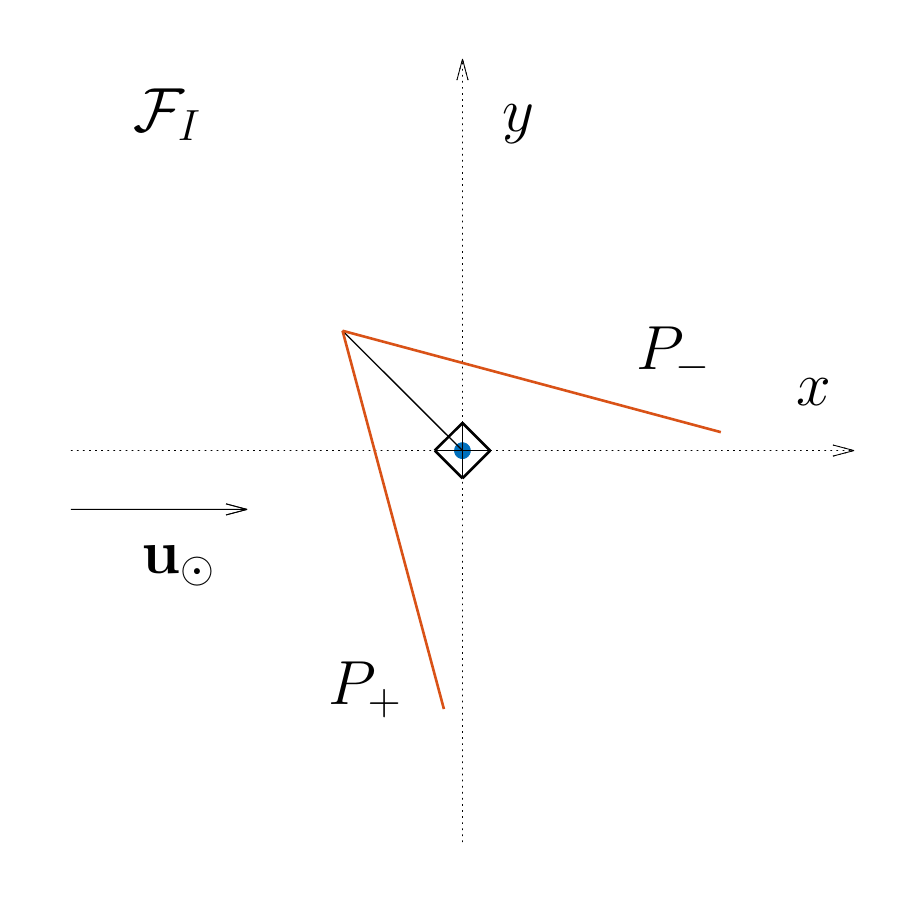}&
\includegraphics[width = 0.32\textwidth]{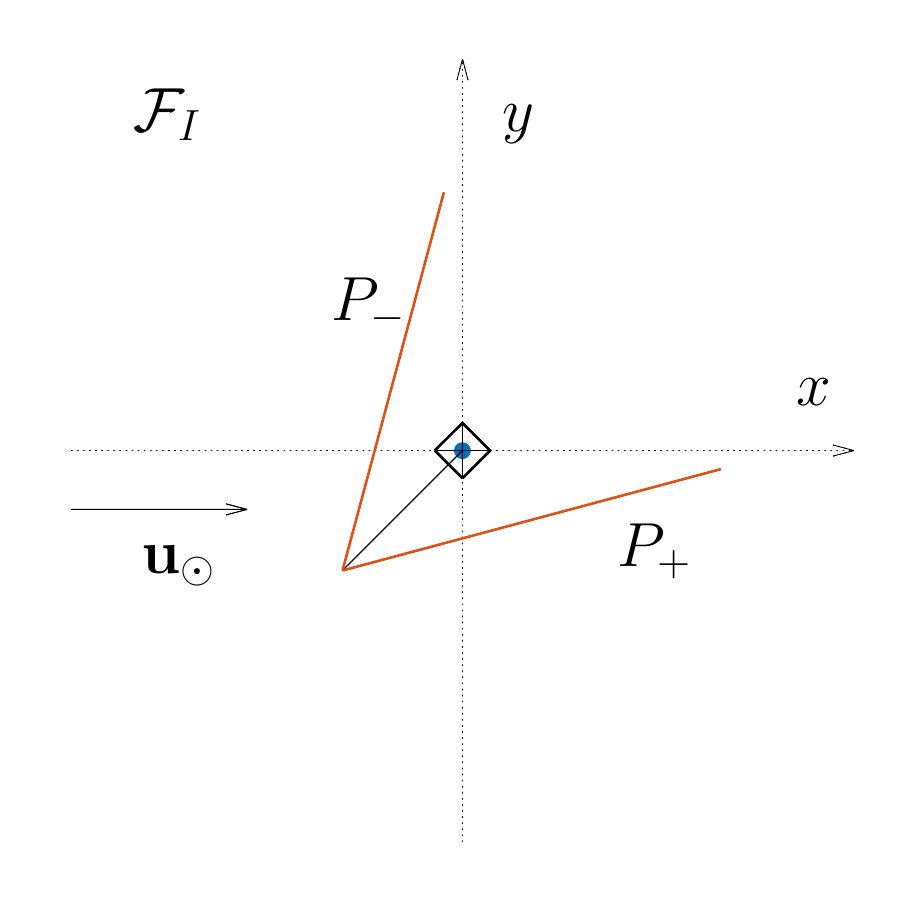}&
\includegraphics[width = 0.32\textwidth]{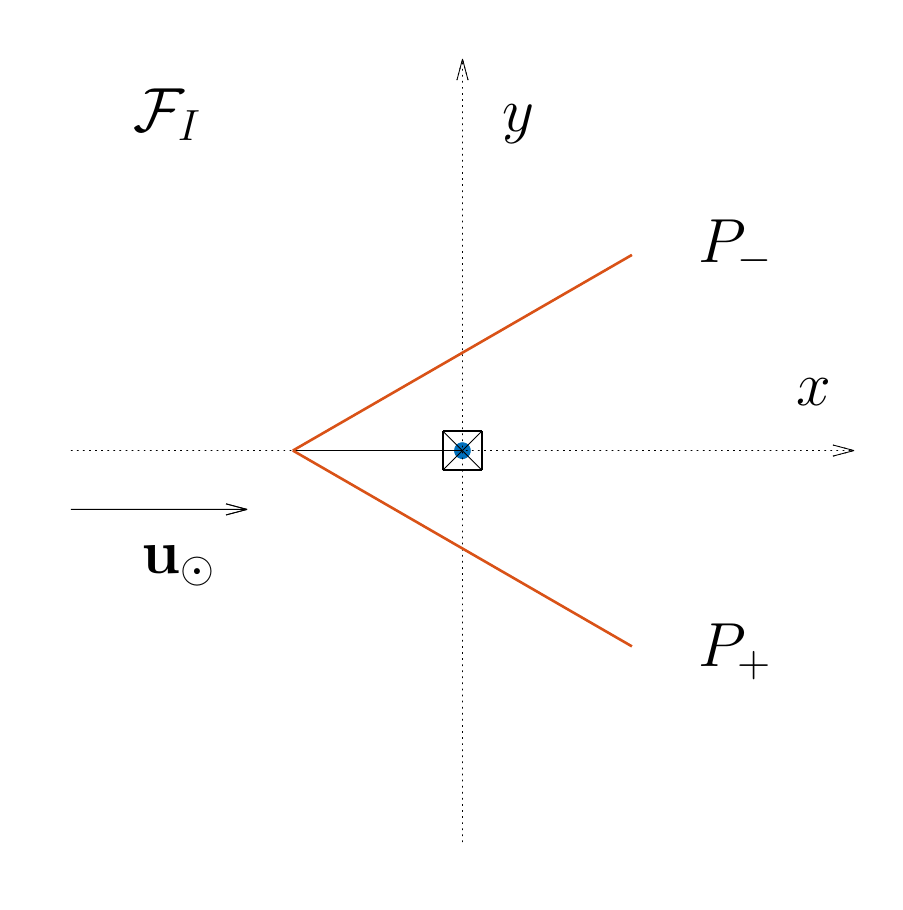}
\end{tabular}
\end{center}
\caption{
Sketches of different orientations of spacecraft in the $\Fcal_I$
frame.  In all three panels, $\lambda = 180^\circ$, see
Fig.~\ref{fig:sketchangles}.
Case (a): only $P_+$ produces torque $\alpha\leq\vp-\lambda \leq\pi-\alpha$;
case (b): only $P_-$ produces torque $-\pi +\alpha\leq \vp-\lambda\leq-\alpha$;
case (c): both panels produce torque $|\vp-\lambda|\leq\alpha$.
}
\label{fig:numpanels}
\end{figure}

If $\phi\in[-\pi+\alpha,\alpha]$, the panel $P_+$ produces torque, see
Fig.~\ref{fig:numpanels}, (a); and if $\phi\in[-\alpha,\pi-\alpha]$, it is the
panel $P_-$ who produces torque, see Fig.~\ref{fig:numpanels}, (b); in
particular, if $|\phi|\leq\alpha$, both panels face the Sun, see
Fig.~\ref{fig:numpanels} (c).  In all panels of Fig.~\ref{fig:numpanels}, the
sunlight direction is depicted as if it was in the direction $(-1,0)^\top$,
hence $\lambda = 180^\circ$, see Fig.~\ref{fig:sketchangles}.

Assume that the bus is symmetric in the sense that $I_{\xi,b} = I_{\nu,b} =
I_{\zeta,b}$. In this case, as derived in~\cite{MC18}, the attitude equations
of motion reduce to the following second order Ordinary Differential Equation
(ODE)
\begin{eqnarray}\label{eq:dynatt-dim}
\ddot{\vp} & = & \frac{A_s}{m_b+m_s}\frac{p_{\rm SR}k_{1,1}}{2C} M_1(\vp-\lambda) +  
\displaystyle    \frac{3\mu}{r^3}\frac{D(\alpha, d)}{C}
                 \sin\left(2(\theta+\omega+\Omega-\vp)\right), 
\end{eqnarray}
where $D(\alpha,d)$ is defined in Eq.~\ref{eq:inertiamom}, $\theta$, $\omega$
and $\Omega$ are the true anomaly, argument of perigee and Right Ascension of
the Ascending Node (RAAN) of the osculating orbit; the last is being considered
as it precesses due to the $J_2$ effect considered in \S~\ref{subsubsect:OD},
and 
\begin{subequations}\label{eq:SRPtorque}
\begin{eqnarray}
M_1(\psi) & = & M_0^-(\psi)X_{[-\alpha,\pi-\alpha]}(\psi) +
                M_0^+(\psi)X_{[-\pi+\alpha,\alpha]}(\psi)\nonumber,\\
          & = & M_0^-(\psi)X_{[-\alpha,\pi-\alpha]}(\psi)  
               -M_0^-(-\psi)X_{[-\pi+\alpha,\alpha]}(\psi)
                \label{eq:SRPtorque_easy}\\
M_0^\pm(\psi)
          & = & -\frac{1}{2}\sin(2\psi)\mp\frac{k_{2,0}}{k_{1,1}}\cos^2\psi
                                     \mp\frac{k_{0,2}}{k_{1,1}}\sin^2\psi,
                \label{eq:M0}
\end{eqnarray}
\end{subequations}
where $X_J$ is the characteristic function of the interval $J$, 
$$
X_J:\Rset\to\{0,1\},
\qquad
X_J(\psi) = \left\{
\begin{array}{rcl}
1 & \mbox{if} & \psi\in    J\\
0 & \mbox{if} & \psi\notin J\\
\end{array}
\right..
$$
The parameter $\mu=GM=3.986\times10^{14}\;{\rm m}^3/{\rm s}^2$ is the
gravitational parameter of the Earth. The rest of the coefficients are physical
parameters that depend on the geometry of the spacecraft and on the reflectance
parameter $\eta$, and read
\begin{subequations}\label{eq:coeftorque}
\begin{eqnarray}
k_{1,1} & = & \sin\alpha\left[2 d m_b (2\eta\cos(2\alpha) + \eta + 1) + 
w (m_b + m_s)(\cos\alpha - \eta\cos(3\alpha))\right], \label{eq:a11}\\
k_{2,0} & = & \sin^2\alpha\left[4 d \eta m_b\cos\alpha + w(m_b + m_s)(1-\eta\cos(2\alpha))\right],\mbox{ and}\label{eq:a20}\\
k_{0,2} & = & \cos\alpha\left[2dm_b(\eta\cos(2\alpha)+1)+\eta w(m_b+m_s)\sin\alpha\sin(2\alpha)\right].\label{eq:a02}
\end{eqnarray}
\end{subequations}

The function $M_0$ in Eq.~\ref{eq:M0} can be interpreted as the scaled SRP
torque due to a single panel. Since both panels are equal but oriented in a
different way, one expects the expressions for both panels to be similar.
Namely these can be found to differ only by a sign. The function $M_1$ in
Eq.~\ref{eq:SRPtorque_easy}, on the other hand, represents the joint SRP
torque, that takes in consideration that the panels have to face the sunlight
to produce torque; this is why $M_1$ is a piece-wise defined function. 

\subsubsection{Orbit dynamics}\label{subsubsect:OD}

Inspired by previous studies on the effect of SRP for the design of end-of-life
disposals, 
the considered orbit dynamics has been the two-body problem, perturbed by the
$J_2$ term and the SRP acceleration due to the sail structure~\cite{LCI12,
CLI12, LCI13, CBF16}.
The (dimensional) equations of motion of the orbit dynamics read
\begin{eqnarray}\label{eq:dyntra-dim}
\left\{
\begin{array}{rcl}
\ddot{x} & = &\displaystyle -\frac{\mu x}{r^3} - \frac{3R^2\mu J_2}{2}\frac{x}{r^5} + 
                \frac{A_sp_{\rm SR}}{m_b+m_s}a_x,\\
\ddot{y} & = &\displaystyle -\frac{\mu y}{r^3} - \frac{3R^2\mu J_2}{2}\frac{y}{r^5} + 
                \frac{A_sp_{\rm SR}}{m_b+m_s}a_y,
\end{array}
\right.
\end{eqnarray}
where the first summand is the Keplerian term the second is the $J_2$ effect
and the third one is the SRP acceleration. The constant
$J_2=1.082\times10^{-3}$ is the adimensional coefficient of the second order
term in the expansion of the perturbing potential in spherical harmonics, $R$
is the radius of the planet, and $r = \sqrt{x^2+y^2}$.
 
It is important to remark that the factors $a_x$ and $a_y$ in the third
summands of the right hand side of Eq.~\ref{eq:dyntra-dim} are piecewise
defined, as the SRP acceleration depends on $\vp - \lambda$. Namely, $a_x$ and
$a_y$ can be expressed as follows:
\begin{eqnarray}\label{eq:axay}
(a_x, a_y)^\top &\!\! =\!\! & \left[ 
(a_x^+, a_y^+)^\top X_{[-\alpha,\pi-\alpha]}  (\vp-\lambda) + 
(a_x^-, a_y^-)^\top X_{[-\pi + \alpha,\alpha]}(\vp-\lambda)
\right],
\end{eqnarray}
where
\begin{eqnarray}\label{eq:accpanel}
\begin{array}{rcl}
a_x^\pm & = &\displaystyle\sin(\alpha\pm\lambda\mp\vp)
        (   \eta\cos(2\alpha\pm\lambda\mp 2\vp)-\cos\lambda))\\
a_y^\pm & = &\displaystyle\sin(\alpha\pm\lambda\mp\vp)
        (\mp\eta\sin(2\alpha\pm\lambda\mp 2\vp)-\sin\lambda))
\end{array}
\end{eqnarray}
are the expressions of the adimensional factor of the accelerations due to panel
$P_+$ (subscript~$+$) and due to panel $P_-$ (subscript~$-$).
Equations~\ref{eq:accpanel} are obtained from Eq.~\ref{eq:forcesrp}, with 
$$
\bfn_\pm=R_3(\vp)(\sin\alpha,\pm\cos\alpha)^\top,
\qquad
\bfu_\odot=(\cos\lambda, \sin\lambda)^\top.
$$ 

In the particular case that $\alpha=\pi/2$ and $\vp = \lambda$, that is, when
the sail is completely opened and hence it is a rectangular flat panel with
width $2w$ and height $h$, and the direction of the normal to the surface of
the sail is parallel to the Sun-spacecraft direction, the SRP acceleration
reads 
$$
-2\frac{A_sp_{\rm SR}}{m_b+m_s}(1+\eta)(\cos\lambda,\sin\lambda)^\top,
$$
twice the acceleration of a flat panel of area $A_s = hw$ always oriented
towards the Sun, maximizing the SRP acceleration.\\ 

The assumption that the Sun-spacecraft distance is constant and equal to the
Sun-Earth distance is equivalent to considering a linear approximation of the
potential of the Sun~\cite{JCFJ16}.

\subsection{Multiple time scale dynamics}\label{subsect:prepeq}

The full set of 6 coupled differential equations, Eq.~\ref{eq:dynatt-dim} and
Eq.~\ref{eq:dyntra-dim}, have two different time scales, attitude being faster
than orbit dynamics. Hence these equations fit within the context of fast-slow
dynamical systems. In this section we choose adequate variables to put them in
a standard form for their treatment as a fast-slow system. 

Appendix~\ref{sect:adim} is devoted to the adimensionalisation of the equations
of motion, where the adimensionalisation factors $L$ (longitude) and $T$ (time)
are introduced. For the sake of lightening the notation, we use the same
notation for the adimensional variables, as the dimensional analogues will not
be further used in this contribution. Also $(\dot{\,})$ denotes the derivative
with respect to the adimensional time, $\tau$. The full set of equations read
\begin{eqnarray}\label{eq:fullsetadim}
\left\{
\begin{array}{rcl}
\ddot{\phi} &=&\displaystyle c_1 M_1(\phi) +  \frac{c_2}{r^3}\sin(2\arctan(y/x)-2(\phi+\lambda)),\\
\ddot{x}    &=&\displaystyle -\frac{x}{r^3} - c_3\frac{x}{r^5} + c_4 a_x,\\ 
\ddot{y}    &=&\displaystyle -\frac{y}{r^3} - c_3\frac{y}{r^5} + c_4 a_y,
\end{array}
\right.
\end{eqnarray}
where $\phi = \vp - \lambda$ (depending on adimensional time), $a_{x}$ and
$a_{y}$ are as in Eq.~\ref{eq:axay}, and the constants are the adimensional
quantities
\begin{eqnarray}\label{eq:fullconstants}
c_1 = \frac{A_s}{m_b+m_s}\frac{p_{\rm SR} k_{1,1} L^3}{2C\mu},\quad
c_2 = 3\frac{D(\alpha,d)}{C},\quad
c_3 = \frac{3R^2J_2}{2L^2},\quad
c_4 = \frac{A_s}{m_b+m_s}\frac{p_{\rm SR}L^2}{\mu}.
\end{eqnarray}
Written like this, the problem can be put in the form of a fast-slow system.
To write Eq.~\ref{eq:fullsetadim} as an ODE of first order, one must introduce
$\Phi=\dot{\phi}=\dot{\vp}-\dot{\lambda}$, and $v_x=\dot{x}, v_y=\dot{y}$. Let
us denote $\bm{\phi}=(\phi,\Phi)$ and $\bm{x}=(x,y,v_x,v_y)$
\footnote{This notation will also be used for other sets of variables that are going to be introduced later: 
$\hat{\bm{\phi}}=(\hphi,\hPhi)$,
$\bar{\bm{\phi}}=(\bphi,\bPhi)$,
$\tilde{\bm{\phi}}=(\tphi,\tPhi)$,
$\tilde{\tilde{\bm{\phi}}}=(\tilde{\tphi},\tilde{\tPhi})$,
$\hat{\bm{x}}=(\hat{x},\hat{y},\hat{v}_{\hat{x}},\hat{v}_{\hat{y}})$,
$\bar{\bm{x}}=(\bar{x},\bar{y},\bar{v}_{\bar{x}},\bar{v}_{\bar{y}})$.
$\tilde{\bm{x}}=(\tilde{x},\tilde{y},\tilde{v}_{\tilde{x}},\tilde{v}_{\tilde{y}})$, 
$\tilde{\tilde{\bm{x}}}=(\tilde{\tilde{x}},\tilde{\tilde{y}},\tilde{\tilde{v}}_{\tilde{\tilde{x}}},\tilde{\tilde{v}}_{\tilde{\tilde{y}}})$.
}. 
Then the following holds. 

\begin{prop} \label{prop:fastslow} Assume that $\dot{\lambda} = n_\odot = {\rm
constant}$, and that the torque coefficient $k_{1,1}>0$ (Eq.~\ref{eq:a11}).
There exists $\eps = \eps(c_1)>0$ and a phase scaling
$(\bm{\phi},\bm{x})\to(\hat{\bm{\phi}},\hat{\bm{x}})$ such that in the
$(\hat{\bm{\phi}},\hat{\bm{x}})$ variables Eq.~\ref{eq:fullsetadim} has the
following form
\begin{eqnarray}\label{eq:dynfastslow}
\left\{
\begin{array}{rcl}
\eps \frac{{\rm d}\hat{\bm{\phi}}}{{\rm d}\tau} & = & f(\hat{\bm{\phi}}, \hat{\bm{x}}, \eps)\\
     \frac{{\rm d}\hat{\bm{x}}}{{\rm d}\tau}   & = & g(\hat{\bm{\phi}}, \hat{\bm{x}})
\end{array}
\right.,
\end{eqnarray}
for adequate $f$ and $g$. 
\end{prop}

\noindent{\sl Proof.} If we define
\begin{eqnarray}\label{eq:epsilon}
\eps^2 & := & \frac{m_b + m_s}{A_s}\frac{2C\mu}{p_{\rm SR}k_{1,1}L^3} 
          =   \frac{1}{c_1},
\end{eqnarray}
the proposition follows by considering the scaling
$
\hat{\phi}  = \phi, \quad
\hat{\Phi} = \eps\Phi, \quad
\hat{x}    = x, \quad
\hat{y}    = y, \quad
\hat{v}_x  = v_x, \quad
\hat{v}_y  = v_y.
$
The maps $f$ and $g$ correspond to the right hand side of the resulting
equations of motion for $\hat{\bm{\phi}}$ and $\hat{\bm{x}}$, respectively.
\fin

The parameter $\eps$ in Eq.~\ref{eq:epsilon} depends solely on physical
quantities of the system and on the choice of the longitude scaling factor $L$
in the adimensionalisation procedure, see App.~\ref{sect:adim}. If $L$ is
appropriately chosen, see \S~\ref{subsec:physparam}, then $\eps$ is small, and
Eq.~\ref{eq:dynfastslow} is a fast-slow system, written in the slow time scale
$\tau$. One can consider a time scaling $t=\tau/\eps$, the fast time scale, in
which Eq.~\ref{eq:dynfastslow} reads 
\begin{eqnarray}\label{eq:dynfastslow2}
\left\{
\begin{array}{rcl}
\frac{{\rm d}\hat{\bm{\phi}}}{{\rm d}\tau} & = & f(\hat{\bm{\phi}}, \hat{\bm{x}}, \eps)\\
\frac{{\rm d}\hat{\bm{x}}}{{\rm d}\tau}   & = & \eps g(\hat{\bm{\phi}}, \hat{\bm{x}})
\end{array}
\right..
\end{eqnarray}
Hence, $\eps$ is the ratio between the time scales in which the two
characteristic motions, orbit and attitude, take place. 

The change to $\hat{\cdot}$ variables can be extended to the argument of
latitude of the Sun $\hat{\lambda} = \lambda$ which, in the fast time scale
$t$, varies as ${\rm d}\hat{\lambda}/{\rm d}t = \eps n_\odot$, similarly as for
the orbital dynamics in Eq.~\ref{eq:dynfastslow}. That is, from the point of
view of the attitude, the position of the Sun varies in a slower constant rate,
with factor $\eps$.

\subsubsection{Fast equations}
The components of the vector field in Eq.~\ref{eq:fast} of the fast variables are
those representing the evolution of the sail attitude
\begin{eqnarray}\label{eq:fast}
f(\hat{\bm{\phi}},\hat{\bm{x}}, \eps) & = &
\left(
\hat{\Phi},
\displaystyle
\quad M_1(\hat{\phi}) + \eps^2\frac{c_2}{r^3}
                       \sin(2\arctan(\hat{y}/\hat{x}) - 2(\hat{\phi}+\hat{\lambda}))
\right)^\top,
\end{eqnarray}
where, if we denote
\begin{eqnarray}\label{eq:M0fast}
M_0(\psi) & = & 
-\frac{1}{2}
\left[\sin(2\psi) - 
\frac{1}{k_{1,1}}\left((k_{2,0}-k_{0,2})\cos(2\psi) + (k_{2,0}+k_{0,2})\right)
\right],
\end{eqnarray}
we can write
\begin{eqnarray}\label{eq:MSRP}
M_1(\hat{\phi}) & = & 
M_0(\hat{\phi})X_{[-\alpha,\pi-\alpha]}(\hat{\phi}) 
-
M_0(-\hat{\phi})X_{[-\pi+\alpha,\alpha]}(\hat{\phi}). 
\end{eqnarray}
To obtain these expressions one has to use the double angle formulas in
Eq.~\ref{eq:SRPtorque}, using the symmetries of the involved trigonometric
functions, arranged as in Eq.~\ref{eq:SRPtorque_easy}. 

\subsubsection{Slow equations}

Concerning the components of the vector field in Eq.~\ref{eq:fast} of the slow
variables, the scaling of the proof of Prop.~\ref{prop:fastslow} is such that
$\hat{\bm{x}}=\bm{x}$, and hence the form of the equations does not change.
The vector field of the slow subsystem reads
\begin{eqnarray}\label{eq:slow}
g(\hat{\bm{\phi}}, \hat{\bm{x}}) & = &
\left(
\begin{array}{c}
\hat{v}_x\\\hat{v}_y\\
-\hat{x}/\hat{r}^3\\
-\hat{y}/\hat{r}^3
\end{array}
\right)
-
c_3\left(
\begin{array}{c}
0\\0\\
\hat{x}/ \hat{r}^5\\ 
\hat{y}/ \hat{r}^5
\end{array}
\right)
+
c_4\left(
\begin{array}{c}
0\\0\\
\hat{a}_x\\
\hat{a}_y
\end{array}
\right)
\end{eqnarray}
where 
$\hat{r} = \sqrt{\hat{x}^2+\hat{y}^2}$ and 
\begin{eqnarray}\label{eq:acchphi}
(\hat{a}_x, \hat{a}_y)^\top &\!\! =\!\! & \left[ 
(\hat{a}_x^+, \hat{a}_y^+)^\top X_{[-\alpha,\pi-\alpha]}(\hat{\phi}) + 
(\hat{a}_x^-, \hat{a}_y^-)^\top X_{[-\pi + \alpha,\alpha]}(\hat{\phi})
\right],
\end{eqnarray}
being
\begin{eqnarray*}
\hat{a}_{\hat{x}}^\pm & = &
       \sin(\alpha\mp\hat{\phi}) 
      (\eta\cos(2\alpha\mp\hat{\lambda}\mp 2\hat{\phi})-\cos\hat{\lambda})\\
\hat{a}_{\hat{y}}^\pm & = &
       \sin(\alpha\mp\hat{\phi}) 
      (\mp\eta\sin(2\alpha \mp\hat{\lambda}\mp 2\hat{\phi})-\sin\hat{\lambda}).
\end{eqnarray*}

Written as in Eq.~\ref{eq:dynfastslow} and Eq.~\ref{eq:dynfastslow2}, the
equations of motion can be studied by separately considering only the attitude
or the orbit dynamics. This is possible by dealing with the limit case $\eps =
0$, that can be interpreted as the spacecraft having an infinitely large
area-to-mass ratio. Setting $\eps = 0$ has two different meanings in each
equivalent formulation, Eq.~\ref{eq:dynfastslow} and Eq.~\ref{eq:dynfastslow2}:
\begin{itemize}
	\item In the slow time scale, Eq.~\ref{eq:dynfastslow},  the dynamics
are the slow (orbit) system, constrained to the zeros of the function
$f$. In practice, in the context of this paper, this means that the dynamics is
constrained to a specific attitude. In fact, in this case the dynamics is that
of a time-dependent Hamiltonian, and hence the components of $g$ in
Eq.~\ref{eq:slow} can be obtained as the derivatives of
\begin{eqnarray}\label{eq:Hamslow}
\Hcal & = & \frac{1}{2}(\hat{v}_x^2+\hat{v}_y^2)
        -   \frac{1}{\hat{r}}-\frac{c_3}{3\hat{r}^3}
        -    c_4(\hat{x}\hat{a}_{\hat{x}}+\hat{y}\hat{a}_{\hat{y}}).
\end{eqnarray}
The time dependency comes from the SRP acceleration, as it depends on the
position of the Sun.
	\item In the fast time scale, Eq.~\ref{eq:dynfastslow2}, the position
of the spacecraft is assumed to be constant and only the attitude evolves. The
position of the spacecraft is, hence, a parameter of the system.
\end{itemize}

Moreover, for $\eps=0$ both the slow and fast time scale dynamics are
Hamiltonian: in the slow time scale this holds for any fixed attitude, and in
the fast time scale it is not trivial and is justified in
\S~\ref{subsect:dynfrozen}. However, the whole problem is not Hamiltonian due
to the SRP and gravity gradient coupling.

\section{Dynamical aspects of the coupled system}\label{sect:dynasp}

In this section we highlight the most dynamically relevant aspects of the
coupled adimensional attitude and orbit model. Written as a fast-slow problem,
we can deal with the description of the dynamics as it is customary in this
field: after splitting the equations of motion into fast and slow components as
in Eq.~\ref{eq:dynfastslow2}, written in the fast time scale, the parameter
$\eps$ is set to $0$, leaving only as non-trivial the equations that correspond
to the fast dynamics. These are usually referred to as the fast subsystem, or
frozen system. The dimension of the resulting problem is smaller than the
original, and hence easier to study. The core idea is that, as there are two
time scales, for small values of $\eps$ the variables that are frozen for
$\eps=0$ evolve slower than the fast variables, so the dynamics of the full
fast-slow system is expected to be close to the dynamics of the fast subsystem.
More concretely, one can explain the dynamics of the full system by proving
that some invariant manifolds of the frozen system are preserved when
considering $\eps>0$, via the so-called Geometric Singular Perturbation Theory,
based on Fenichel's theory on the preservation of normally hyperbolic invariant
manifolds under perturbation.  The expository text~\cite{Kue15} is strongly
suggested for a global overview on the field. 

In the context of this contribution, the stability of the Sun-pointing attitude
in the fast subsystem can be translated into the oscillatory motion of the
spacecraft close to the Sun-pointing attitude, that is expected to be preserved
for long periods of time. Moreover, the fast dynamics can be averaged out.\\ 

First, \S~\ref{subsect:dynfrozen} is devoted to the study of the dynamics of the
fast subsystem assuming that the slow subsystem is frozen. After this, the
averaging of the fast small oscillations of the sail around the Sun-pointing
direction is studied in \S~\ref{subsect:sfandav}. This allows to obtain a
physical interpretation of the results that are exposed in
\S~\ref{subsect:physinter}.

\subsection{Dynamics of the fast subsystem}\label{subsect:dynfrozen}

Consider Eq.~\ref{eq:dynfastslow2}, the full system written in the fast time
scale $t$. By setting $\eps=0$ the orbit (slow) vector field vanishes so the
only nontrivial equations are those corresponding to the attitude (fast)
dynamics, that read
\begin{eqnarray}\label{eq:eqnsfast}
\frac{{\rm d}}{{\rm d} t} \hat{\bm{\phi}}
& = &
f(\hat{\bm{\phi}},\hat{\bm{x}}, \eps),
\end{eqnarray}
recall Eq.~\ref{eq:fast}.  This is not equivalent to the simplified model found
and studied in~\cite{MC18}, where the orbit dynamics was assumed to happen on a
fixed Keplerian orbit.  Recall also that the SRP acceleration due to the back
part of the panels is neglected, as this effect is not relevant for the
purposes of this paper. 

The first relevant property of the fast subsystem is that it has Hamiltonian
structure.

\begin{prop}\label{prop:fastham} 
The system given in Eq.~\ref{eq:eqnsfast} for $\eps = 0$ is Hamiltonian with
some Hamiltonian function $\Kcal_0$ and hence can be written as 
\begin{eqnarray}\label{eq:fastsubsys}
\frac{\rm d}{{\rm d}t}
(\hat{\phi},\hat{\Phi})^\top = 
f(\hat{\bm{\phi}}, \hat{\bm{x}}, 0)=(\hat{\Phi}, M_1(\hphi))^\top = 
\left(
 \frac{\partial\Kcal_0}{\partial\hat{\Phi}},
-\frac{\partial\Kcal_0}{\partial\hat{\phi}}
\right)^\top.
\end{eqnarray}
\end{prop}

\noindent{\sl Proof}.  If Eq.~\ref{eq:eqnsfast} was a Hamiltonian system with
Hamiltonian $\Kcal_0$, the equations of motion would be obtained as derivatives
of $\Kcal_0$ as indicated in Eq.~\ref{eq:fastsubsys}. As the first component of
$f$ is $\hPhi$, $\Kcal_0$ must have $\hPhi^2/2$ as summand; and as the second
component of $f$ is $M_1$, another summand of $\Kcal_0$ would be an appropriate
primitive of $M_1$, see Eq.~\ref{eq:MSRP}. The function $M_1$ is $\Ccal^0$, so
one has to make sure that this primitive is $\Ccal^1$. 

If we denote
\begin{eqnarray*}
K_0(\psi) & = & - 
\left[\frac{1}{4}\cos(2\psi) +
\frac{1}{4k_{1,1}}
\left((k_{2,0}-k_{0,2})\sin(2\psi) +
     2(k_{2,0}+k_{0,2})\psi
\right)
\right],\\
K_1(\psi) & = & K_0(\psi) - K_0(-\alpha),
\end{eqnarray*}
then a solution is
\begin{eqnarray}\label{eq:hamfast}
\begin{array}{rcl}
\displaystyle
\Kcal_0(\hat{\bm{\phi}},\hat{\bm{x}}) = \frac{\hat{\Phi}^2}{2}
 & + &
 K_1(\hat{\phi}) \chi_{[-\alpha,\pi-\alpha]}(\hat{\phi})
+K_1(-\hat{\phi})\chi_{[-\pi+\alpha,\alpha]}(\hat{\phi})\\
 & + &
 K_1(\pi-\alpha)\chi_{[-\pi,-\pi+\alpha)\cup(\pi-\alpha,\pi]}(\hat{\phi}).
\end{array}
\end{eqnarray}\fin

\begin{rema}\label{rema:hamstruct}\begin{enumerate}
	\item For $|\hat{\phi}|\leq\alpha$, $\Kcal_0 = \hat{\Phi}^2/2 -
\cos(2\hat{\phi})/2 -2K_0(-\alpha)$ is the Hamiltonian function of a pendulum. 
	\item For $\eps>0$, the attitude equations of motion can also be
written as the derivatives of a function $\Kcal_\eps$ with respect to $\hphi$ and
$\hPhi$. Namely, if we extend Eq.~\ref{eq:hamfast}
\begin{eqnarray}\label{eq:Keps}
\begin{array}{rcl}
\Kcal_\eps(\hat{\bm{\phi}},\hat{\bm{x}}) & = &\displaystyle \frac{\hat{\Phi}^2}{2}
  +
 K_1(\hat{\phi}) \chi_{[-\alpha,\pi-\alpha]}(\hat{\phi})
+K_1(-\hat{\phi})\chi_{[-\pi+\alpha,\alpha]}(\hat{\phi})\\
 & + &\displaystyle 
 K_1(\pi-\alpha)\chi_{[-\pi,-\pi+\alpha)\cup(\pi-\alpha,\pi]}(\hat{\phi})
  +  \eps^2\frac{c_2}{2r^3}
       \cos(2\arctan(\hat{y}/\hat{x}) - 2(\hphi+\hat{\lambda}))
\end{array}
\end{eqnarray}
then for $\eps>0$ we recover the vector field Eq.~\ref{eq:fast} via 
$$
f(\hat{\bm{\phi}},\hat{\bm{x}}, \eps) = 
\left(
 \frac{\partial\Kcal_\eps}{\partial\hat{\Phi}},
-\frac{\partial\Kcal_\eps}{\partial\hat{\phi}}
\right)^\top.
$$
\end{enumerate}
\end{rema}

\begin{figure}[h!]
\begin{center}
\includegraphics[width = 0.45\textwidth]{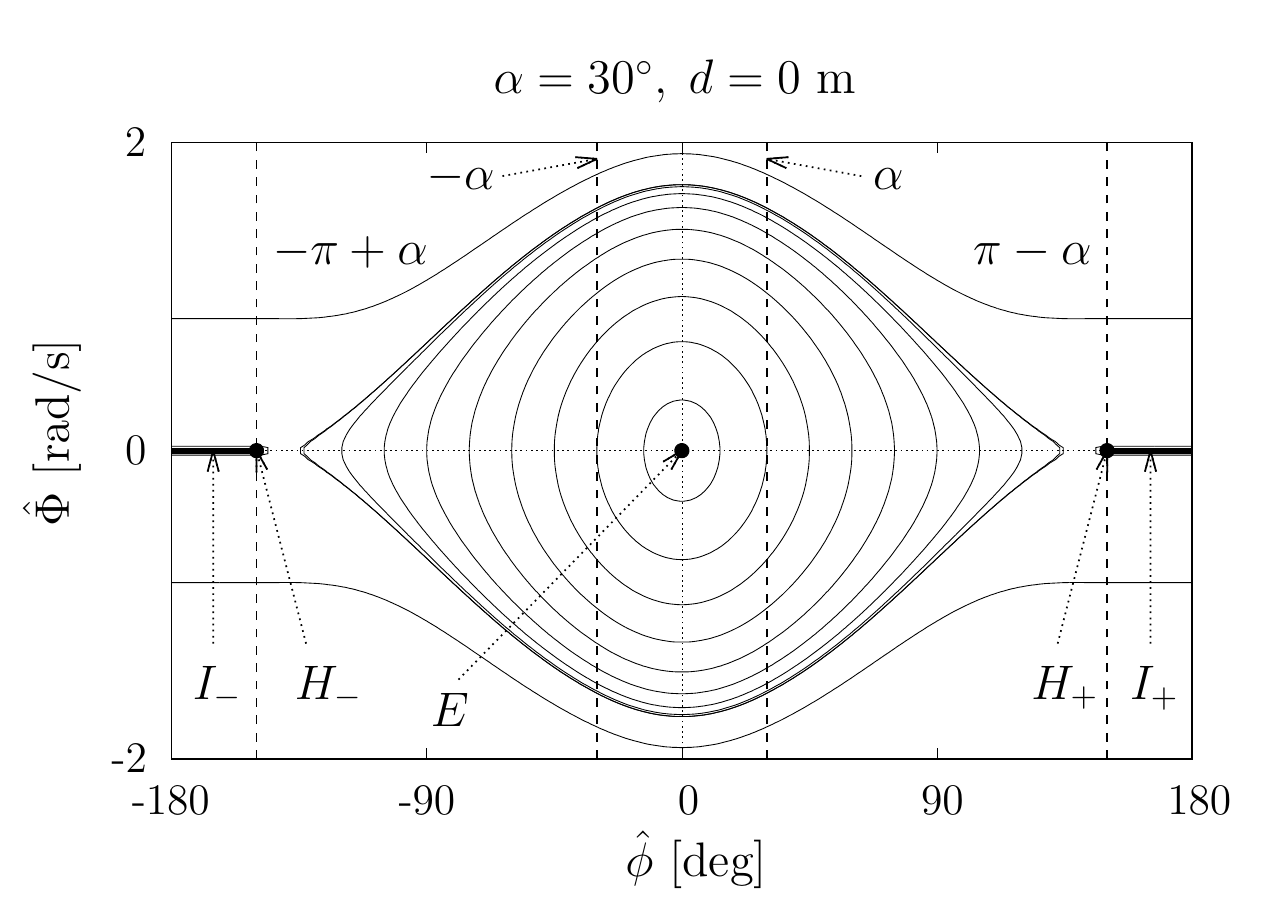}
\includegraphics[width = 0.45\textwidth]{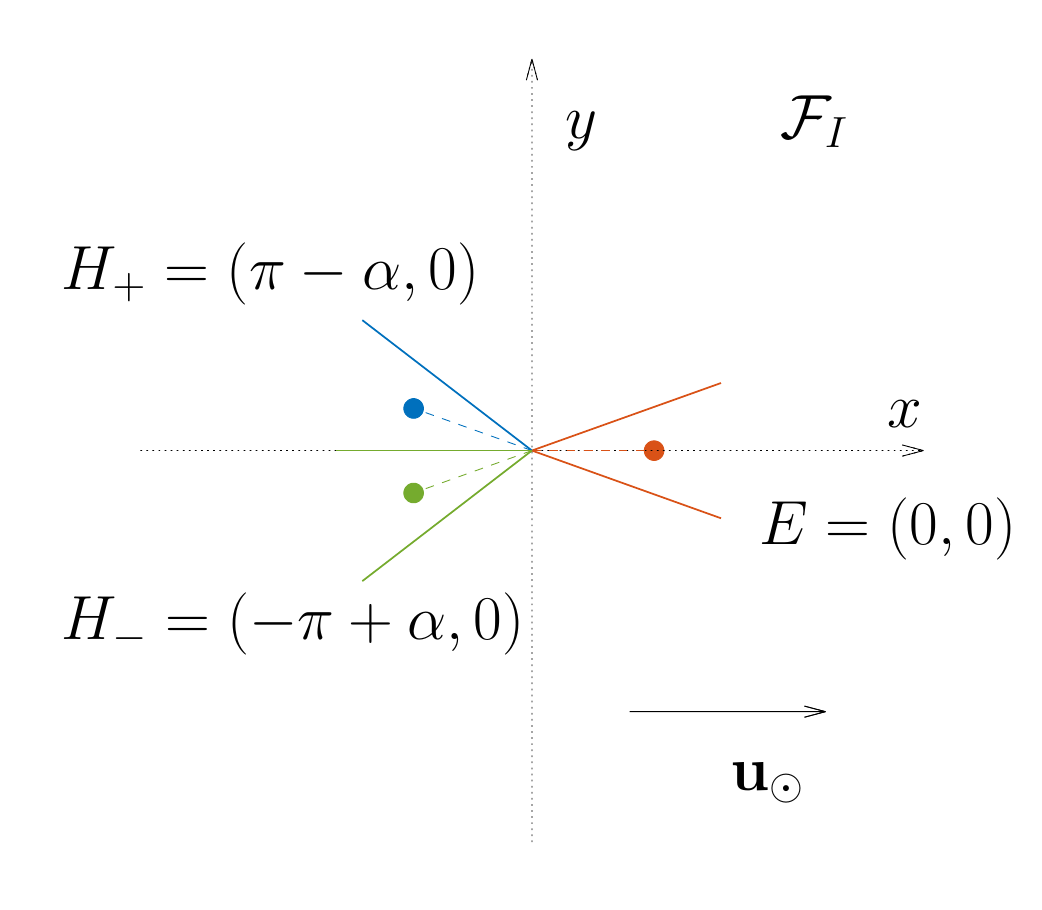}
\end{center}
\caption{Dynamics of the fast subsystem, Eq.~\ref{eq:fastsubsys}. Left: Phase
space, switching manifolds (vertical dashed lines) and equilibria. Right:
Sketch of equilibrium orientations of the sail.} 
\label{fig:phaseSRP}
\end{figure}

The most relevant properties of the fast vector field $f$ in
Eq.~\ref{eq:fastsubsys} are: it is $\mathcal{C}^0$, the differentiability being
lost at the switching manifolds $\hat{\phi} = \pm\pi\mp\alpha,\pm\alpha$; it is
$2\pi$-periodic in $\hat{\phi}$; and it is also symmetric with respect to
$\hat{\phi}=0$ as $\Kcal_\eps((\hat{\phi},\hat{\Phi}),\hat{\bm{x}}) =
\Kcal_\eps((-\hat{\phi},\hat{\Phi}),\hat{\bm{x}})$. An example of the phase space
can be seen in the left panel of Fig.~\ref{fig:phaseSRP}, for $\alpha=30^\circ$
and $d =0$. In this Figure, each curve represents an orbit of the fast
subsystem Eq.~\ref{eq:eqnsfast} that is obtained as a level set of $\Kcal_0$,
Eq.~\ref{eq:hamfast} (that is, points for which $\Kcal_0={\rm constant}$). The
origin $E$ represents the Sun-pointing attitude and the orbits around it
represent oscillatory motion. The vertical dashed lines represent switching
manifolds, that in this case represent physically that a panel either starts or
ceases to face sunlight and hence starts or ceases to produce torque. For
$|\hphi|<30^\circ$ both panels face sunlight, for $30^\circ < |\hphi| <
180^\circ-30^\circ$ only one of them do (as explained in
\S~\ref{subsubsect:attdyn}) and for $|\hphi|>180^\circ-30^\circ$ the motion is
completely rotational, i.e. the spacecraft tumbles, as no panel is assumed to
produce torque. 

The set of equilibria are an isolated point $E$ and a continuum: at all points
$(\hat{\phi},0)$ with $\hat{\phi}\in I_-\cup \{0\}\cup I_+$, where $I_-=[-\pi,
-\pi + \alpha]$ and $I_+=[\pi - \alpha, \pi)$ the vector field vanishes. Among
these the most dynamically relevant are $E = (0,0)$, which is stable (provided
$k_{1,1}>0$ in the original coordinates, see Eq.~\ref{eq:coeftorque}), and
$H_\pm = (\pm \pi \mp \alpha,0)$, that are saddles whose invariant manifolds
coincide,  $W^u(H_+) = W^s(H_-)$ and $W^u(H_-) = W^s(H_+)$. The equilibria are
indicated in the left panel of Fig.~\ref{fig:phaseSRP}, and in the right panel
the physical meaning of $E,H_{\pm}$ is sketched: $E$ is the Sun-pointing
attitude and $H_{\pm}$ represent the angles of transition to from no reflective
panel facing sunlight to one, or vice-versa.  The rest of equilibria, those
whose abscissa is $\hat{\phi}\in I_-\cup I_+\setminus\{-\pi+\al,\pi-\al\}$ have
0 as double eigenvalue.

\subsection{Fast-slow dynamics and averaging}\label{subsect:sfandav}

This section is devoted to the study of the dynamics of the full system for
$\eps>0$. For small values of $\eps$ one expects the structure found in
\S~\ref{subsect:dynfrozen} to be close to conserved in a sense that is made
explicit here. More concretely, the function $\Kcal_\eps$ is a first integral
of the frozen fast subsystem for $\eps=0$, so for $\eps>0$ and for adequate
initial conditions, it is expected to vary, but slowly, along orbits. Hence,
one expects the system to admit an analogue of an adiabatic invariant (recall
that the full system in Eq.~\ref{eq:dynfastslow} is not Hamiltonian): there is an
equivalent formulation of the system in which one of the variables experiences
oscillations of at most $\Ocal(\eps)$ for time intervals of length
$\Ocal(1/\eps)$. The goal of this section is to find the adequate change of
variables that translates our problem in this context.

Attitudes close to the Sun-pointing direction $\hat{\phi}=0$ are those for
which there is numerical evidence of helio-stability properties of the
spacecraft, see~\cite{MC18}, and the dynamics in this regime is that of a
mathematical pendulum, recall item $1$ in Rem.~\ref{rema:hamstruct}. The
following study is restricted to this situation.

For the full set of equations of motion written in the fast time scale in
Eq.~\ref{eq:dynfastslow2}, let $G$ be a bounded subset of the phase space such
that, for all $(\hat{\bm{\phi}},\hat{\bm{x}})\in G$, $|\hphi|<\alpha$ (that is,
both panels face the sunlight) where all trajectories lie wholly in $G$. Note
that, in $G$, for fixed values of $\hat{\bm{x}}$, the condition $\Kcal_\eps =
k$ defines one and only one trajectory of Eq.~\ref{eq:dynfastslow2}, and the
restriction of $G$ onto the $(\hphi, \hPhi)$ variables consists of a continuum
of closed and nested periodic orbits.

\begin{lema}\label{lema:averaging}
For small enough $\eps>0$ and small enough $\hPhi$ (that is, close enough to
the Sun-pointing attitude) there exists a real analytic change of variables
defined in the interior of $G$  
$$
C:
(\hat{\phi},\hat{\Phi},\hat{x},\hat{y},\hat{v}_{\hat{x}},\hat{v}_{\hat{y}})
\mapsto
(\tilde{\phi},\tilde{\Phi},\tilde{x},\tilde{y},\tilde{v}_{\tilde{x}},\tilde{v}_{\tilde{y}})
$$
that transforms Eq.~\ref{eq:dynfastslow2}, with $f$ and $g$ as given in
Eq.~\ref{eq:fast} and Eq.~\ref{eq:slow}, respectively, into a system of the
form
\begin{eqnarray}\label{eq:befav}
\left\{
\begin{array}{rcl}
\dot{\tphi} & = & \beta + \tilde{f}(\tPhi,\tilde{\bm{x}}) 
                     + \eps^2R_1(\tilde{\bm{\phi}},\tilde{\bm{x}}),\\
\dot{\tPhi} & = & \eps^2R_2(\tilde{\bm{\phi}},\tilde{\bm{x}}),\\
\dot{\tilde{\bm{x}}} & = & \eps (\tilde{g}(\tPhi,\tilde{\bm{x}}) 
                     +  \bm{R}(\tilde{\bm{\phi}},\tilde{\bm{x}})),
\end{array}
\right.
\end{eqnarray}
where $\tilde{f},\tilde{g}$ have zero average with respect to $\tphi$.
\end{lema}

\noindent{\sl Proof}. This result fits within the scope of averaging theory,
and the proof can be sketched as consisting of the two following steps.
\begin{enumerate}
	\item Consider for the moment the dynamics of the fast subsystem
Eq.~\ref{eq:fastsubsys}. It depends on $\hphi$ in a periodic way, but for each
initial condition in $G$, the period and the range of $\hphi$ of points in this
orbit are different. Restricting to the behaviour close to the Sun-pointing
direction, we can substitute the equations of motion by the Taylor series
around $\hphi=0$. Since the frozen subsystem is Hamiltonian, one can have
$\Kcal_0$ in mind for the moment.  The leading non-constant terms are 
$$
\frac{1}{2}(\hat{\Phi}^2+2\hat{\phi}^2).
$$

This suggests to consider a first change of variables ($\hat{\cdot}\to\bar{\cdot}$)
$$
C_1:
(\hat{\phi},\hat{\Phi},\hat{x},\hat{y},\hat{v}_{\hat{x}}, \hat{v}_{\hat{y}})
\mapsto
(\bar{\phi},\bar{\Phi},\bar{x},\bar{y},\bar{v}_{\bar{x}},\bar{v}_{\bar{y}})
$$
where $(\bar{\Phi},\bar{\phi})$ are the usual Poincar\'e action-angle variables
\begin{eqnarray}\label{eq:Poinlike}
\begin{array}{rclcrcl}
\hat{\phi}          & = & \sqrt{2\bar{\Phi}/\beta}\sin \bar{\phi} ,         & & 
\hat{\Phi}          & = & \sqrt{2\bar{\Phi}\beta} \cos \bar{\phi},\\
2\bar{\Phi}\beta   & = & \hat{\Phi}^2 + \beta^2\hat{\phi}^2, & &
\bar{\phi}          & = & \arctan(\beta\hat{\phi}/\hat{\Phi}),
\end{array}
\end{eqnarray}
the frequency being $\beta=\sqrt{2}$. The change is extended to the rest of
variables by simply choosing $\bar{x}=\hat{x}$, $\bar{y}=\hat{y}$,
$\bar{v}_{\bar{x}}=\hat{v}_{\hat{x}}$, and
$\bar{v}_{\bar{y}}=\hat{v}_{\hat{y}}$.

After this change of variables, the obtained equations of motion are $2\pi$
periodic in $\bphi$, which is still fast with respect to the rest of the
variables. 

It is worth noting that as the fast subsystem has Hamiltonian structure and the
change of variables $C_1$ restricted to the attitude motion is canonical, one
can still deal with these two equations as if they were a Hamiltonian system
with Hamiltonian function $\Kcal_\eps$.  So, for orbits in $G$, we can obtain
the equations of motion in the $\bar{\cdot}$ variables as follows. Start by
considering
\begin{eqnarray}
\Kcal_\eps & = & \frac{\hPhi^2}{2} - \frac{1}{2}\cos(2\hphi) 
          + \eps^2\frac{c_2}{r^3}\cos(2\vth - 2\hphi) \nonumber \\
      & = & \frac{\hPhi^2}{2} - \frac{1}{2}\cos(2\hphi) 
          + \eps^2\frac{c_2}{r^3}
          \left(\cos(2\vth)\cos(2\hphi)+\sin(2\vth)\sin(2\hphi)\right),\label{eq:befexp}
\end{eqnarray}
where $\vth := \arctan(\hat{y}/\hat{x})-\hat{\lambda}$. Now, the expansion
around $\hphi=0$ reads
\begin{eqnarray*}
\Kcal_\eps & = & \frac{\hPhi^2}{2} - \frac{1}{2}
            \sum_{j\geq0}\frac{(-1)^j}{(2j)!}(2\hphi)^{2j}\\
      & + & \eps^2\frac{c_2}{r^3}
            \left(
            \cos(2\vth)\sum_{j\geq0}\frac{(-1)^j}{(2j)!}(2\hphi)^{2j} + 
            \sin(2\vth)\sum_{j\geq0}\frac{(-1)^j}{(2j+1)!}(2\hphi)^{2j+1} 
            \right),
\end{eqnarray*}
where, if we introduce the $\bar{\cdot}$ variables Eq.~\ref{eq:Poinlike} with $\beta = \sqrt{2}$, it reads
\begin{subequations}\label{eq:TaylorKbar}
\begin{eqnarray}
\Kcal_\eps & = & \sqrt{2}\bPhi +
            \sum_{j\geq2}\frac{(-1)^{j+1}2^{5j/2-1}}{(2j)!}\bPhi^j\sin^{2j}\bphi
            \label{eq:frompen}\\
      & + & \eps^2\frac{c_2}{r^3}
            \left(
          \cos(2\vth)
          \sum_{j\geq0}\frac{(-1)^{j}2^{5j/2}}{(2j)!}\bPhi^{j}\sin^{2j}\bphi
            \right.\label{eq:fromGGcos}\\
      & + & \left.
          \hspace{1.2cm}\sin(2\vth)
          \sum_{j\geq0}\frac{(-1)^j2^{5(2j+1)/4}}{(2j+1)!}\bPhi^{j+1/2}\sin^{2j+1}\bphi
            \right)\label{eq:fromGGsin}.
\end{eqnarray}
\end{subequations}
Now the attitude equations in $\bar{\cdot}$ variables can be recovered via
$$
\frac{{\rm d}\bphi}{{\rm d}t} =  \frac{\partial\Kcal_\eps}{\partial\bPhi},\qquad
\frac{{\rm d}\bPhi}{{\rm d}t} = -\frac{\partial\Kcal_\eps}{\partial\bphi}.
$$
Concerning the orbital motion (or slow subsystem), it only depends on
$\hphi$ via the term due to SRP acceleration, see the rightmost summand in
Eq.~\ref{eq:slow}. In the region $G$, this acceleration (see
Eq.~\ref{eq:acchphi}) reads
\begin{eqnarray}\label{eq:accSRP2Pan}
\left(
\begin{array}{c}
\hat{a}_x\\
\hat{a}_y
\end{array}
\right)
\!\!=\!\!
\left(
\begin{array}{c}
-(2+\eta)\sin\alpha\cos\lambda\cos\hphi 
 - \eta\sin\alpha\sin\lambda\sin\hphi 
 + \eta\sin(3\alpha)\cos(\lambda-3\hphi)\\
-\sin\alpha((1+\eta)\sin(\lambda-\hphi) \sin(\lambda+\hphi)
 - \eta(1+2\cos(2\alpha))\sin(\lambda-3\hphi))
\end{array}
\right).
\end{eqnarray}
This can be treated as we did for $\Kcal_\eps$ above, first separating the
dependence on $\hphi$, substituting the sine and cosine functions by their
Taylor expansions, and then introducing the $\bar{\cdot}$ coordinates. These
expressions are not added as only a part of them are useful for the next step,
and we can refer to the expansions above to justify the form they have.

	\item Let us focus for the moment on the differential equation of
$\bPhi$, that is, after performing the first change of variables $C_1$. This
equation is obtained as $-\partial\Kcal_\eps/\partial\bphi$. Note that all
summands except $\beta\bPhi$ in the right hand side of Eqs.~\ref{eq:TaylorKbar}
have a factor that depends on $\bphi$, only as an argument of a power of a sine
function, so after deriving with respect to $\bphi$, the derivative of all
terms appears in the right hand side of the equation of motion of $\bPhi$.
Moreover, in each summand there is a factor of the form of the right hand side
of 
$$
\frac{\rm d}{{\rm d}\bphi}\sin^l\bphi = l\cos\bphi\sin^{l-1}\bphi,
\qquad l\in\Zset,\quad l > 0
$$
that have zero average with respect to $\bphi$. Hence, one can get rid of the
dependence on $\bphi$ via some averaging steps as done, for instance, in the
classical reference~\cite{Nei84}. After ordering the terms (by orders in
$\bar{\Phi}$, for instance), the $j$th step would consist of two substeps.
Firstly the terms of the right hand side we want to get rid of have to be
separated as the sum of periodic (with respect to $\bar{\phi}$) plus average
parts. Secondly, a change of variables has to be constructed in a way that the
non-periodic part remains intact and the periodic targeted terms do not appear
in the equations written in the new variables.  This is done at the expense of
more terms appearing in higher orders that have to be dealt with in subsequent
averaging steps. 

To be able to perform these changes we have to require that, on the one hand,
the frequency $\beta>0$ has to be bounded away from zero. This is satisfied as
$\beta=\sqrt{2}$. On the other hand, the leading terms of the expansion are
$\Ocal(\bPhi^2)$ (see Eq.~\ref{eq:frompen}), hence we have to assume that
$\bPhi$ is small enough so that the successive changes of variables to be
performed are, in fact, invertible. 

To reach the claimed form of the equations, Eq.~\ref{eq:befav}, one has to
perform a second change of variables
$$
C_2:
(\bar{\phi},\bar{\Phi},\bar{x},\bar{y},\bar{v}_{\bar{x}},\bar{v}_{\bar{y}})
\mapsto
(\tilde{\phi},\tilde{\Phi},\tilde{x},\tilde{y},\tilde{v}_{\tilde{x}},\tilde{v}_{\tilde{y}})
$$
that is a composition of $N\geq0$ averaging steps. The number of steps $N$
depends on the size of $|\bPhi|$ relative to $\eps$. Note that there are terms
in the expansions in Eqs.~\ref{eq:TaylorKbar} that do not have $\eps$ as
factor, namely those in the summatory in Eq.~\ref{eq:frompen}, which start with
$\bPhi^2$. It may happen, if we are far enough from the equilibrium, that
$\Ocal(\bPhi^2)>\Ocal(\eps^2)$, that is, these terms dominate over those in
Eqs.~\ref{eq:fromGGcos} and \ref{eq:fromGGsin}. Let $l\geq2, l\in\Zset$ be the
first power such that $\Ocal(\bPhi^l)<\Ocal(\eps^2)$.  To reach the desired
form in the equations of motion in Eq.~\ref{eq:befav}, we have to get rid of
all terms in Eq.~\ref{eq:frompen} that have $\bPhi^j$, $j<l$ as factor. This is
done via a sequence of averaging steps, each dedicated to eliminate the lowest
power appearing (see, e.g., \cite{Eft11}). As the first term has order $2$ in
$\bPhi$, the number of averaging steps to perform is $N=l-2$. In particular, if
$\Ocal(\bPhi^2)$ is already smaller than $\Ocal(\eps^2)$ all terms are small
enough and then $C_2$ is the identity.

For small enough $\bPhi$, $C_2$ is the composition of near-the-identity changes
of variable such that, in the new variables, the lowest order term in $\bPhi$
in the Taylor expansion that depends on $\bphi$ is exchanged by a zero-average
term with respect to $\bphi$ of the same order in $\bPhi$.  This can be done in
such a way that after the $N$ steps the differential equations consist of the
average of the original equation written in the $\bar{\cdot}$ variables, at the
expense of having a change of variables that does not have zero average, plus
higher order terms that have $\eps$ as factor.
\end{enumerate}

After performing $C_1$ and then $C_2$, Eq.~\ref{eq:dynfastslow} has, in the
$\tilde{\cdot}$ variables, the claimed form Eq.~\ref{eq:befav}, where
$\tilde{f}$ is the derivative with respect to $\bPhi$ of the average with
respect to $\bphi$ of the terms in the sum Eq.~\ref{eq:frompen} up to the first
order comparable with $\eps$ in magnitude, with $\tPhi$ in the place of
$\bPhi$; $\tilde{g}$ consists of the slow equations of motion plus the average
up to the same order of Eq.~\ref{eq:accSRP2Pan}; and the functions $R_1$, $R_2$
and the map $\bm{R}$ contain the non averaged terms plus extra terms that
appear as a byproduct of the successive averaging changes of variables.\fin\\

Now we are in position of using the averaging principle, that consists of
replacing the right hand side of Eq.~\ref{eq:befav} with its average with
respect to $\tphi$ and to neglect the terms $\Ocal(\eps^2)$. Let us refer to
the averaged variables as $\tilde{\tilde{\cdot}}$. The averaged equations of
motion read
\begin{eqnarray}\label{eq:averaged}
\dot{\tilde{\tilde{\Phi}}}=0,\quad
\dot{\tilde{\tilde{\bm{x}}}} = \eps(\tilde{g} + \left<\bm{R}\right>),
\end{eqnarray}
where the angle brackets denote average with respect to $\tphi$. Recall that
$\tphi$ indicates the average angle between the $\xi$ axis in $\Fcal_b$ (i.e.
the orientation of the sail) and the direction of sunlight.  Consider the
solutions of Eq.~\ref{eq:befav} and Eq.~\ref{eq:averaged} starting at the same
initial conditions, and denote them by $\tilde{\tilde{\Phi}}(t)$,
$\tilde{\tilde{\bm{x}}}(t)$ and $\tilde{\Phi}(t)$, $\tilde{\bm{x}}(t)$,
respectively.

\begin{teor}\label{teo:averaging} Assume $\eps>0$ is small enough. Then, on the
one hand, $\tPhi$ is an adiabatic invariant, that is,
$$
|\tPhi(t)-\tPhi(0)| = \Ocal(\eps)
\quad\mbox{for}\quad 
0\leq t\leq1/\eps.
$$
On the other hand, starting with the same initial conditions, the solution of
the slow variables in Eq.~\ref{eq:befav} is described by the solution of
Eq.~\ref{eq:averaged} with an accuracy of $\Ocal(\eps)$ over time intervals of
length $\Ocal(1/\eps)$, that is,
$$
|\tilde{\tilde{\bm{x}}}(t) - \tilde{\bm{x}}(t)| = \Ocal(\eps)
\quad\mbox{up to}\quad
t=\Ocal(1/\eps).
$$
\end{teor}

\noindent{\sl Proof.} This follows from the theorem about the accuracy of the
averaging method, see~\cite{Arn88}, Chapter 10, \S~52. This applies as the
equations of motion are analytic, so they meet the minimum (finite)
differentiability requirements, and $|\beta|=\sqrt{2}>0$ is bounded away from
zero. \fin\\

\begin{rema}The averaging steps performed have been done in such a way that the
resulting equations of motion after the successive changes of variables were
the average with respect to the fast angle $\bphi$.  Although this is the
expected observed mean motion, it adds difficulties in the equations as it
forces the successive changes to not have zero average and makes the remainders
of the expression Eq.~\ref{eq:befav} more involved. If instead one performs
these steps getting rid of all possible periodic terms, the changes and the
remainders remain $2\pi$ periodic through the whole process.  Moreover, in this
situation, the classical theorem of Neishtadt~\cite{Nei84} is applicable, so
there exists a change of variables that separates the phase $\hphi$ from the
rest of the variables except from a remainder that has exponentially small
bounds.
\end{rema} 
 
\subsection{Explicit equations and physical interpretation}
\label{subsect:physinter}

The benefits of the analysis performed in this section is that one can extract
physical interpretations of interest for prospective applications: spacecraft
design guidelines, simplification of the equations of motion and the
interpretation of the averaged problem as an equivalent already studied
problem. 

\subsubsection{Stability of the Sun-pointing direction}

The first practical benefit can be extracted from the study of the fast
dynamics. The condition $k_{1,1}>0$, that is a hypothesis of
Prop.~\ref{prop:fastslow} is actually equivalent to 
\begin{eqnarray}\label{eq:condstab}
d > \frac{w(m_b+m_s)}{2m_b}K(\alpha, \eta),\qquad
K(\alpha, \eta) := \frac{\eta\cos(3\alpha) - \cos(\alpha)}{2\eta\cos(2\alpha)+\eta+1}.
\end{eqnarray}

This is a necessary constraint between the physical parameters of the system,
$\alpha$ and $d$. Note that since $\alpha\in[0,\pi/2]$ and $\eta\in(0,1)$, the
function $K(\alpha,\eta)<0$ we have the following result.

\begin{corol}\label{prop:mind}\cite{MC18} For each aperture
angle $\alpha\in(0,\pi/2)$ there exists $d_{\min}$ such that if $d>d_{\min}$
the origin of Eq.~\ref{eq:fastsubsys} is locally stable.
\end{corol}

This local stability can be physically understood as when the motion starts
close enough to the Sun-pointing attitude the attitude librates around this
state, as seen in Fig.~\ref{fig:phaseSRP}.

\begin{figure}[h!]
\begin{center}
\includegraphics[width = 0.45\textwidth]{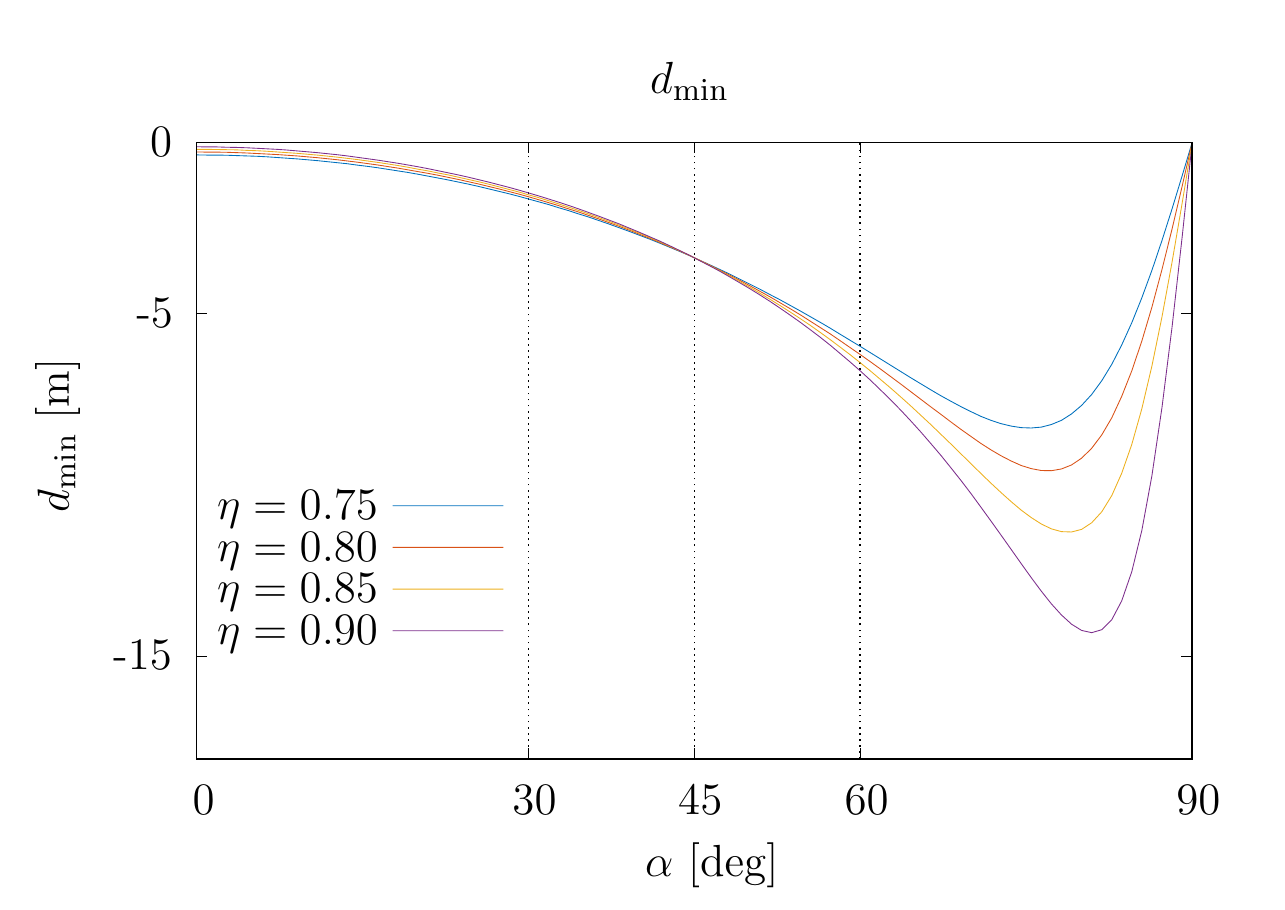}
\end{center}
\caption{Minimum value of the MPO $d$ for which the Sun-pointing attitude is a
stable equilibrium of the the fast frozen subsystem, for reflectance
coefficients $\eta = 0.75,0.80,0.85, 0.90$.}
\label{fig:stabsunpoint}
\end{figure}

This result relates the aperture angle with the MPO and can be understood as a
guideline for construction and can be used for future control-related studies.
Note that, in particular, $d_{\min}<0$ for $\al\in(0,\pi/2)$ and $d_{\min}=0$
for $\al=\pi/2$, the flat plate case; recall Fig.~\ref{fig:shapesc}. In
Fig.~\ref{fig:stabsunpoint} the quantity $d_{\min}$ is displayed as a function
of $\al$, for different values of the reflectance coefficient $\eta$.

\subsubsection{Averaged equations of motion}

The previous analysis justifies the averaging procedure with respect to the
oscillations around the Sun-pointing direction. The main hypotheses for the
theorems were, on the one hand, the closeness to the Sun-pointing direction;
and on the other hand, the smallness of $\eps$, that actually measures the
separation of time scales between the attitude and orbit components. This was
already observed in the numerical study performed in~\cite{MC18}.

In the proof of Lemma~\ref{lema:averaging}, the changes of variable were chosen
to justify the applicability of the well-known averaging results, but from a
practical perspective, provided $\eps$ and $\bPhi$ are small, the averaging
process can be carried out up to any required order. Namely, the larger $\bPhi$
is, the higher the order of the averaging procedure has to be. 

Concerning the attitude equations, in practice one can obtain the differential
equations by first averaging the Hamiltonian $\Kcal_\eps$, and then computing
the derivatives with respect to the new averaged action. Namely, using that
$$
\frac{1}{2\pi}\int_0^{2\pi}\sin^n\psi\,{\rm d}\psi = 
\left\{
\begin{array}{ll}
\frac{1}{2^n}{n\choose n/2} & \mbox{ if $n$ is even}\\
0                           & \mbox{ if $n$ is odd}
\end{array}
\right.,
$$
the average of Eq.~\ref{eq:TaylorKbar} reads (using the same names for the
variables)
\begin{eqnarray}\label{eq:averagedKeps}
\left<\Kcal_\eps\right> & = & \beta \bPhi + 
\sum_{j\geq2}\frac{(-1)^{j+1}2^{j/2-1}}{(j!)^2}\bPhi^j +
\eps^2\frac{c_2}{r^3}\cos(2\vartheta)
\sum_{j\geq0}\frac{(-1)^{j}2^{j/2}}{(j!)^2}\bPhi^j.
\end{eqnarray}
The derivative of $\left<\Kcal_\eps\right>$ with respect to $\bPhi$ gives the
approximation of the average frequency of rotation of $\bphi$.

Concerning the orbital dynamics, after the averaging procedure they are
approximately decoupled from the attitude. The averaged effect of the SRP
acceleration is found by computing the average of Eq.~\ref{eq:accSRP2Pan}. 

Let us proceed as done from Eq.~\ref{eq:befexp} to Eq.~\ref{eq:TaylorKbar},
that is, first separating the dependence on $\hphi$ and then expanding the sine
and cosine functions around $\hphi=0$ in Eq.~\ref{eq:accSRP2Pan}.  After this
procedure, the only terms that will contribute to the average are with those
that have $\cos(l\hphi)$, $l\in\Zset$ as factor. These can be written in the
following compact form,
$$
-\left(
\begin{array}{c}
\cos\lambda\\
\sin\lambda
\end{array}
\right)
\left(
(2+\eta)\sin\alpha\cos\hphi
-\eta\sin(3\alpha)\cos(3\hphi)
\right),
$$
where, if we expand the cosine terms and introduce the $\bar{\cdot}$ variables,
we obtain
$$
-\left(
\begin{array}{c}
\cos\lambda\\
\sin\lambda
\end{array}
\right)
\sum_{j\geq0}\frac{(-1)^j2^{j/2}}{(2j)!}\bPhi^j\sin^{2j}\bphi
\left[
(2+\eta)\sin\alpha-\eta\sin(3\alpha)3^{2j}
\right],
$$
whose average reads, re-using again the $\bar{\cdot}$ variables for their
averaged analogue,
\begin{eqnarray}\label{eq:averagedacceleration}
\left(
\begin{array}{c}
\left<a_x\right>\\
\left<a_y\right>
\end{array}
\right)
=
-\left(
\begin{array}{c}
\cos\lambda\\
\sin\lambda
\end{array}
\right)
\sum_{j\geq0}\frac{(-1)^j2^{-3j/2}}{(j!)^2}\bPhi^j
\left[
(2+\eta)\sin\alpha-\eta\sin(3\alpha)3^{2j}
\right].
\end{eqnarray}
Note that the sums in both Eq.~\ref{eq:averagedKeps} and
Eq.~\ref{eq:averagedacceleration} are convergent.

\subsubsection{Interpretation as an equivalent flat sail}

After the averaging procedure, the attitude and orbit dynamics become decoupled
since the acceleration due to SRP becomes constant in the direction of
Eq.~\ref{eq:averagedacceleration}, that in fact depends only on the attitude's
initial condition of the integration through the mean amplitude of the
oscillations, as will be seen in \S~\ref{sect:numerics}.

The first benefit of the analysis is that the equations of motion become
Hamiltonian again as the averaging acts as fixing the attitude (recall the end
of \S~\ref{subsect:prepeq}). In fact, one can interpret the averaging as follows:
\begin{quote}
For each initial condition (that fixes the initial value of $\bPhi$) the
dynamics of the averaged equations is that of a spacecraft with the same mass
$m_b+m_s$ with a flat panel of effective area $A_s\cdot A_{\rm eff}$, always
perpendicular to the Sun-spacecraft direction. 
\end{quote}

The expression of this effective area is obtained from
Eq.~\ref{eq:averagedacceleration}. Comparing the acceleration due to SRP in our
problem and the acceleration due to SRP that a sail with the same mass with
area $A_sA_{\rm eff}$ would have yields
$$
-\frac{A_sp_{\rm SR}}{m_b+m_s}
\left(
\begin{array}{c}
\cos\lambda\\
\sin\lambda
\end{array}
\right)
\sum_{j\geq0}\frac{(-1)^j2^{-3j/2}}{(j!)^2}\bPhi^j
\left[
(2+\eta)\sin\alpha-\eta\sin(3\alpha)3^{2j}
\right]=-\frac{A_sA_{\rm eff}p_{\rm SR}}{m_b+m_s}
\left(
\begin{array}{c}
\cos\lambda\\
\sin\lambda
\end{array}
\right),
$$
and from this we get that
\begin{eqnarray}\label{eq:Aeff}
A_{\rm eff} = 
A_{\rm eff}(\bPhi, \al) & = &
\sum_{j\geq0}\frac{(-1)^j2^{-3j/2}}{(j!)^2}\bPhi^j
\left[
(2+\eta)\sin\alpha-\eta\sin(3\alpha)3^{2j}
\right].
\end{eqnarray}
This quantity is referred to as the area factor.

\section{Numerical test cases}\label{sect:numerics}

This section is devoted to exemplify numerically the analysis performed in
\S~\ref{sect:dynasp}. Since the results refer strictly to the case in which
both panels face sunlight, that is, $|\hphi|<\alpha$ in
Eq.~\ref{eq:fullsetadim}, the simulations are restricted to this case. The
equations of motion used in this section are those obtained after the change of
variables of Prop.~\ref{prop:fastslow}, that in the fast time scale $t$
(denoting $(\;')={\rm d}/{\rm d}t$) read
\begin{eqnarray}\label{eq:tointegrate}
\left\{
\begin{array}{lcl}
\hphi'   &=& \displaystyle \hPhi,\\
\hPhi'   &=& \displaystyle -\sin(2\hphi) +  \eps^2\frac{c_2}{\hat{r}^3}\sin(2\arctan(\hat{y}/\hat{x})-2(\hphi+\hat{\lambda})),\\
\hat{x}' &=& \eps \hat{v}_{\hat{x}},\\
\hat{y}' &=& \eps \hat{v}_{\hat{y}},\\
\hat{v}'_{\hat{x}}    &=&\displaystyle \eps\left(-\frac{\hat{x}}{\hat{r}^3} - c_3\frac{\hat{x}}{\hat{r}^5} + c_4 \hat{a}_{\hat{x}}\right),\\ 
\hat{v}'_{\hat{y}}    &=&\displaystyle \eps\left(-\frac{\hat{y}}{\hat{r}^3} - c_3\frac{\hat{y}}{\hat{r}^5} + c_4 \hat{a}_{\hat{y}}\right),
\end{array}
\right.
\end{eqnarray} 
where $(\hat{a}_{\hat{x}},\hat{a}_{\hat{y}})^\top$ read are given in
Eq.~\ref{eq:accSRP2Pan}. 

Note that the theoretical restriction gives a stopping condition for the
numerical integrations: if $|\hphi|>\alpha$ simulations are stopped, as in this
case the sail is expected to tumble.

\subsection{Physical parameters and adimensionalisation
factors}\label{subsec:physparam}

The system of ODE in Eq.~\ref{eq:tointegrate} has its own interest for arbitrary
choices of the parameters $c_1,c_2,c_3$ and $c_4$. Despite this, to justify the
usefulness of the analysis performed in prospective real applications we are
lead to choose values of the parameters that correspond to a structure that is
constructible according to current technological boundaries.  So, as done
in~\cite{MC18} and following the guidelines in~\cite{DV18}, the physical
parameters of the structure are chosen to be $\eta=0.8$, $m_b=100$ kg, $w = h =
9.20$ m and $m_s=3.60$ kg, that corresponds to an area-to-mass ratio of
$A_s/(m_b+m_s)=0.75$ ${\rm m}^2$/kg.

The results are exemplified for spacecraft with $\al=35^\circ, 40^\circ,
45^\circ, 60^\circ$, all of them with $d=0$, as for these parameters the
Sun-pointing attitude is helio-stable, see Fig.~\ref{fig:stabsunpoint}.
It is worth noting that smaller aperture angles $\alpha$ allow smaller
oscillation amplitude and hence smaller angular velocity, and any value of $d$
with $|d|>0$ would give rise to a larger size of the gravity gradient
perturbation.

\begin{figure}[h!]
\begin{center}
\includegraphics[width = 0.45\textwidth]{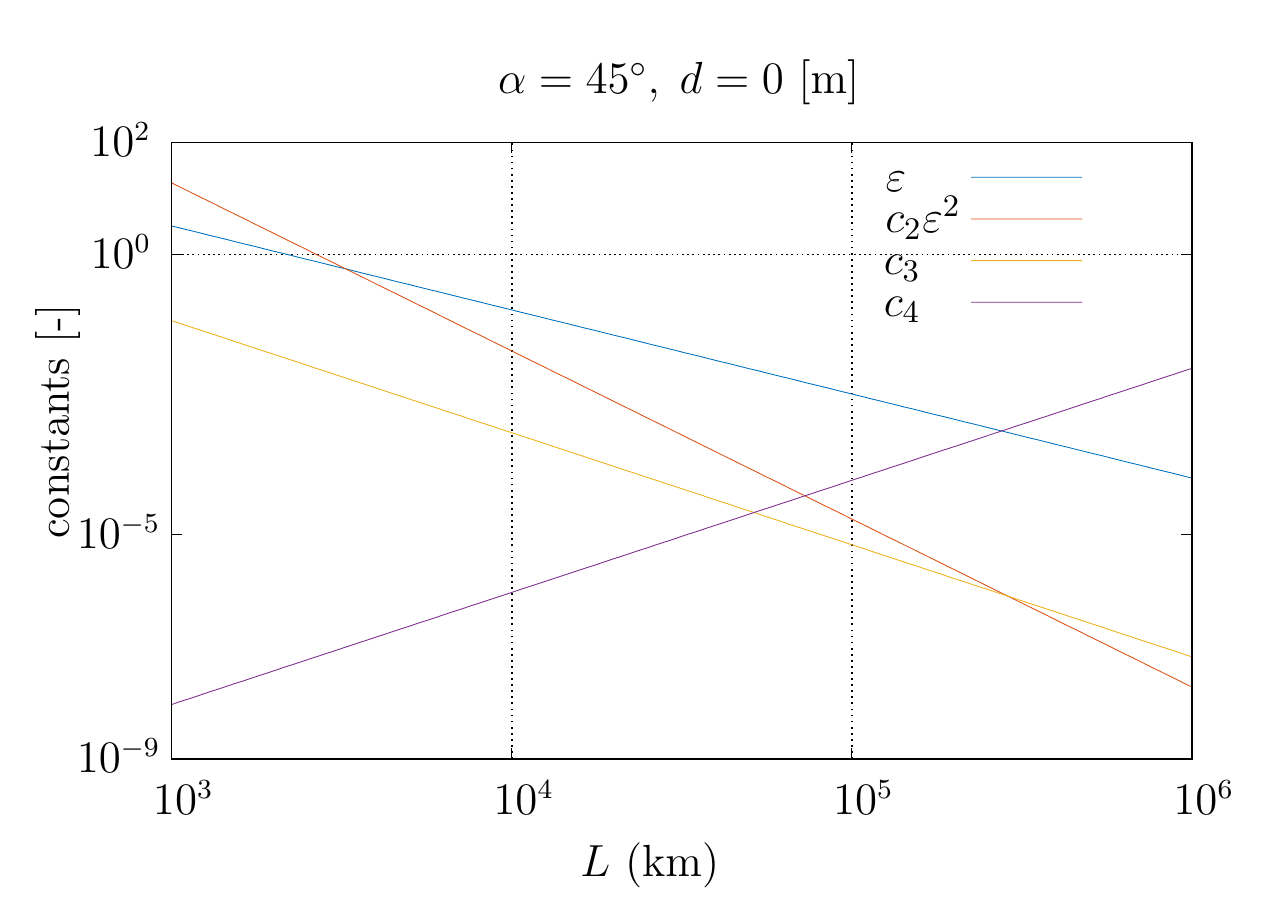}
\includegraphics[width = 0.45\textwidth]{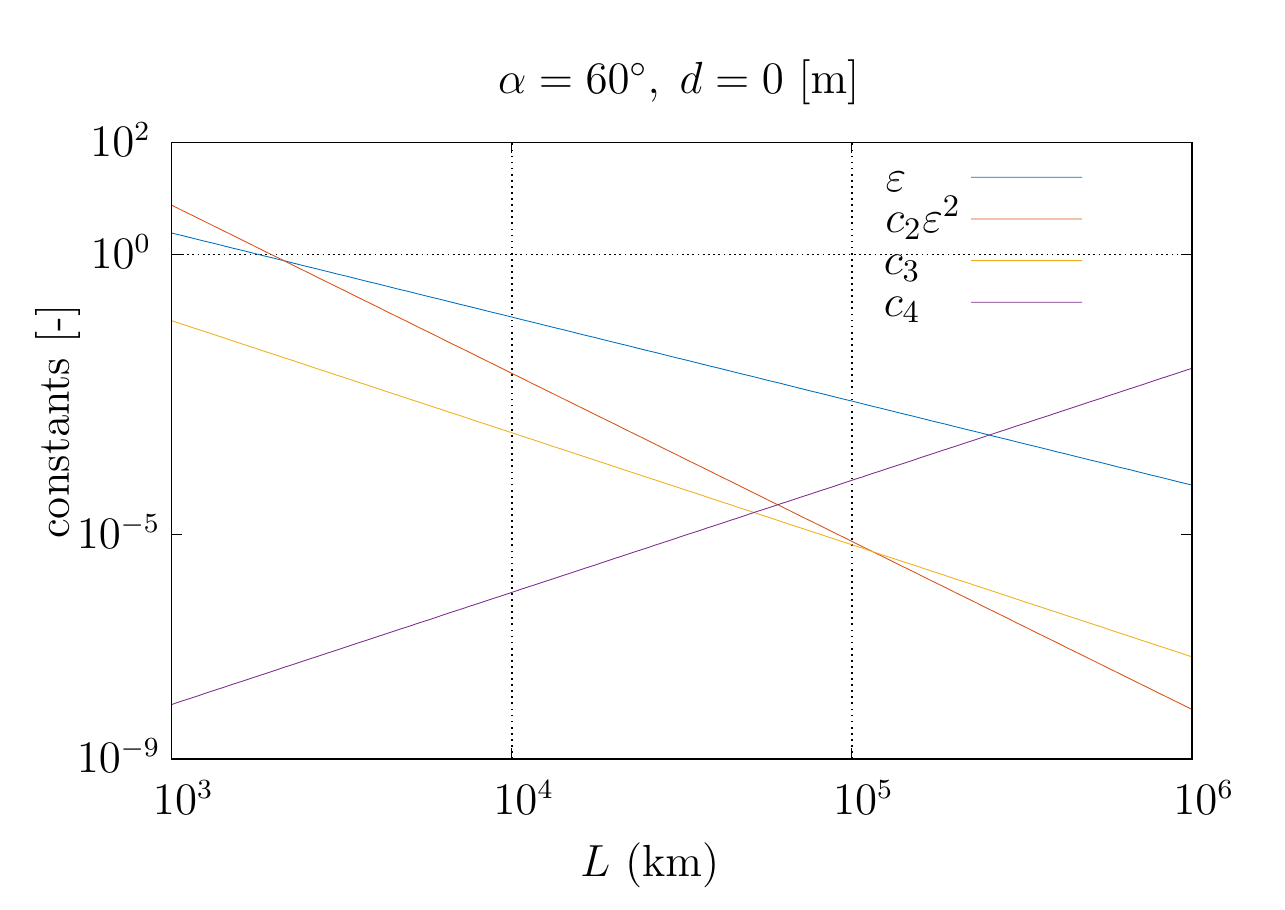}
\end{center}
\caption{Constants $c_1^{-1/2},\eps^2 c_2,c_3$ and $c_4$
(Eq.~\ref{eq:fullconstants}) as a function of $L$, for the cases,
$\al=45^\circ$ (left panel) and $\al=60^\circ$, (right panel), both with $d=0$
m.}
\label{fig:choiceLT}
\end{figure}

Now, the values of the parameters still depend on the adimensionalisation
quantities, $L$ for length and $T$ for time, see App.~\ref{sect:adim}. There it
is justified that to obtain Eq.~\ref{eq:tointegrate} one has to choose
$T=\sqrt{L^3/\mu}$ so only $L$ has to be chosen. It has to be done in a way
that $\eps=1/c_1^{-1/2}$ is small, the gravity gradient torque has to have
smaller size than the SRP torque, and the $J_2$ and SRP accelerations have to
be also smaller than the term of the Kepler problem that is $\Ocal(1)$. In
Fig.~\ref{fig:choiceLT} the values of $\eps=1/c_1^{1/2}$, $c_2\eps^2$ - as it
appears as prefactor in the equation of $\hPhi'$ in Eq.~\ref{eq:tointegrate} -,
$c_3$ and $c_4$ are shown as a function of the adimensionalisation length
factor $L$, recall Eq.~\ref{eq:fullconstants}, for $\al=45^\circ, d=0$ m (left)
and $\al=60^\circ, d=0$ m (right), as examples.  From the two panels in
Fig.~\ref{fig:choiceLT} one infers that $L=20\,000$ km (that is highlighted as
a vertical dashed line) is a proper choice for the purposes of this
contribution according to the requirements listed above.  The figures
corresponding to $\al=35^\circ$ and $\al=40^\circ$ are qualitatively the same
and quantitatively very similar.

With all these choices, the physical constants of the system that are used in
the following sections are summarized in Tabs.~\ref{tab:valuesconstants1} and
\ref{tab:valuesconstants2}. It is worth noting that, in both cases, $\eps =
\Ocal(10^{-2})$ and that one unit of time $t$ of Eq.~\ref{eq:tointegrate} is
equivalent to $T\eps\approx2-3$ min.

\begin{table}[h!]
\begin{center}
\begin{tabular}{c|c|c}
& $\al=35^\circ,d=0$ m & $\al=40^\circ,d=0$ m\\ \hline
$c_1$ &   $4.133317062536305\times10^{2}$     & $5.747509656406245\times10^{2}$\\
$c_2$ &   $2.014647115843597\times10^{0}$     & $1.923989341570575\times10^{0}$\\ 
$c_2$ &   $1.650597476175750\times10^{-4}$    & $1.650597476175750\times10^{-4}$\\
$c_2$ &   $3.738547970136426\times10^{-6}$  & $3.738547970136426\times10^{-6}$\\ \hline
$\eps$&   $4.918703449585804\times10^{-2}$    & $4.171191657263433\times10^{-2}$\\
Time unit & $2.203569462524180\times10^{2}$ s & $1.868685651104933\times10^{2}$ s 
\end{tabular}
\end{center}
\caption{Values of the physical parameters for $\alpha=35^\circ$ and
$\alpha=40^\circ$.}
\label{tab:valuesconstants1}
\end{table} 

\begin{table}[h!]
\begin{center}
\begin{tabular}{c|c|c}
& $\al=45^\circ,d=0$ m & $\al=60^\circ,d=0$ m\\ \hline
$c_1$ &   $7.624959636935995\times10^{2}$   & $1.366246396170031\times10^{3}$\\
$c_2$ &   $1.811184377377631\times10^{0}$   & $1.297157388066479\times10^{0}$\\ 
$c_2$ &   $1.650597476175750\times10^{-4}$  & $1.650597476175750\times10^{-4}$\\
$c_2$ &   $3.738547970136426\times10^{-6}$  & $3.738547970136426\times10^{-6}$\\ \hline
$\eps$&   $3.621439426788271\times10^{-2}$  & $2.705424915355282\times10^{-2}$\\
Time unit & $1.622397734086550\times10^{2}$ s & $1.212025036217823\times10^{2}$ s 
\end{tabular}
\end{center}
\caption{Values of the physical parameters for $\alpha=45^\circ$ and
$\alpha=60^\circ$.}
\label{tab:valuesconstants2}
\end{table} 

\subsection{Main numerical experiment}\label{subsec:numexp}

The attitude and orbit coupling and averaging results of \S~\ref{sect:dynasp}
are exemplified using one single orbit initial condition on which all the
considered disturbing effects play a strong role. Namely, all motions
considered start at the perigee of a Keplerian orbit characterized by 
$$
a_0 = 9000\;{\rm km}, \quad e_0 =0.25, \quad \omega_0 = 0^\circ,\quad \Omega_0 = 0^\circ.
$$

Note that, in particular, the perigee radius is above the surface of planet
Earth (with minimal altitude $350$ km) and this remains true along all
propagation performed. Despite the low altitude, atmospheric drag is not
considered as it was not taken into account in the study of
\S~\ref{sect:dynasp}. Such considerations are left for future contributions.

Concerning the attitude initial conditions, a sampling between the Sun-pointing
attitude and close to the limit of the validity of the theoretical results,
$|\hphi|<\alpha$, has been performed. More concretely, for
each aperture angle $\alpha$, 480 initial conditions of the form
\begin{eqnarray}\label{eq:jalpha}
\hphi_0 & = & \frac{0.9 (j_\alpha+1)}{480}\alpha, \quad
\hPhi_0  = 0, \qquad
j_\alpha = 0,1,\ldots,479,
\end{eqnarray} 
have been considered. That is, $0<\hphi\leq0.90\alpha$ is sampled. This is to make
sure that most trajectories do not reach $|\hphi|>\alpha$ before the maximal
integration time. But this may also happen before for these initial conditions,
as will be observed later in the case $\al=35^\circ$.

Finally, the maximal integration time for all considered initial conditions has
been 1 year (that corresponds to a different maximal $t$ for each value of
$\alpha$). 

The theoretical study of \S~\ref{sect:dynasp} only justifies $\eps$-closeness
of the original and averaged system in intervals of length $\Ocal(1/\eps)$ in
the adimensional scale $t$. In actual time units, using the data in
Tabs.~\ref{tab:valuesconstants1} and \ref{tab:valuesconstants2} this length is
$\Ocal(10^2)$ times the time unit, that accounts for $10^4\;{\rm s}\approx
3\;{\rm h}$ in the studied 4 cases. Despite this, in the performed numerical
experiments one observes that for the considered orbit initial condition this
interval is larger. The study of the time intervals where such $\eps$-closeness
holds in a large family of orbits is out of the scope of the present paper and
hence it is not addressed here.

As a final consideration, as the flow is always transversal to $\hat{x} = 0$,
instead of the full 6D ODE we have considered the Poincar\'e section defined by 
\begin{eqnarray}\label{eq:poinsec}
\Sigma & = & \left\{\hat{x} = 0,\;\hat{y} < 0 \right\}.
\end{eqnarray}
Hence, the following is a study of a 5D discrete map on this surface. The
reason for this choice is twofold: on the one hand, this is a reduction of the
dimension of the phase space by one and this eases the analysis; and on the
other hand it allows for better comparison when studying the system in
Eq.~\ref{eq:tointegrate} and its averaged analogue: non-averaged and averaged
equations have been integrated with different numerical schemes (an implicit
Runge-Kutta Gauss of 2 stages and order 4 and a Runge-Kutta-Fehlberg 7(8),
respectively) with automatic stepsize control. The Poincar\'e section sets a
fixed position where to compare the integrated orbits of both equations. Note
that the integrated equations are different, and hence the times where the
orbits intersect $\Sigma$ do not necessarily coincide, even if the integration
starts at the same initial condition.

\subsection{Numerical results}\label{subsec:numres}

A selection of numerical results are shown and described here. These are
related to the shape, size and orientation of the osculating orbit along the
integration, to be able to compare between different apertures and oscillation
amplitudes; to the assessment of the applicability of the equivalent flat sail
in the averaged equations; and finally to the assessment of the difference
between the original and averaged equations.

\subsubsection{Shape, size and orientation of the osculating
orbit}\label{subsubsec:shapesize}

The shape, size and orientation of the orbit of the spacecraft are studied via
the Keplerian elements semi-major axis $a$, eccentricity $e$ and
$\gamma:=\omega + \Omega$, the sum of the argument of the perigee and the RAAN.
Recall that the latter has to be considered as the $J_2$ effect makes the
ascending node precess.  On a scale where the whole evolution along one
complete year of integration is displayed, the differences between these three
observables is qualitatively the same for all the considered values of
$\alpha$.

\begin{figure}[h!]
\begin{center}
\includegraphics[width = 0.45\textwidth]{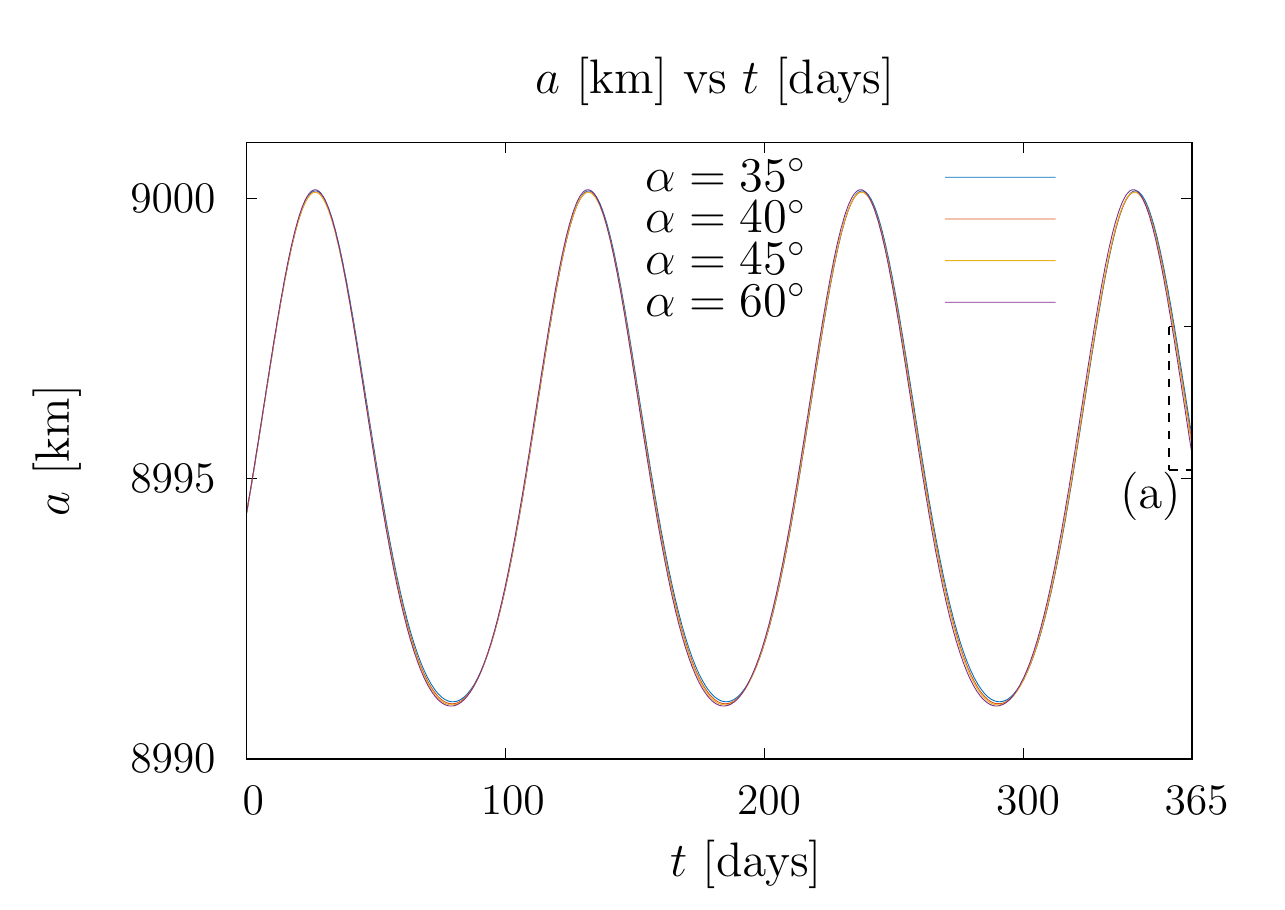}
\includegraphics[width = 0.45\textwidth]{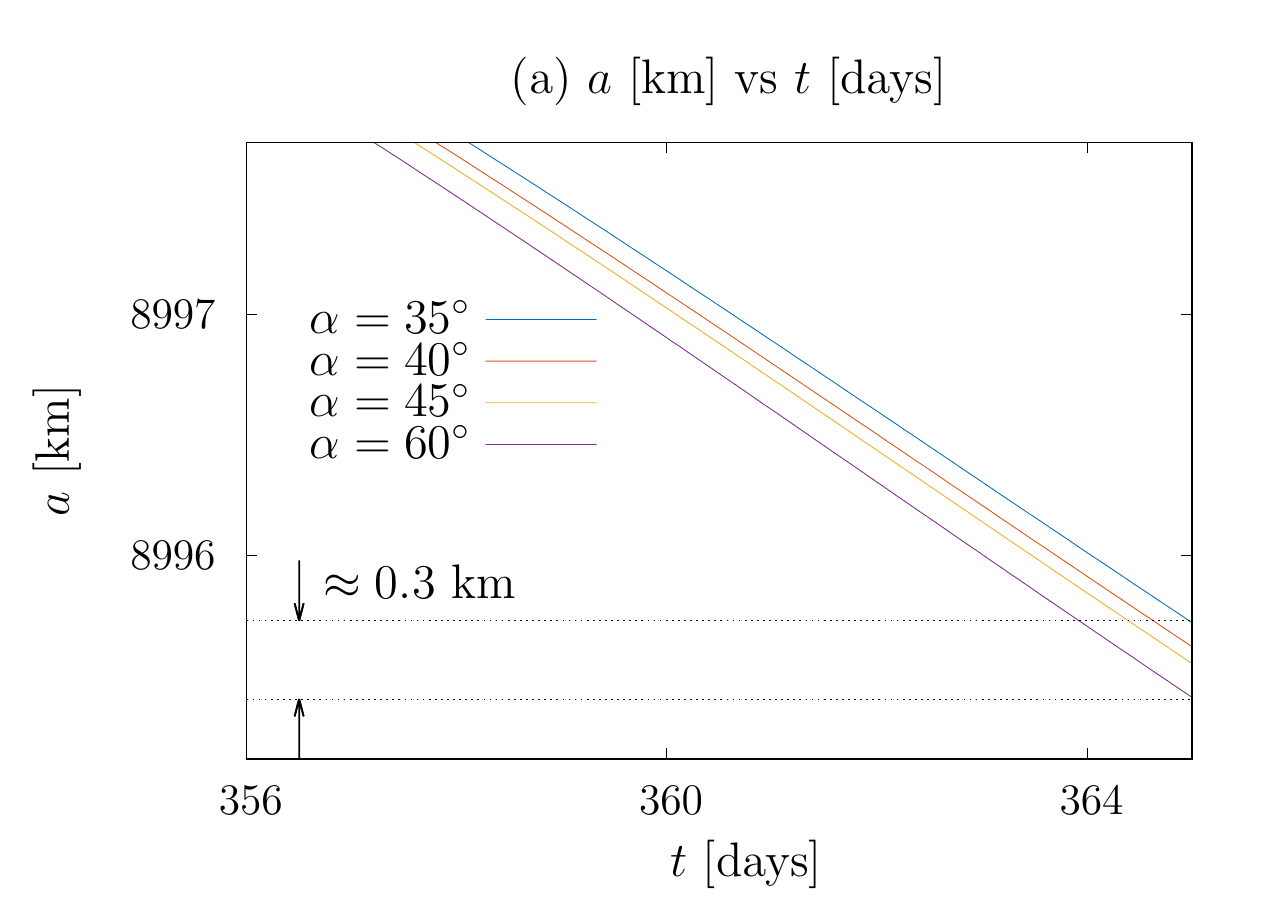}\\
\includegraphics[width = 0.45\textwidth]{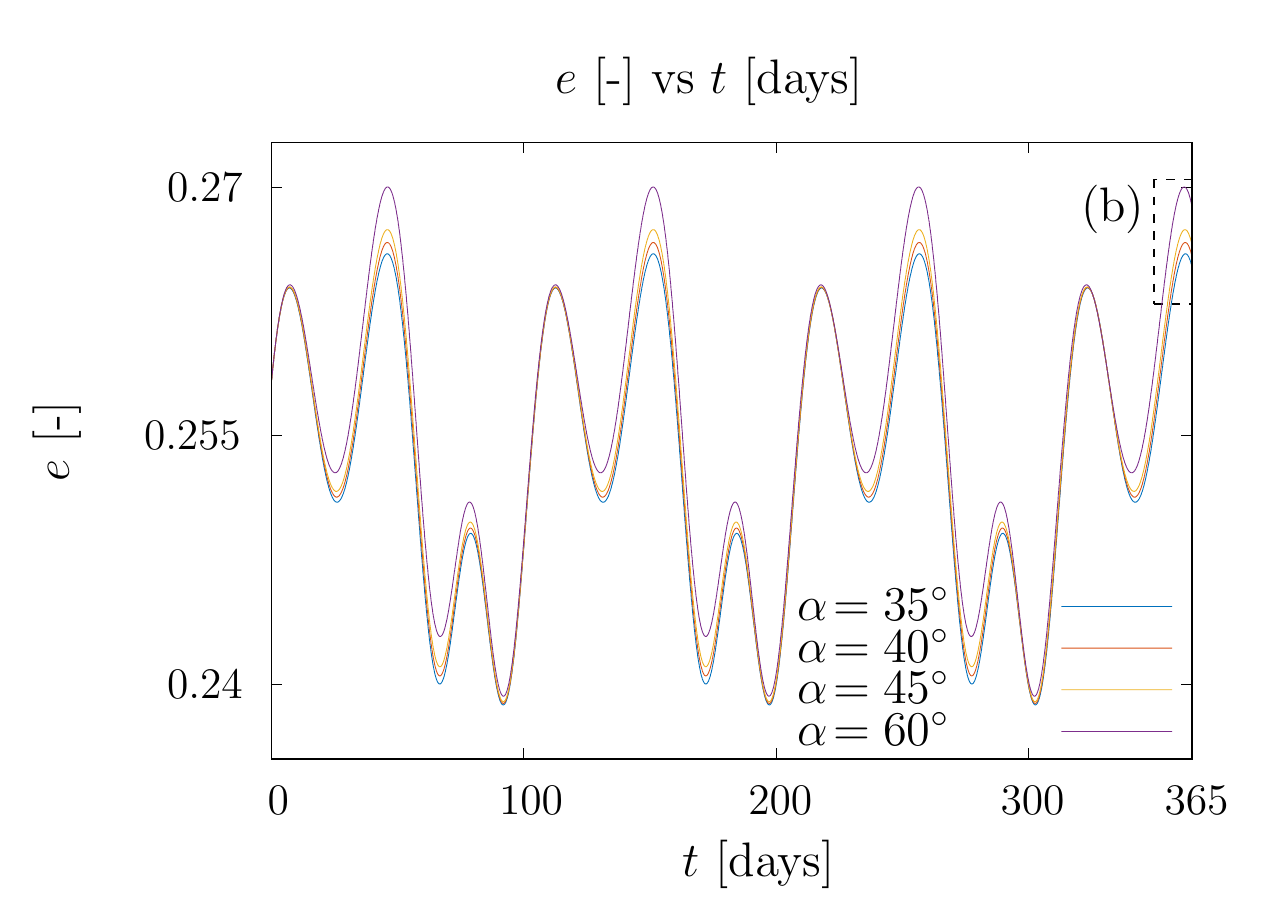}
\includegraphics[width = 0.45\textwidth]{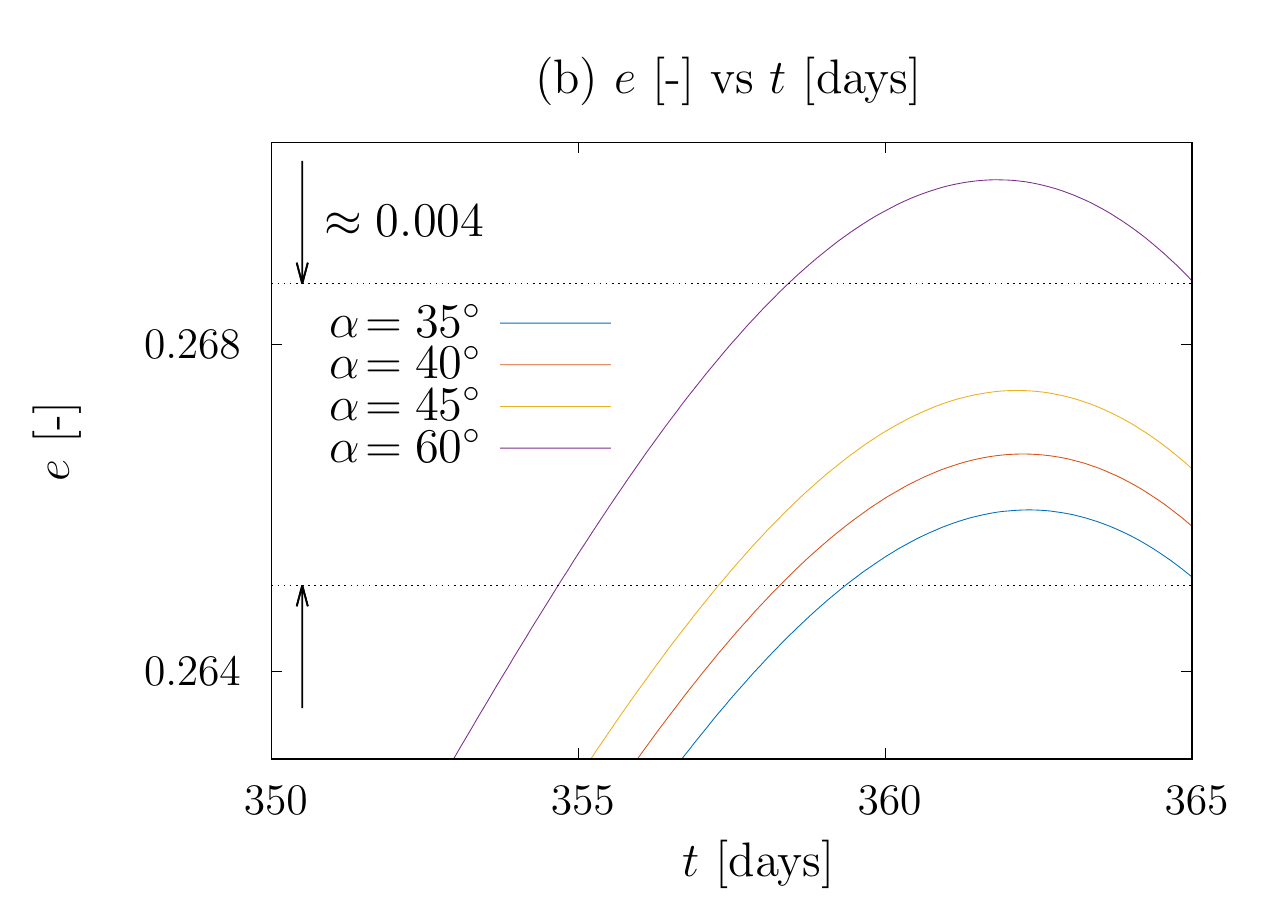}\\
\includegraphics[width = 0.45\textwidth]{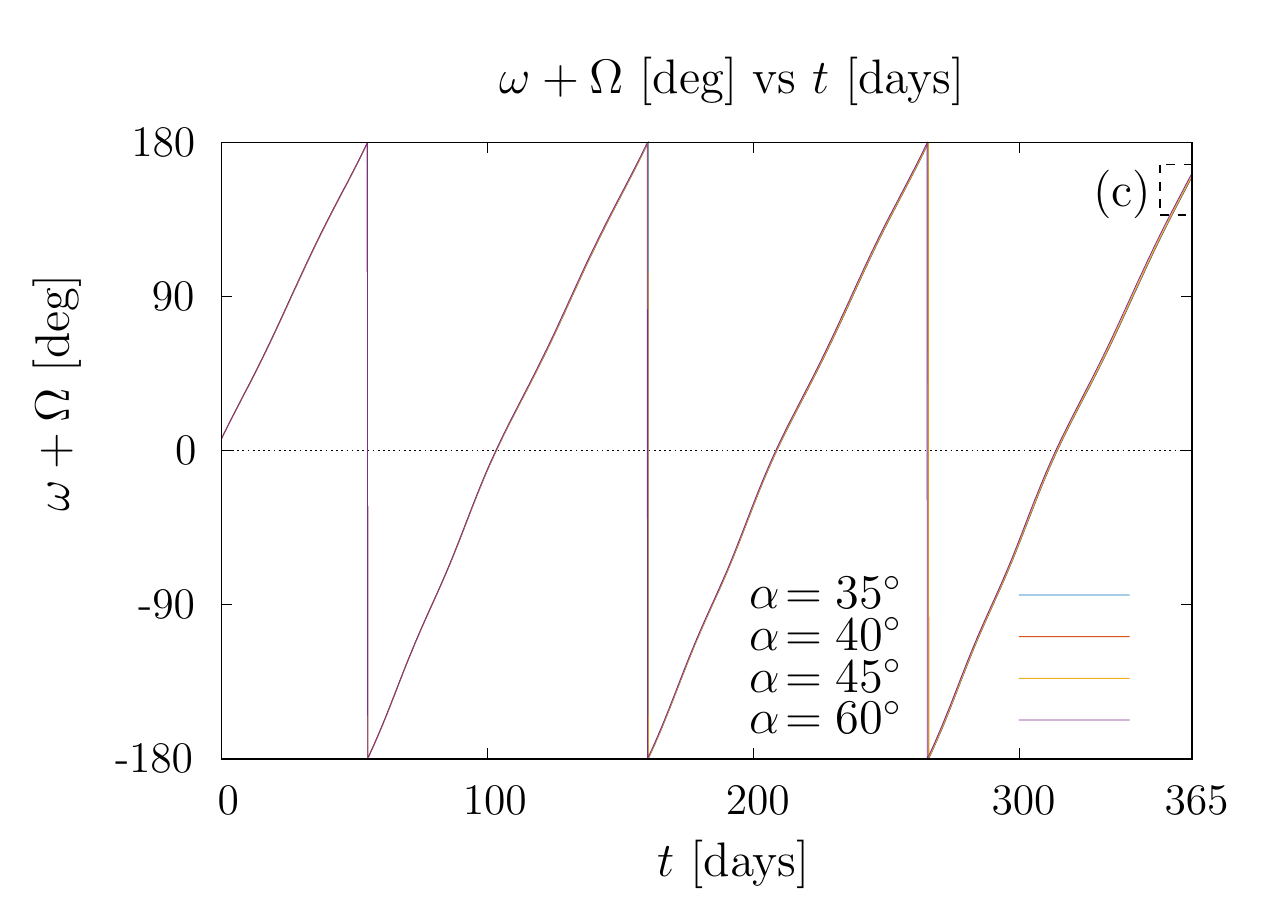}
\includegraphics[width = 0.45\textwidth]{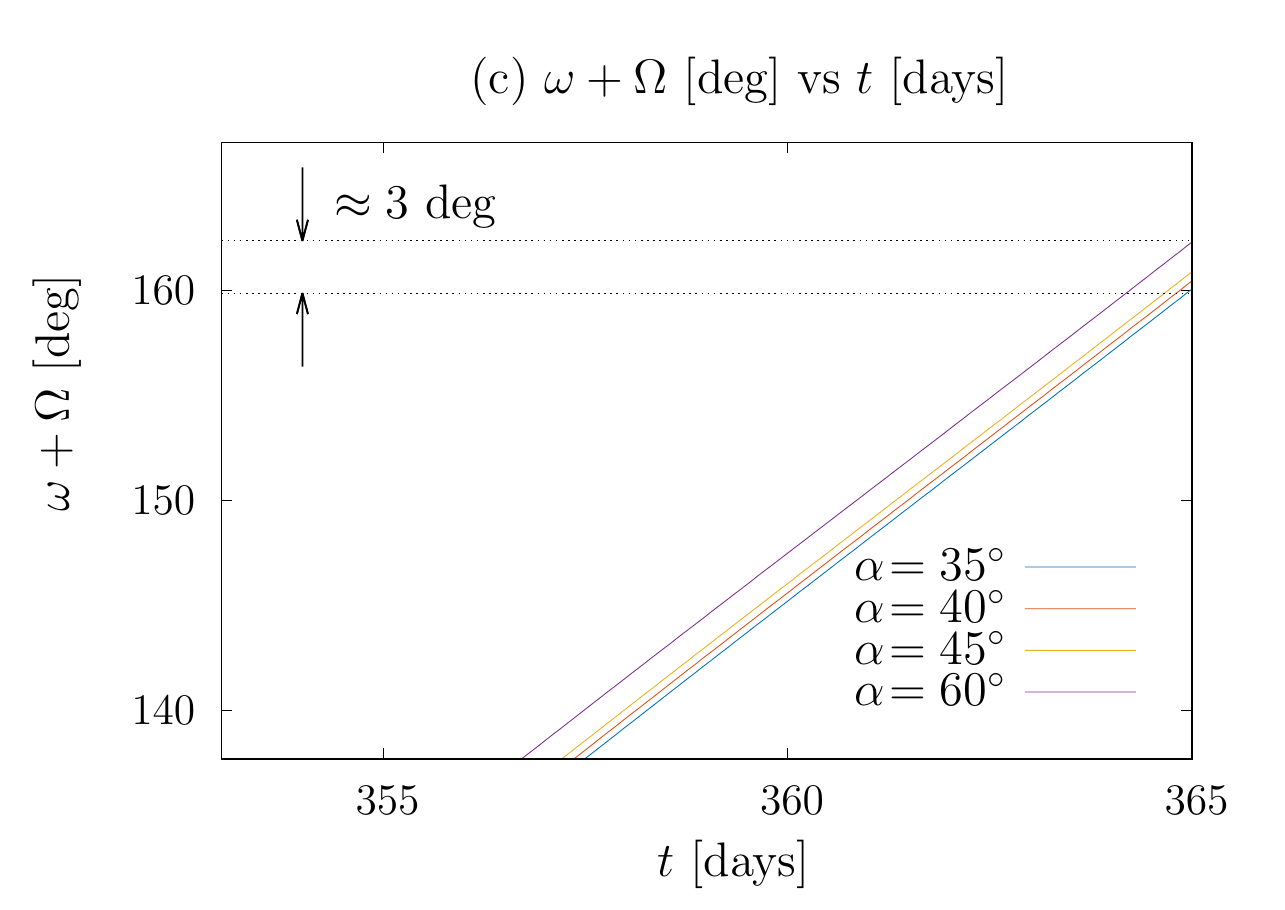}\\
\end{center}
\caption{
Evolution of the semi-major axis $a$ (top row), eccentricity $e$ (middle row)
and $\gamma=\omega + \Omega$ (bottom row) , the sum of the argument of the
perigee and the RAAN of the osculating orbit for the initial condition
$j_\alpha=0$ in Eq.~\ref{eq:jalpha}.}
\label{fig:samecc}
\end{figure}

In Fig.~\ref{fig:samecc} an example of such evolution is displayed. The shown
evolution is obtained with the attitude initial condition with $j_\alpha = 0$
in Eq.~\ref{eq:jalpha}, that is, the initial attitude closest to the
Sun-pointing direction considered. Top, middle and bottom panels show the
evolution of $a$, $e$ and $\gamma$ respectively. The differences due to
choosing different aperture angles are highlighted in the right column zooms. 

This shows that despite the sail has the same shape and size in all cases, the
aperture angle of the sail, even if it oscillates mildly around its
helio-stable attitude, produces a non-negligible impact on the orbit. Namely in
this case we observe $\Ocal(0.1)$ variations of $a$ measured in km, that is,
$\Ocal(10^{-2})$ in the adimensional variables; $\Ocal(0.001)$ variations in
$e$; and $\Ocal(1)$ variations in $\gamma$, measured in degrees. Other attitude
initial conditions show differences of the same orders of magnitude.

\subsubsection{The definition of equivalent effective area}

From the results concerning the variation of osculating orbit one reads that
the fact that the oscillations close to the Sun-pointing direction are fast
with respect to the orbit dynamics implies that the sail produces an average
uniform effect. This can be interpreted as the QRP sail behaving as a flat sail
with area $A_s\cdot A_{\rm eff}$, recall Eq.~\ref{eq:Aeff}.

\begin{figure}[h!]
\begin{center}
\includegraphics[width = 0.45\textwidth]{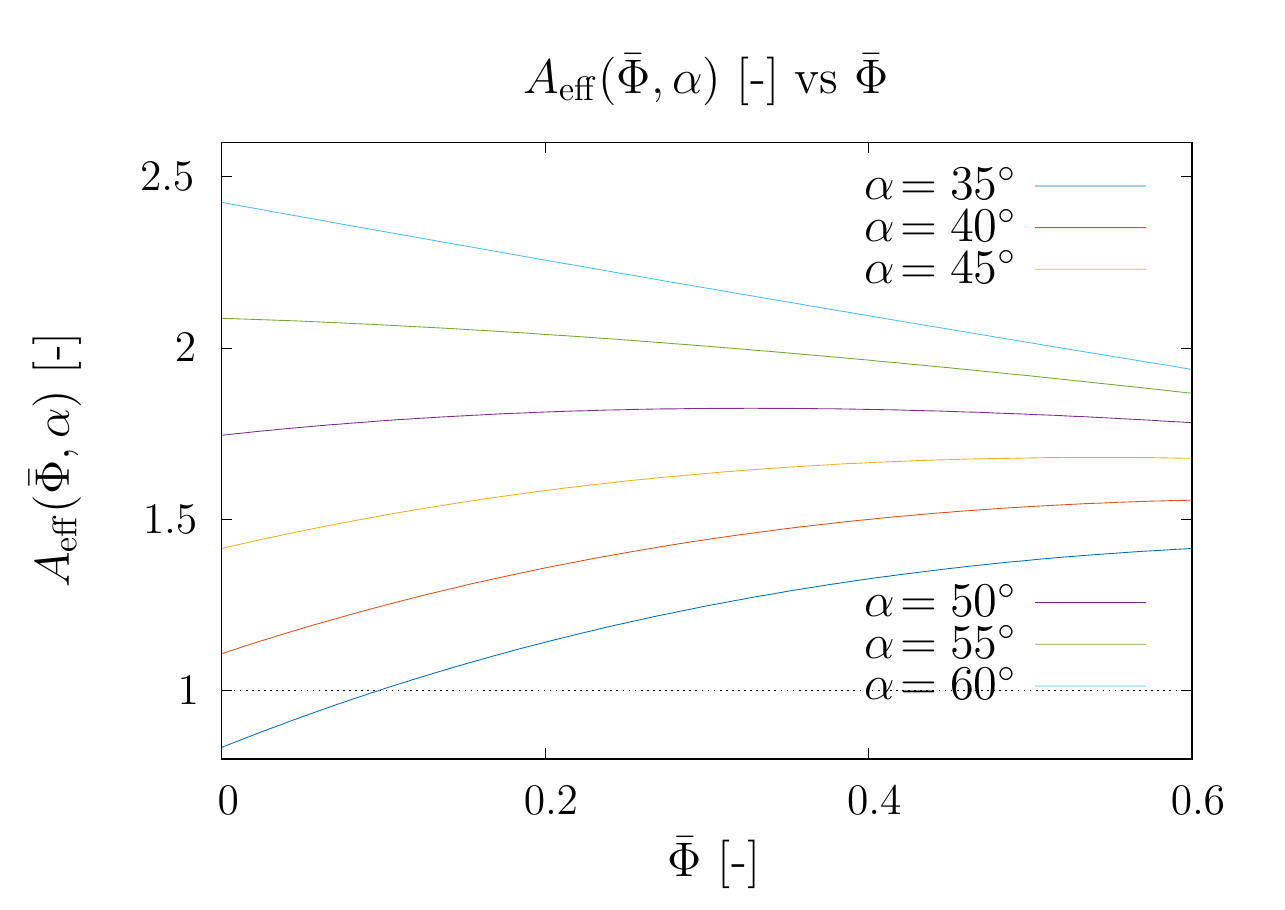}
\end{center}
\caption{Theoretical factor of the effective area of the averaged sail dynamics
$A_{\rm eff}$.}
\label{fig:Aeffteor}
\end{figure}

In Fig.~\ref{fig:Aeffteor} the area factor $A_{\rm eff}$ is plotted, as a
function of $\bPhi$ for different values of $\alpha$ that include the study
cases. 

Recall that $\bPhi = (2\hphi^2+\hPhi^2)/(2\sqrt{2})$, so this action gives
information about the amplitude of the oscillations around the Sun-pointing
direction. Figure~\ref{fig:Aeffteor} shows that one can choose the aperture
angle in such a way that for small oscillations the effect of the QRP on the
SRP acceleration magnitude is, on average, that of more than 2 flat panels of
area $A_s$. In other words, for adequate $\alpha$, the oscillations are
equivalent as to having a larger flat panel always oriented towards the Sun.

The question now is how to measure if this effect can be recovered in the
attitude dynamics and SRP perturbation in the performed simulations. Recall
that the results of the simulations are data on $\Sigma$ of the propagation
of Eq.~\ref{eq:tointegrate}. To test the formulas provided in
\S~\ref{sect:dynasp}, one can proceed as follows:
\begin{enumerate}
	\item {\sf Measure of the average of the action}. The quantity $\bPhi =
(2\hphi^2+\hPhi^2)/(2\sqrt{2})$ evaluated along the data appears to oscillate
in what seems a quasi-periodic manner. Since we are interested in the average,
for each initial condition, we denote $\left<\bPhi\right>$ the time average of
$\bPhi$ along the whole 1-year long integration.
	\item {\sf Evaluation of the area factor}. With each obtained value of
$\left<\bPhi\right>$ we can evaluate $A_{\rm
eff}(\left<\bPhi\right>,\alpha)$, the area factor, see Eq.~\ref{eq:Aeff}. This
is referred to as the ``theoretical" value.
	\item {\sf Measure of the average SRP acceleration}. The values of the
components of the SRP acceleration in Eq.~\ref{eq:accSRP2Pan} are also averaged
along the 1-year long data for each initial condition. As given in this
equation, the obtained average values correspond to the area factor divided by
either $\cos(\hat{\lambda})$ (along $\hat{x}$) or $\sin(\hat{\lambda})$ (along
$\hat{y}$). The mean is denoted as $A_{\rm eff}^\star
(\left<\bPhi\right>,\alpha)$, and referred to as the ``numerical" value. 
\end{enumerate}

\begin{figure}[h!]
\begin{center}
\includegraphics[width = 0.45\textwidth]{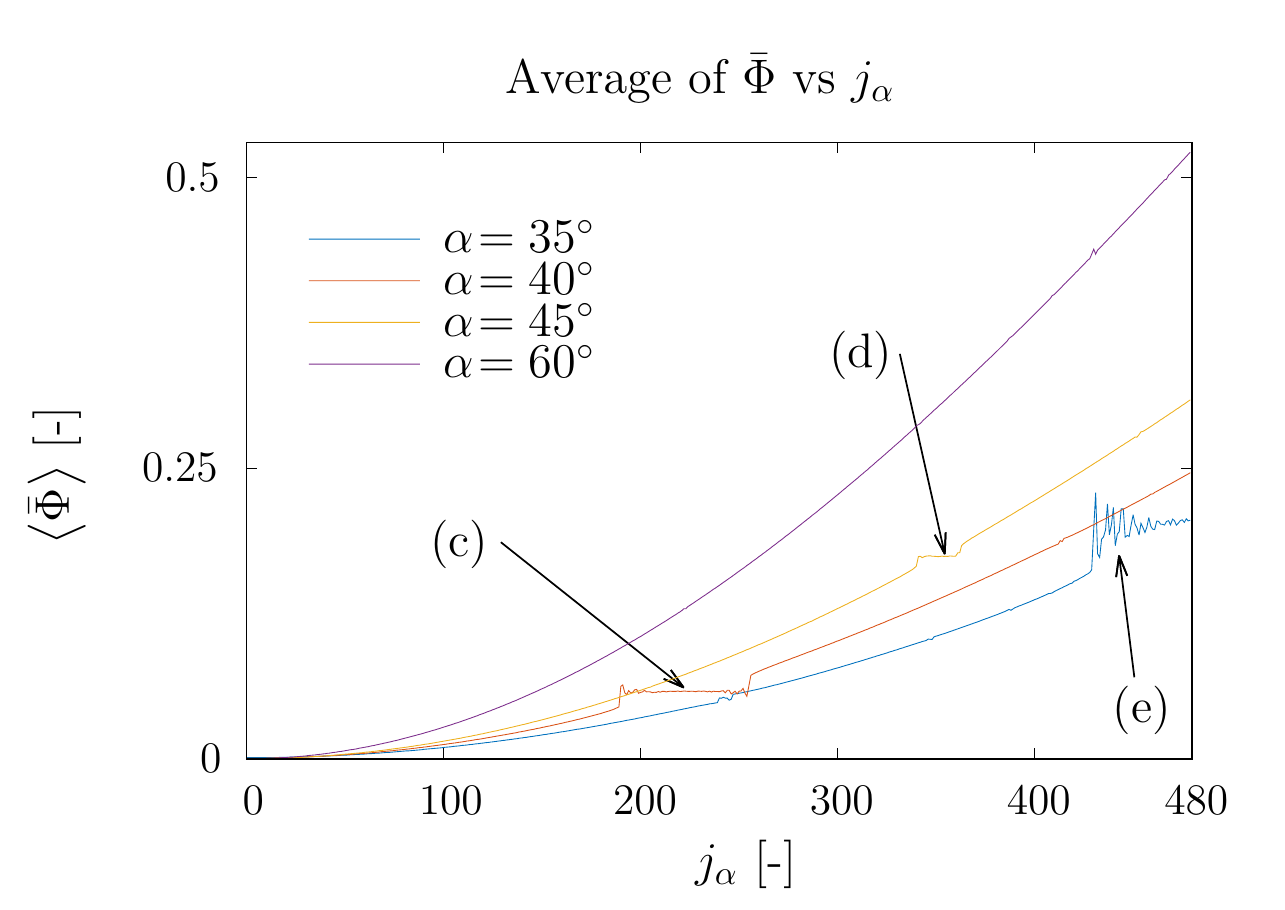}\\
\includegraphics[width = 0.45\textwidth]{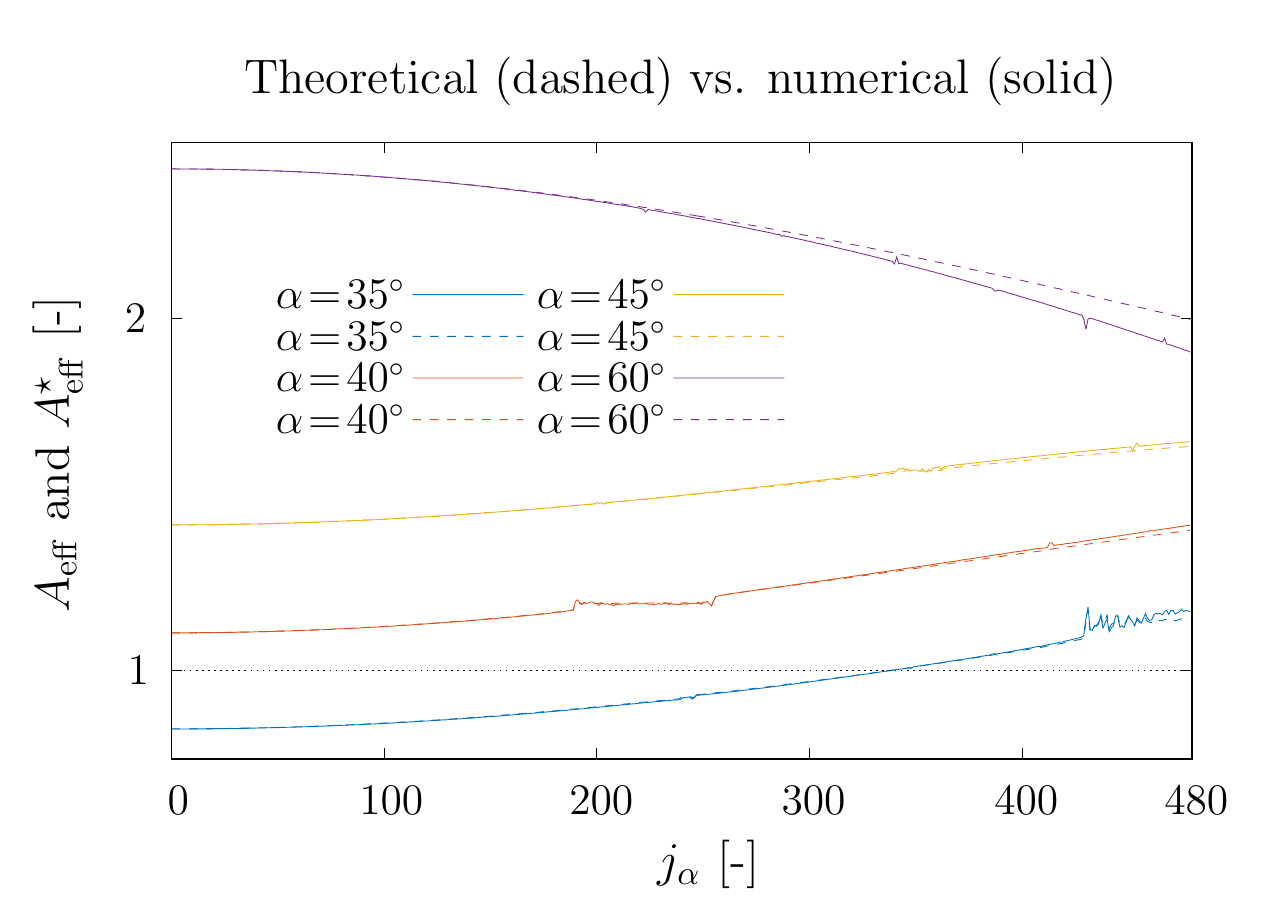}
\includegraphics[width = 0.45\textwidth]{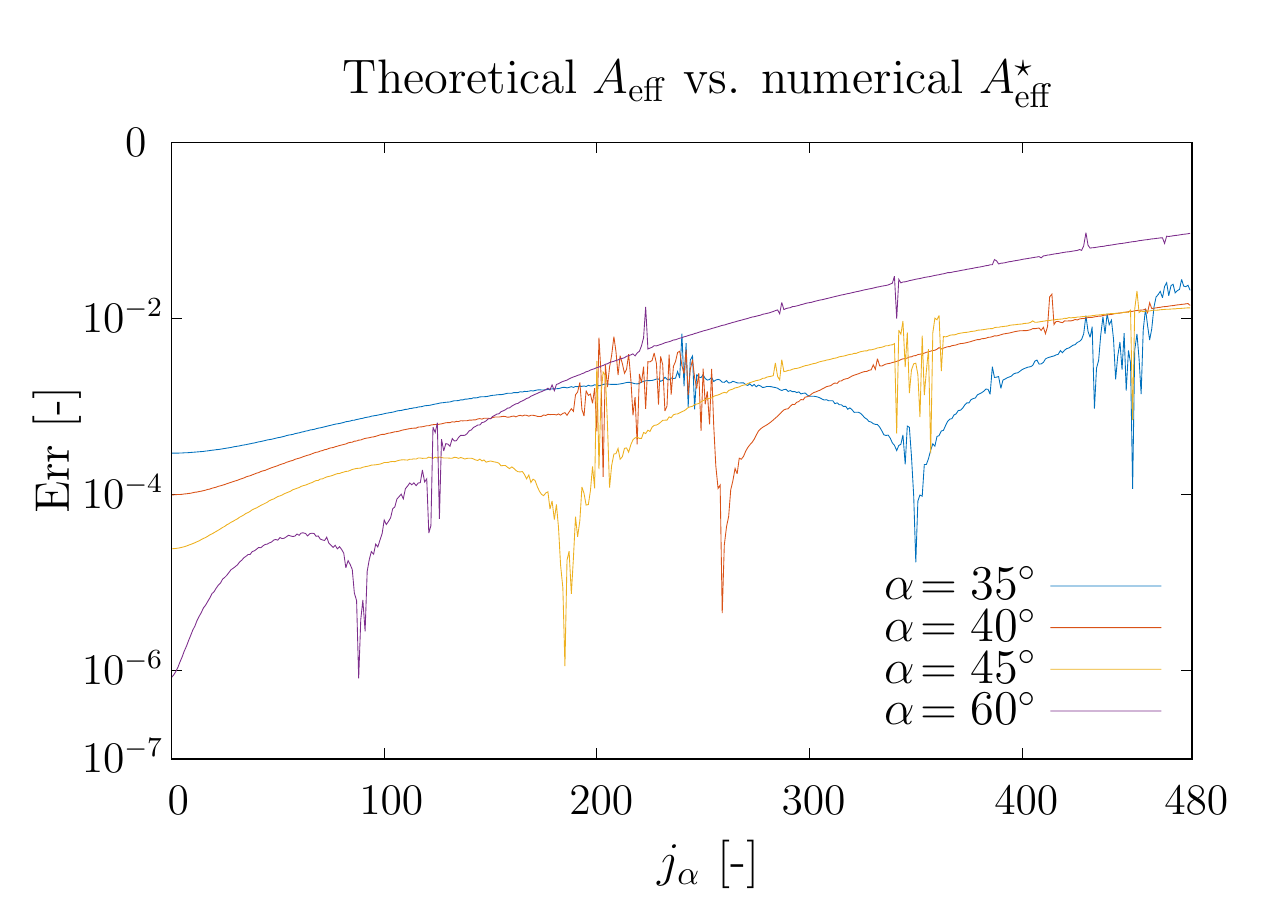}
\end{center}
\caption{Results of the numerical study of the equivalent flat sail area. Top:
time average of $\bPhi = (2\hphi^2+\hPhi^2)/(2\sqrt{2})$. Bottom: theoretical
vs. numerical area factor; left: nominal values; right: error in $\log_{10}$
scale.}
\label{fig:Effarea}
\end{figure}

The results of this study are summarized in Fig.~\ref{fig:Effarea}. To allow
comparison for different aperture angles, since each aperture angle has a
different maximal oscillation, $j_\alpha$ is chosen as abscissa instead of
$\hphi_0$ itself, recall Eq.~\ref{eq:jalpha}. The top panel displays the time
average of $\bPhi = (2\hphi^2+\hPhi^2)/(2\sqrt{2})$. This shows that the
tendency is that the larger $\alpha$ is the faster $\left<\bPhi\right>$
increases in the $|\hphi|<\alpha$ region. In this figure three phenomena are
highlighted as they require further clarification. Both (d) and (e) correspond
to the same phenomenon that occur for $\alpha = 40^\circ$ and $45^\circ$,
respectively, that is a plateau. This can be explained by the fact that the
corresponding initial conditions do not evolve on a torus or close to a torus
but on a large width chaotic region where the attitude can range freely.  Even
though the phase space of our problem is 5D and confining manifolds in this
context have dimension 4, a geometrical analogy would be the dynamics of a
chaotic orbit in a Birkhoff region of an area-preserving map (2D): a region
that is bounded by invariant curves (confining manifolds of dimension 1) with
no other invariant curves inside, where chaotic orbits are confined and their
iterates eventually fill a positive measure set that range the whole width
between the two confining curves.  The variations in (f) that occur for
$\alpha=35^\circ$ are spurious data, as the corresponding sail reached
$|\hphi|>\alpha$ before 1 year.

The bottom left panel of Fig.~\ref{fig:Effarea} shows the theoretical and
numerical values of the area factor. The bottom right panel shows the
difference between the theoretical and numerical values of $A_{\rm eff}$ (as
defined in the enumeration in the beginning of this section)
$$
{\rm Err} = \left|A_{\rm eff}^\star - A_{\rm eff}\right|.
$$
On the left the theoretical and numerical values are displayed in dashed and
solid lines, respectively. As expected from the hypotheses of the theory, the
fit is better the closer to the Sun-pointing direction we are, but even for
larger values of $\left<\bPhi\right>$ Eq.~\ref{eq:Aeff} still gives a good
first approximation of the area factor.

\subsubsection{Full system versus averaged system}

The application of the averaging method converts Eq.~\ref{eq:tointegrate} into
the simplified version that only contains orbit dynamics whose equations read 
\begin{eqnarray}\label{eq:tointegrate-av}
\left\{
\begin{array}{lcl}
\dot{\tilde{\tilde{x}}} &=& \tilde{\tilde{v}}_{\tilde{\tilde{x}}},\\
\dot{\tilde{\tilde{y}}} &=& \tilde{\tilde{v}}_{\tilde{\tilde{y}}},\\
\dot{\tilde{\tilde{v}}}_{\tilde{\tilde{x}}}    &=&\displaystyle -\frac{\tilde{\tilde{x}}}{\tilde{\tilde{r}}^3} - c_3\frac{\tilde{\tilde{x}}}{\tilde{\tilde{r}}^5} - c_4 A_{\rm eff} \cos(\lambda),\\ 
\dot{\tilde{\tilde{v}}}_{\tilde{\tilde{y}}}    &=&\displaystyle -\frac{\tilde{\tilde{y}}}{\tilde{\tilde{r}}^3} - c_3\frac{\tilde{\tilde{y}}}{\tilde{\tilde{r}}^5} - c_4 A_{\rm eff} \sin(\lambda), 
\end{array}
\right.
\end{eqnarray}
recall Eq.~\ref{eq:averaged}. Note that here the motion can be integrated in
the slow time scale $\tau$ so the notation $(\dot{\;})={\rm d}/{\rm d}\tau$ has
been used. To test numerically how close the orbits of
Eq.~\ref{eq:tointegrate-av} are to those of Eq.~\ref{eq:tointegrate}, we use
the numerically evaluated area factor, that is, we use
Eq.~\ref{eq:tointegrate-av} with $A^\star_{\rm eff}$ instead of $A_{\rm eff}$.

The initial conditions for this orbit propagation are the same ones as
explained in \S~\ref{subsec:numexp}.  That is, for each value of $\alpha$
considered, we integrate 480 orbits (one per value of $j_\alpha$), for 1 year,
keeping track of the intersections of each orbit in $\Sigma$, see
Eq.~\ref{eq:poinsec}.

\begin{figure}[h!]
\begin{center}
\includegraphics[width = 0.45\textwidth]{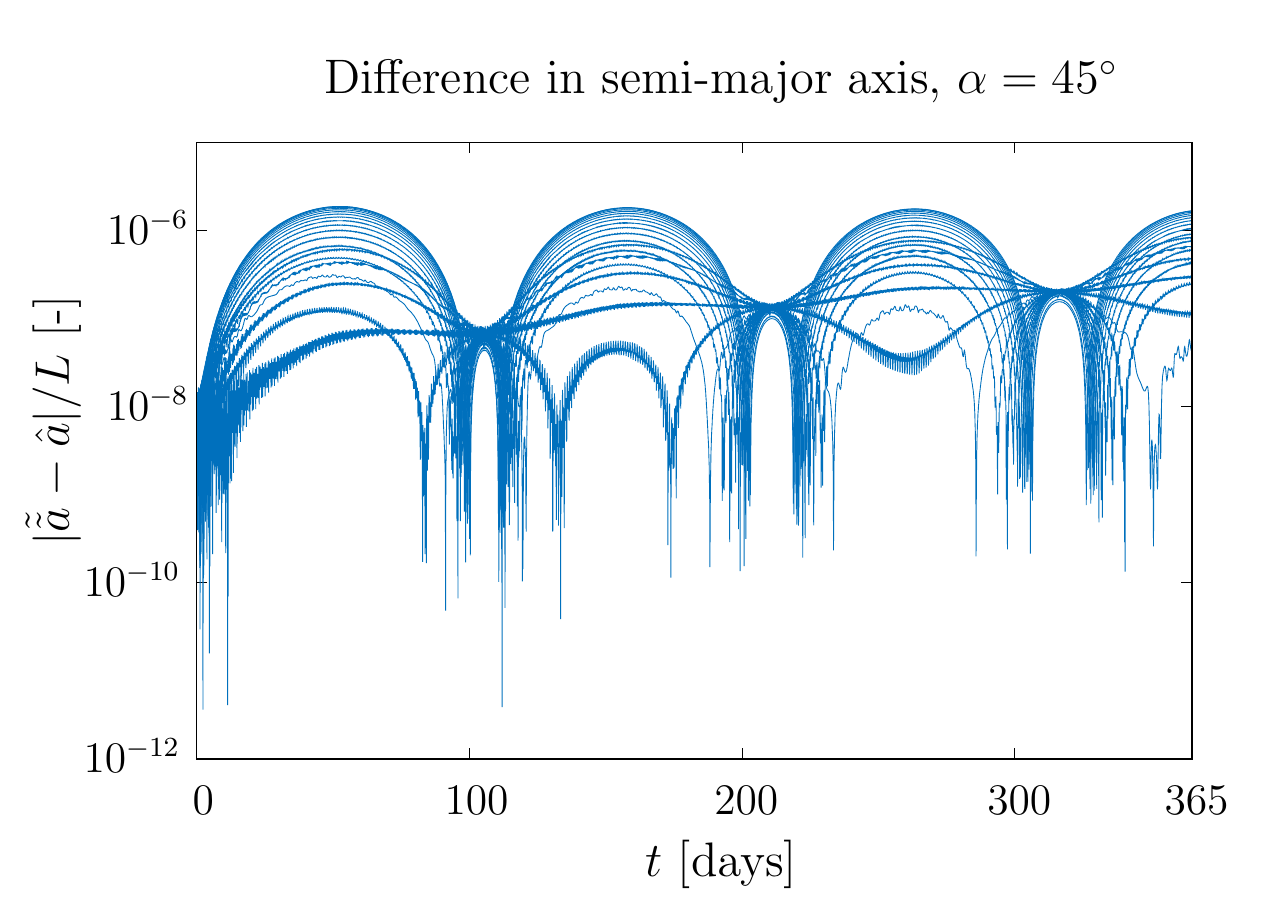}
\includegraphics[width = 0.45\textwidth]{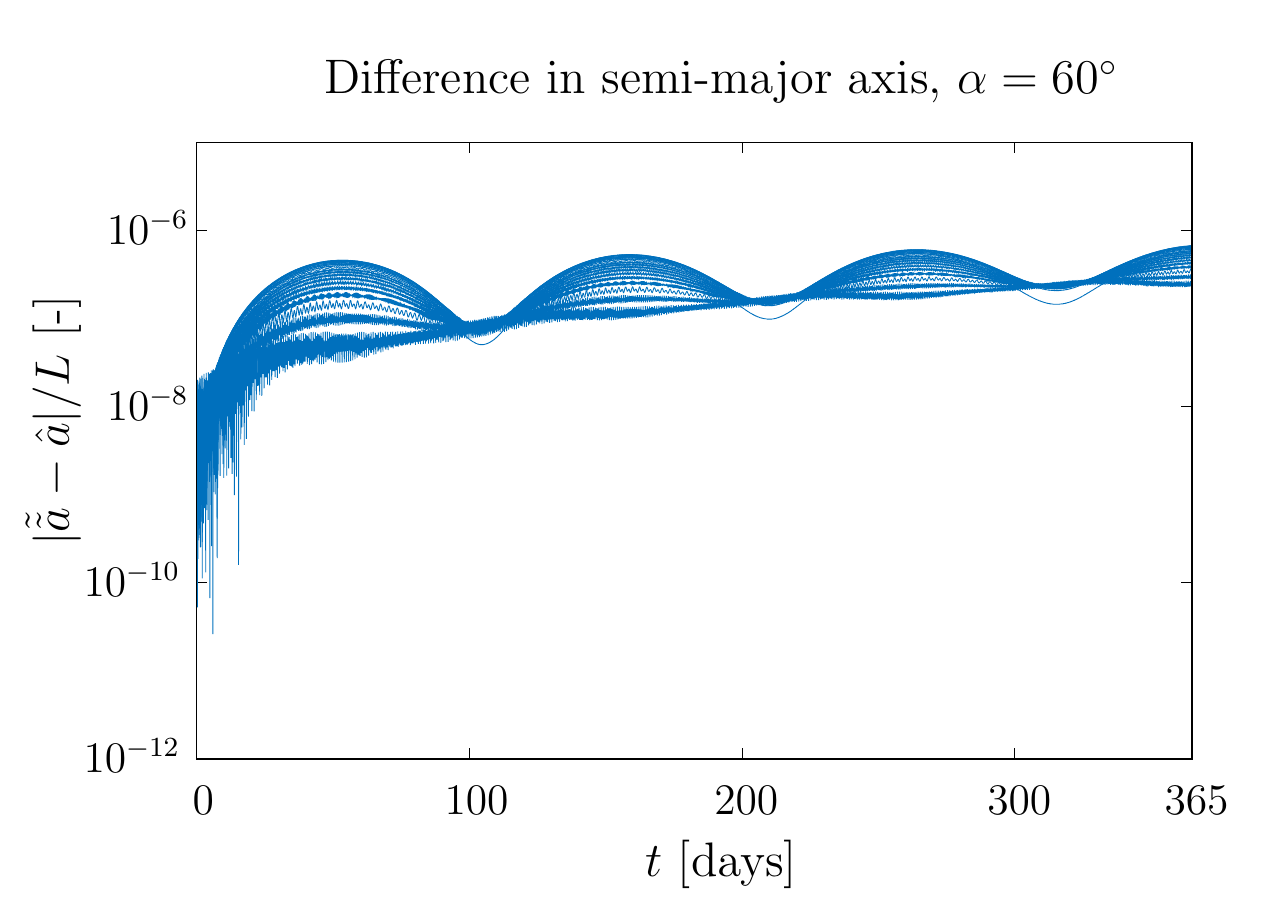}\\
\includegraphics[width = 0.45\textwidth]{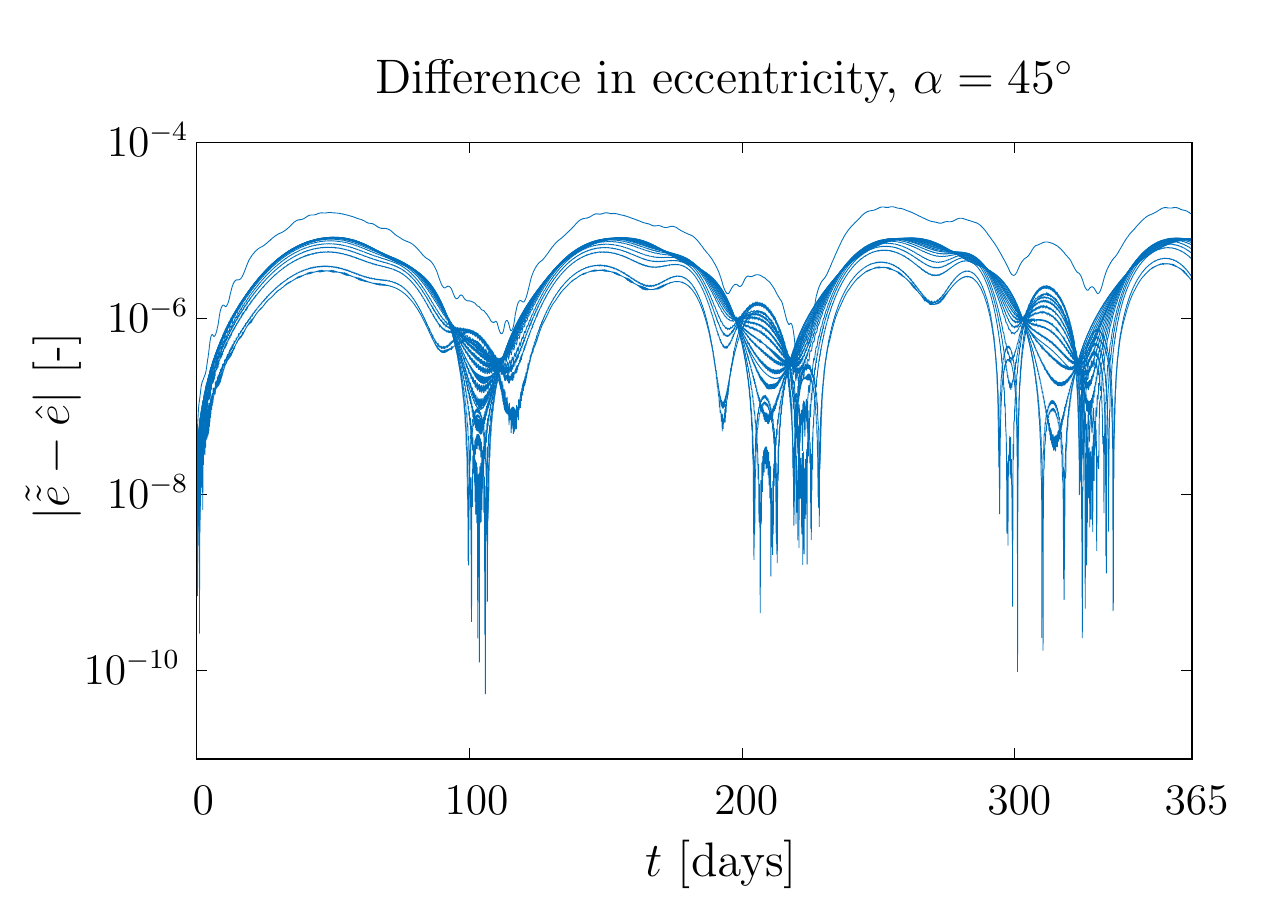}
\includegraphics[width = 0.45\textwidth]{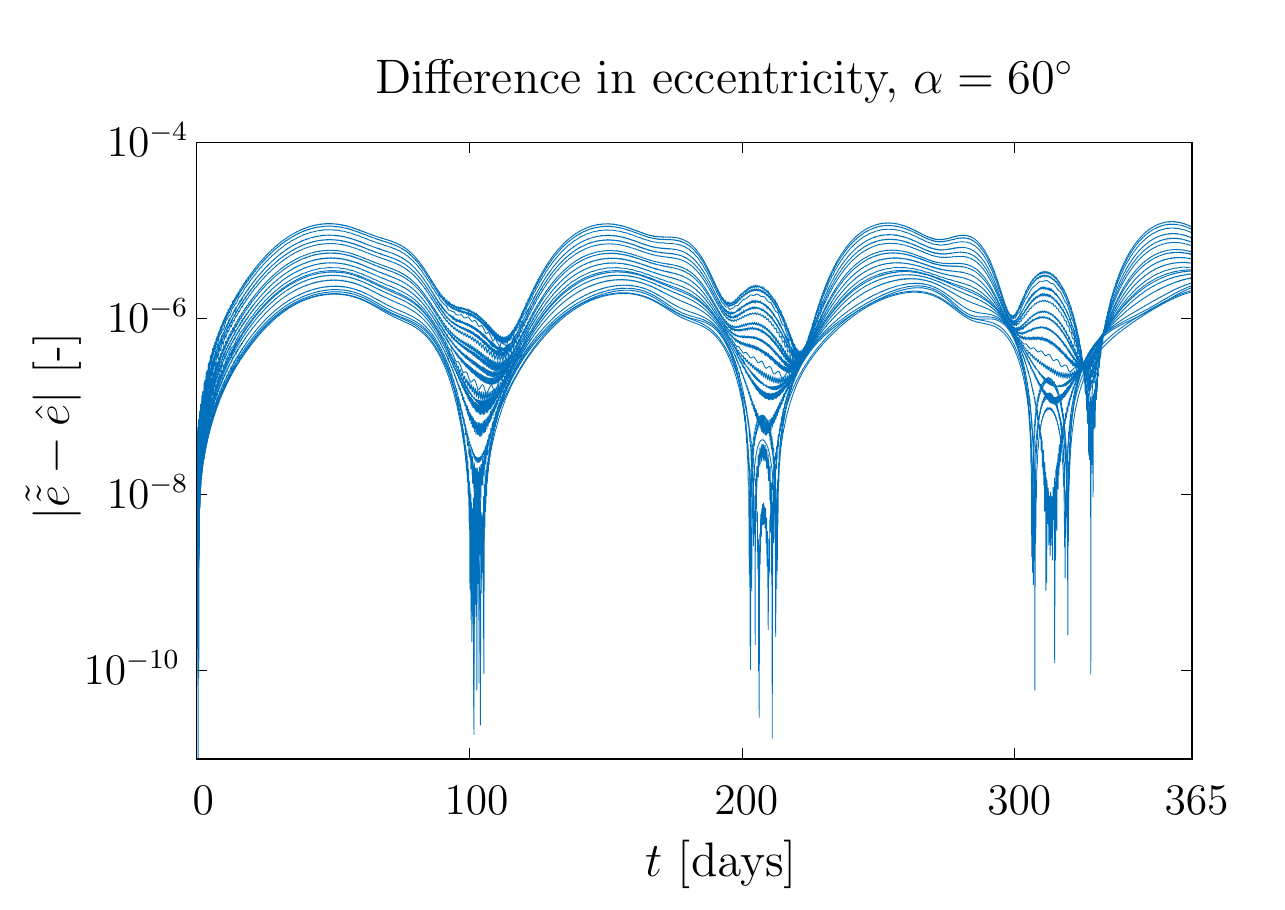}\\
\includegraphics[width = 0.45\textwidth]{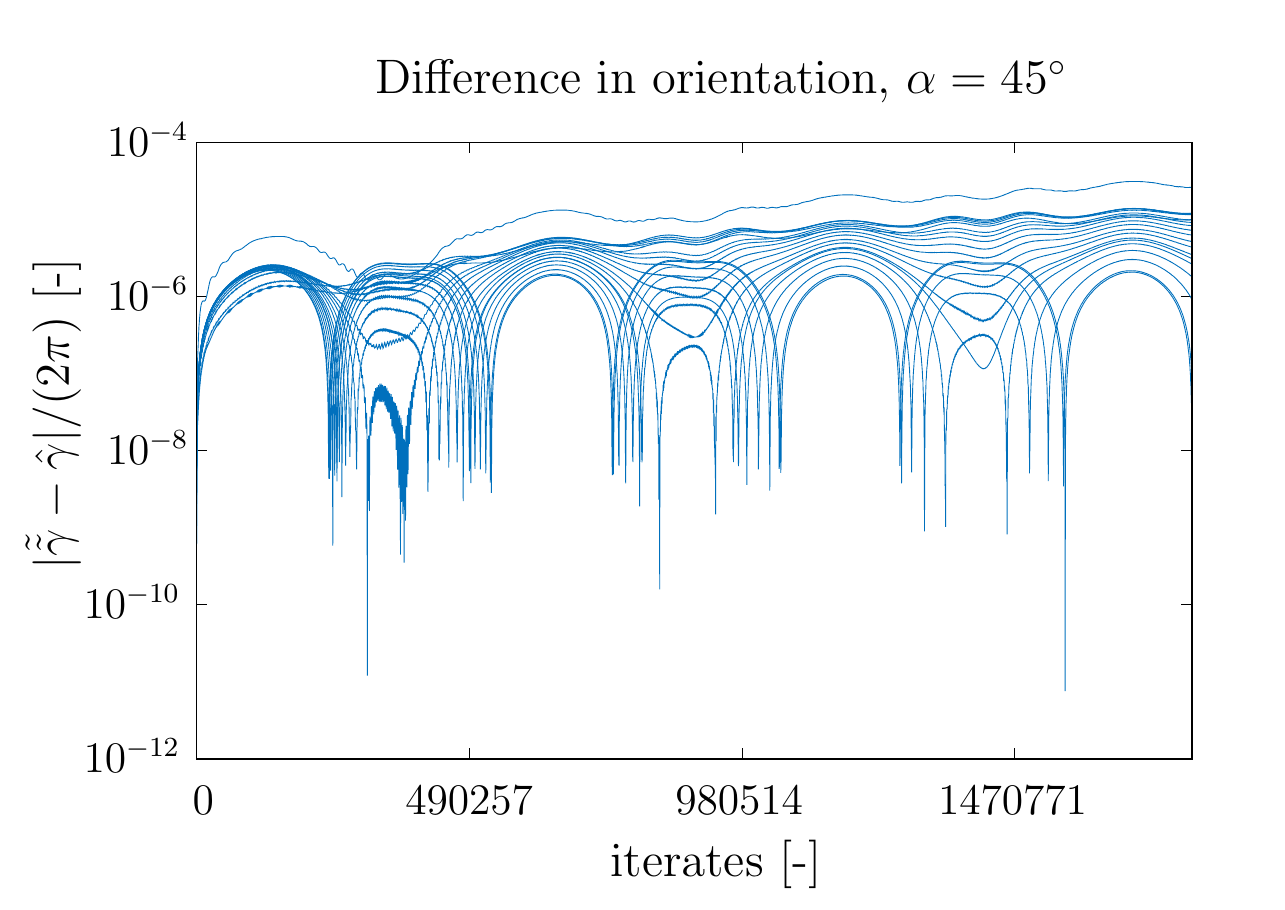}
\includegraphics[width = 0.45\textwidth]{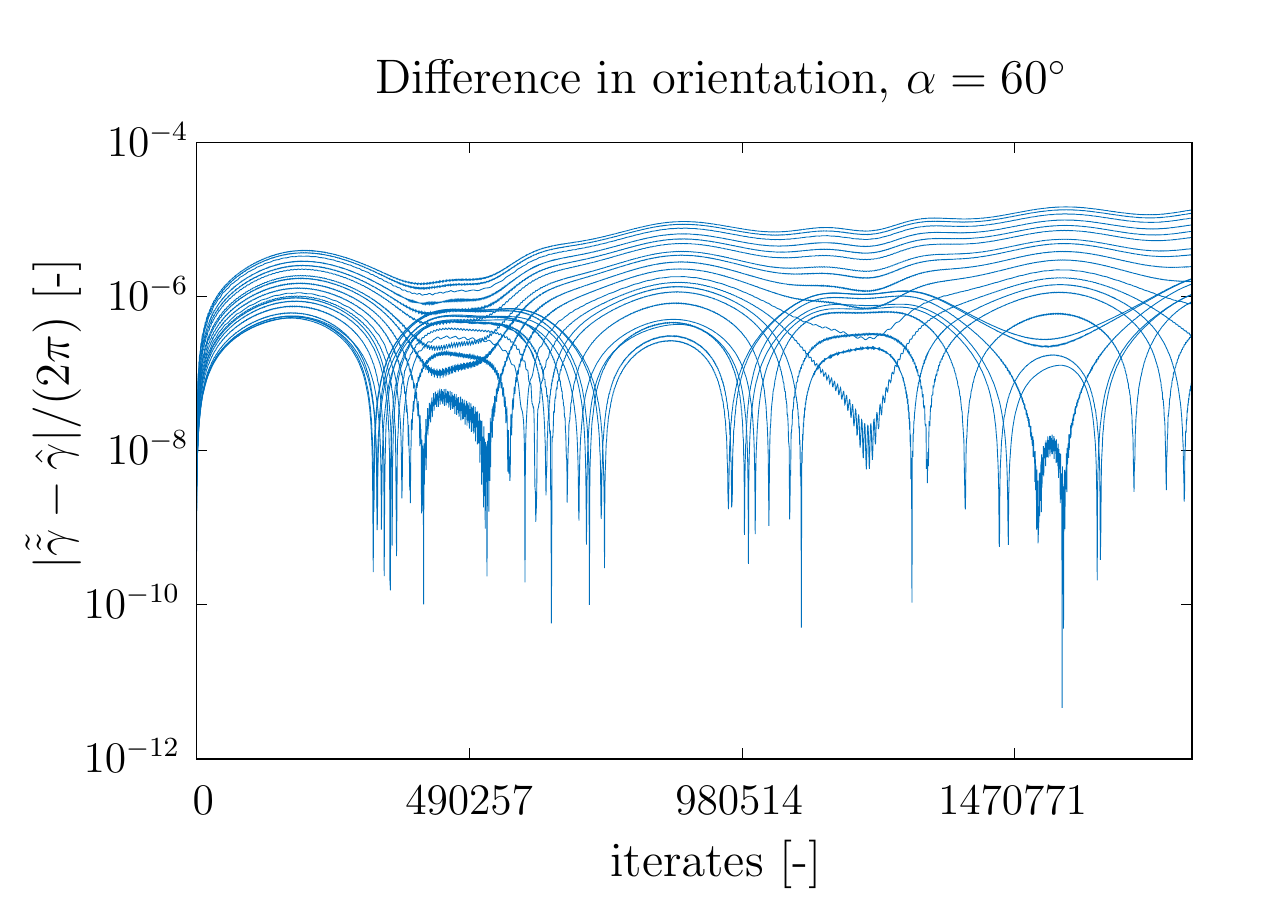}\\
\end{center}
\caption{Difference in adimensional semi-major axes (top) and eccentricity
(middle) and orientation (bottom) of orbits of Eq.~\ref{eq:tointegrate-av} and
Eq.~\ref{eq:tointegrate}, for $\alpha=45^\circ$, $d=0$ m (left) and for
$\alpha=60^\circ$, $d=0$ m (right).}
\label{fig:comparison}
\end{figure}

As above, we study the shape, size and orientation of the osculating ellipse
via $a$, $e$ and $\gamma$ and the results are compared with the orbits
explained in \S~\ref{subsubsec:shapesize}.  On a scale where the whole
evolution along one complete year of integration is seen, the difference
between the evolution of $a,e$ and $\gamma$ in the averaged and original
equations are not noticeable. Their behaviour is such as that shown in
Fig.~\ref{fig:samecc}.  To study the discrepancies between the numerical
results of these two equations, it is convenient to compare the adimensional
semi-major axis, eccentricity and orientation, that shall be denoted as
$\tilde{\tilde{a}}$, $\tilde{\tilde{e}}$ and $\tilde{\tilde{\gamma}}$ for the
averaged equation Eq.~\ref{eq:tointegrate-av} and $\hat{a}$, $\hat{e}$ and
$\hat{\gamma}$ for the original one Eq.~\ref{eq:tointegrate}.  A sample of the
numerical results obtained can be seen in Fig.~\ref{fig:comparison}. It is
important to stress that the comparison is done at the intersection of the
orbits with $\Sigma$. Hence the abscissas are the number of iterates in
$\Sigma$ (a total of $1789440$ in all cases), not the integration time as the
intersection times in $\Sigma$ are not necessarily the same in the two compared
problems. 

The shown results correspond to 19 different values of
$\left<\bPhi\right>$: $j_\al=0:25:480$. Concerning the differences in
semi-major axes, along this first integration year, they are $\Ocal(10^{-2})$;
concerning the differences in eccentricity, they are $\Ocal(10^{-5})$; and
those for the orientation, they are at most $\Ocal(10^{-4})$. Taking into
account that $\eps=\Ocal(10^{-2})$, these numerical results fit within the
accuracy expectations but the time span where these seem to be valid, in the
chosen example, exceed the predictions done by classical theorems that are
summarized in Th.~\ref{teo:averaging}.

As a final remark, note that these numerical results depend strongly on the
chosen orbit, and a specific numerical study to investigate this fact should be
done to assess this dependence. For the studied orbit (as it is indicated in
the numerical results for $A_{\rm eff}$) one can find a non-negligible region
of practical stability around the Sun-pointing attitude so that it makes sense
to consider such structures as potential candidates for passive deorbiting
devices. The studied orbit initial condition example is highly eccentric and
produces a large gravity gradient torque perturbation.  Attitude stability is
expected to be enhanced in high altitude orbits, even all the way down to low
Medium Earth Orbits (MEO), i.e.  2000 km of altitude, and for less eccentric
orbits.

\section{Conclusions}\label{sec:conclusions}

The coupled attitude and orbit motion is known to be a problem with two
characteristic time scales. Although this property is not necessarily obvious
in the form of the equations of motion, in particular for the class of
spacecraft studied in this paper, -a simplified Quasi-Rhombic-Pyramid (QRP)
that consists of a symmetric payload attached to an already deployed sail that
is composed of two reflective panels forming an angle that endows it with
helio-stability properties- the ratio between time scales, $\eps$, can be found
explicitly. An adequate time and phase scaling that depend on $\eps$ can be
performed to highlight the two different characteristic times of the motion.
This parameter, in turn, depends solely on physical quantities that describe
the shape and mass distribution of the spacecraft under consideration.

The attitude dynamics can be understood via a formal treatment of the equations
by studying the fast subsystem of the motion by artificially setting the
parameter $\eps$ to $0$ with equations written in the fast time scale. In the
specific case of the simplified QRP it is a Hamiltonian system of one degree of
freedom with $\Ccal^0$ equations of motion that resembles a pendulum. The
Sun-pointing attitude is a stable equilibrium provided an adequate center of
mass - center of pressure offset and aperture angle are chosen. The phase space
inside the separatrices is foliated by periodic orbits with different periods
but both the Hamiltonian and their equations of motion are only analytic in a
neighborhood of the Sun-pointing attitude. A change of variables of Poincar\'e
type has to be performed inside this neighbourhood to force all attitude
periodic orbits to have the same period. 

The separation of the attitude and orbit dynamics by means of direct averaging
of the orientation of the sail with respect to the sunlight direction provides
a reasonable first approximation of the orbit dynamics. The theoretical
justification of the applicability of averaging results rely mainly on two
independent requirements: smallness of $\eps$ and initial closeness to the
Sun-pointing attitude.  Also, the smoothness hypotheses of classical theorems
only allow to restrict to the case where both panels face sunlight.

Even though in the studied practical examples, where the chosen physical data
is that of a constructible structure, the value of $\eps$ is not necessarily
small, the oscillations around the Sun-pointing direction are fast enough so
that the effect along the whole integration is perceived in the orbit dynamics
as uniform.

This uniform effect can be interpreted as the sail structure consisting of a
flat panel with the same reflectance as the original panels; and whose area is
the area of one of the panels times a factor that depends on the amplitude of
the oscillations and the aperture angle. By tuning these two parameters, this
factor can be made larger than 2, hence giving rise to an effective
area-to-mass ratio that is at least double the amount if we only considered one
of the panels of the structure, but with the advantage of the enhanced
stability provided by the oscillating character.

The formulas for the area factor could be improved by performing the changes of
variable in a way that they have zero average. This would introduce, in
particular, terms due to the gravity gradient in the area factor. The great
disadvantage is that the kind of expressions one would be forced to handle
would be increasingly involved as one goes further in the sequence of averaging
steps. So, one should take into account the trade off between these tedious
computations and the possible improvements in the formula, as the way they are
presented in this contribution may be enough as a first approximation for
practical purposes.\\

This work is intended to be a first step towards the comprehension of the long
term dynamics of QRP. Future lines of research include the extension of the
results of this paper to the 3D structure. In this situation it makes sense to
consider eclipses as a moderate spin along one of the axes of inertia can be
considered to enhance attitude stability, as done in~\cite{FHC17}. Such
assumption would reduce the rotation dynamics from 3 to 2 degrees of freedom.
Other interesting lines of research would include the consideration of damping
effects to enhance the stability properties, and the possible transition
scenarios from SRP dominated to atmospheric drag dominated regions where such
as structure should be used as a drag-sail.

Concerning applications, the long term stability of these families of
spacecraft make them suitable for cargo transportation missions and
interplanetary transfers, see e.g.~\cite{MQ07}.

\subsection*{Acknowledgments}
The research leading to these results has received funding from the European
Research Council (ERC) under the European Unions Horizon 2020 research and
innovation programme as part of project COMPASS (Grant agreement No 679086).
The authors acknowledge the use of the Milkyway High Performance Computing
Facility, and associated support services at the Politecnico di Milano, in the
completion of this work. The datasets generated for this study can be found in
the repository at the link www.compass.polimi.it/$\sim$~publications.
Fruitful conversations with I. Gkolias, M. Lara and C. Sim\'o are also
acknowledged.

\appendix

\section{Adimensional model}\label{sect:adim}

In this section the adimensional set of differential equations are derived.
This is done for the sake of providing a set of equations written in the form
of a fast-slow system where averaging theorems can be formally applied. 

To fix notation, here the superscript $\tilde{\cdot}$ is used to refer to the
original variables with dimensions, not to be confused with those introduced in
the averaging Lemma~\ref{lema:averaging}. The variables without $\tilde{\cdot}$
refer to the adimensional analogues. Let $L$ and $T$ denote the length and time
units, measured in km and sec, respectively. Finally, 
$$
(\dot{\;}):=\frac{\rm d}{{\rm d}\tilde{\tau}},\qquad
('):=\frac{\rm d}{{\rm d}\tau}
$$ 
are used to denote derivatives with respect to dimensional and adimensional
time, respectively.

Consider the relations
\begin{eqnarray}\label{eq:scalingadim}
\tilde{x} = Lx,\quad
\tilde{y} = Ly,\quad\mbox{and}\quad
\tilde{\tau} = T\tau.\quad
\end{eqnarray}

The length $L$ and time $T$ units are chosen to normalize the Kepler problem in
such a way that it simplifies to the vectorial equation
$$
\bm{x}'' = \bm{x}/r^3,\qquad \bm{x} = (x,y)^\top,
                        \quad        r = \sqrt{x^2+y^2}.
$$
The first component of the dimensional Kepler problem reads $\ddot{\tilde{x}} =
- \mu \tilde{x}/\tilde{r}^3$, where $\tilde{r} =
\sqrt{\tilde{x}^2+\tilde{y}^2}$.
From Eq.~\ref{eq:scalingadim},
$\tilde{r} = Lr$, and also
\begin{eqnarray*}
x'' & = & \frac{{\rm d}^2x}{{\rm d}\tau^2} 
      = \frac{T^2}{L}\frac{{\rm d}^2\tilde{x}}{{\rm d}\tilde{\tau}^2} 
      = \frac{T^2}{L}\ddot{\tilde{x}} = - \frac{T^2}{L}\frac{\mu\tilde{x}}{\tilde{r}^3}
      = - \frac{T^2}{L^3}\frac{\mu x}{r^3} \stackrel{!}{=} - \frac{x}{r^3}
\end{eqnarray*}
if we choose
$$
\frac{T^2}{L^3}\mu = 1,
\quad\mbox{or, equivalently}
\quad T = \sqrt{\frac{L^3}{\mu}}.
$$ 
It is clear that for this choice the second component of the adimensional
Kepler problem reads $y'' = -y/r^3$. The choice of $L$ is critical so that the
magnitudes of the constants that appear in the problem are adequate for the
applicability of the known averaging results. This is done in
\S~\ref{sect:numerics}.

Before scaling the equations of motion, it is convenient to shift the angle
$\tilde{\vp}$: let $\tilde{\phi} = \tilde{\vp} - \tilde{\lambda}$, where
$\tilde{\lambda}$ is the argument of latitude the Sun (i.e. the angle between
the $x$ axis and its position vector in Earth-Centered coordinates, measured
counter-clockwise) in its apparent motion around the Earth. 

In the adimensional variables, the attitude equations of motion read
\begin{eqnarray}\label{eq:dynatt-app}
\phi'' & = & \frac{A_s p_{\rm SR} k_{1,1} L^3}{2C\mu(m_b+m_s)}M_1(\phi) 
         +   3\frac{D(\alpha,d)}{C}\frac{1}{r^3}\sin(2\arctan(y/x)-2(\lambda - \phi)),
\end{eqnarray}
as the right hand side of Eq.~\ref{eq:dynatt-dim} has to be multiplied by $T^2$. On the other hand, the translational equations of motion read
\begin{eqnarray}\label{eq:dyntra-app}
\left\{
\begin{array}{rcl}
x'' & = &\displaystyle -\frac{x}{r^3} - \frac{3R^2J_2}{2L^2}\frac{x}{r^5} 
                         + \frac{A_s}{m_b+m_s}\frac{p_{\rm SR}L^2}{\mu}a_x,\\
y'' & = &\displaystyle  -\frac{y}{r^3} - \frac{3R^2J_2}{2L^2}\frac{y}{r^5} 
                         + \frac{A_s}{m_b+m_s}\frac{p_{\rm SR}L^2}{\mu}a_y
\end{array}
\right.,
\end{eqnarray}
where $a_x$ and $a_y$ are given in Eq.~\ref{eq:axay} putting $\vp = \phi + \lambda$.

\subsection*{Conflict of interest}

The authors declare that there are no conflicts of interest regarding the
publication of this paper.

\bibliographystyle{plain}
\bibliography{attorb}
\end{document}